\documentclass[12pt,epsfig]{article}

\catcode`\@=11
\def\@citex[#1]#2{%
\if@filesw \immediate \write \@auxout {\string \citation {#2}}\fi
\@tempcntb\m@ne \let\@h@ld\relax \def\@citea{}%
\@cite{%
  \@for \@citeb:=#2\do {%
    \@ifundefined {b@\@citeb}%
      {\@h@ld\@citea\@tempcntb\m@ne{\bf ?}%
      \@warning {Citation `\@citeb ' on page \thepage \space undefined}}%
      {\@tempcnta\@tempcntb \advance\@tempcnta\@ne%
      \@tempcntb\number\csname b@\@citeb \endcsname \relax%
      \ifnum\@tempcnta=\@tempcntb 
        \ifx\@h@ld\relax%
          \edef \@h@ld{\@citea\csname b@\@citeb\endcsname}%
        \else%
          \edef\@h@ld{\ifmmode{-}\else--\fi\csname b@\@citeb\endcsname}%
        \fi%
      \else
        \@h@ld\@citea\csname b@\@citeb \endcsname%
        \let\@h@ld\relax%
      \fi}%
    \def\@citea{,\penalty\@highpenalty\,}%
  }\@h@ld
}{#1}}

\def\@citeb#1#2{{[#1]\if@tempswa , #2\fi}}
\def\@citeu#1#2{{$^{#1}$\if@tempswa , #2\fi }}
\def\@citep#1#2{{#1\if@tempswa , #2\fi}}

\def\bcites{         
        \catcode`\@=11
        \let\@cite=\@citeb
        \catcode`\@=12
}

\def\upcites{         
        \catcode`\@=11
        \let\@cite=\@citeu
        \catcode`\@=12
}

\def\plaincites{      
        \catcode`\@=11
        \let\@cite=\@citep
        \catcode`\@=12
}

\newcount\hour
\newcount\minute
\newtoks\amorpm
\hour=\time\divide\hour by 60
\minute=\time{\multiply\hour by 60 \global\advance\minute by-\hour}
\edef\standardtime{{\ifnum\hour<12 \global\amorpm={am}%
        \else\global\amorpm={pm}\advance\hour by-12 \fi
        \ifnum\hour=0 \hour=12 \fi
        \number\hour:\ifnum\minute<10 0\fi\number\minute\the\amorpm}}
\edef\militarytime{\number\hour:\ifnum\minute<10 0\fi\number\minute}

\def\draftlabel#1{{\@bsphack\if@filesw {\let\thepage\relax
   \xdef\@gtempa{\write\@auxout{\string
      \newlabel{#1}{{\@currentlabel}{\thepage}}}}}\@gtempa
   \if@nobreak \ifvmode\nobreak\fi\fi\fi\@esphack}
        \gdef\@eqnlabel{#1}}
\def\@eqnlabel{}
\def\@vacuum{}
\def\marginnote#1{}
\def\draftmarginnote#1{\marginpar{\raggedright\scriptsize\tt#1}}
\def\draft{
        \pagestyle{plain}
        \overfullrule=2pt
        \oddsidemargin -.5truein
        \def\@oddhead{\sl \phantom{\today\quad\militarytime} \hfil
        \smash{\Large\sl DRAFT} \hfil \today\quad\militarytime}
        \let\@evenhead\@oddhead
        \let\label=\draftlabel
        \let\marginnote=\draftmarginnote
        \def\ps@empty{\let\@mkboth\@gobbletwo
        \def\@oddfoot{\hfil \smash{\Large\sl DRAFT} \hfil}
        \let\@evenfoot\@oddhead}
        \def\@eqnnum{(\theequation)\rlap{\kern\marginparsep\tt\@eqnlabel}%
        \global\let\@eqnlabel\@vacuum}  }

\def\blackfonts{
        \font\blackboard=msbm10 scaled\magstep1
        \font\blackboards=msbm8
        \font\blackboardss=msbm6
}

\def\nblack{            
        \def\ZZ{{Z \n{10} Z}}
        \def\NN{{N \n{14} N}}
        \def\CC{{C \n{11} C}}
        \def\RR{{R \n{11} R}}
        \def\QQ{{Q \n{12} Q}}
        \def\PP{{P \n{11} P}}
}

\def\prep{         
        \catcode`\@=11
        \input art10.sty
        \catcode`\@=12
        
        \let\small\null
        \def\blackfonts{
                \font\blackboard=msbm10
                \font\blackboards=msbm7
                \font\blackboardss=msbm5
        }
        \let\sl\it
        \twocolumn
        \sloppy
        \voffset=-2.54truecm
        \hoffset=-2.54truecm
        \flushbottom
        \parindent 1em
        \leftmargini 2em
        \leftmarginv .5em
        \leftmarginvi .5em
        \marginparwidth 48pt
        \marginparsep 10pt
        \setlength{\columnsep}{2truecm}
        \setlength{\textwidth}{25.4truecm}
        \setlength{\textheight}{17truecm}
        \baselineskip=16pt
        \oddsidemargin .18truein
        \evensidemargin .17truein
}

\def\eqalign#1{\null\,\vcenter{\openup\jot\m@th
  \ialign{\strut\hfil$\displaystyle{##}$&$\displaystyle{{}##}$\hfil
      \crcr#1\crcr}}\,}
\def\eqalignno#1{\displ@y \tabskip\centering
  \halign to\displaywidth{\hfil$\@lign\displaystyle{##}$\tabskip\z@skip
    &$\@lign\displaystyle{{}##}$\hfil\tabskip\centering
    &\llap{$\@lign##$}\tabskip\z@skip\crcr
    #1\crcr}}

\def\section{\@startsection {section}{1}{\z@}{3.ex plus 1ex minus
 .2ex}{2.ex plus .2ex}{\large\bf}}
\def\subsection{\@startsection{subsection}{2}{\z@}{2.75ex plus 1ex minus
 .2ex}{1.5ex plus .2ex}{\bf}}        

\def\appendix{{\newpage\section*{Appendix}}\let\appendix\section%
        {\setcounter{section}{0}
        \gdef\thesection{\Alph{section}}}\section}

\def\abstract{\if@twocolumn
\section*{Abstract}
\else 
\begin{center}
{\bf Abstract\vspace{-.5em}\vspace{0pt}}
\end{center}
\quotation
\fi}

\catcode`\@=12

\newcommand{\beq}{\begin{equation}}
\newcommand{\eeq}{\end{equation}}
\newcommand{\beqa}{\begin{eqnarray}}
\newcommand{\eeqa}{\end{eqnarray}}
\newcommand{\N}{{\rm N}}
\newcommand{\Z}{{\bf Z}}

\newcommand{\R}{{\bf R}}
\newcommand{\C}{{\bf C}}
\newcommand{\e}{{\rm e}}

\newcommand{\M}{{\rm M}}

\newcommand{\dd}{{\rm d}}

\def\noj#1,#2,{{\bf #1} (19#2)\ }
\def\jou#1,#2,#3,{{\sl #1\/ }{\bf #2} (19#3)\ }
\def\ann#1,#2,{{\sl Ann.\ Physics\/ }{\bf #1} (19#2)\ }
\def\cmp#1,#2,{{\sl Comm.\ Math.\ Phys.\/ }{\bf #1} (19#2)\ }
\def\ma#1,#2,{{\sl Math.\ Ann.\/ }{\bf #1} (19#2)\ }
\def\ng#1,#2,{{\sl Nagoya.\ Math.\ J.\/ }{\bf #1} (19#2)\ }
\def\jd#1,#2,{{\sl J.\ Diff.\ Geom.\/ }{\bf #1} (19#2)\ }
\def\invm#1,#2,{{\sl Invent.\ Math.\/ }{\bf #1} (19#2)\ }
\def\cq#1,#2,{{\sl Class.\ Quantum Grav.\/ }{\bf #1} (19#2)\ }
\def\cqg#1,#2,{{\sl Class.\ Quantum Grav.\/ }{\bf #1} (19#2)\ }
\def\ijmp#1,#2,{{\sl Int.\ J.\ Mod.\ Phys.\/ }{\bf A#1} (19#2)\ }
\def\jmphy#1,#2,{{\sl J.\ Geom.\ Phys.\/ }{\bf #1} (19#2)\ }
\def\jams#1,#2,{{\sl J.\ Amer.\ Math.\ Soc.\/ }{\bf #1} (19#2)\ }
\def\grg#1,#2,{{\sl Gen.\ Rel.\ Grav.\/ }{\bf #1} (19#2)\ }
\def\mpl#1,#2,{{\sl Mod.\ Phys.\ Lett.\/ }{\bf A#1} (19#2)\ }
\def\nc#1,#2,{{\sl Nuovo Cim.\/ }{\bf #1} (19#2)\ }
\def\np#1,#2,{{\sl Nucl.\ Phys.\/ }{\bf B#1} (19#2)\ }
\def\pl#1,#2,{{\sl Phys.\ Lett.\/ }{\bf #1B} (19#2)\ }
\def\pla#1,#2,{{\sl Phys.\ Lett.\/ }{\bf #1A} (19#2)\ }
\def\pr#1,#2,{{\sl Phys.\ Rev.\/ }{\bf #1} (19#2)\ }
\def\prd#1,#2,{{\sl Phys.\ Rev.\/ }{\bf D#1} (19#2)\ }
\def\prl#1,#2,{{\sl Phys.\ Rev.\ Lett.\/ }{\bf #1} (19#2)\ }
\def\prp#1,#2,{{\sl Phys.\ Rept.\/ }{\bf #1C} (19#2)\ }
\def\ptp#1,#2,{{\sl Prog.\ Theor.\ Phys.\/ }{\bf #1} (19#2)\ }
\def\ptpsup#1,#2,{{\sl Prog.\ Theor.\ Phys.\/ Suppl.\/ }{\bf #1} (19#2)\ }
\def\rmp#1,#2,{{\sl Rev.\ Mod.\ Phys.\/ }{\bf #1} (19#2)\ }
\def\yadfiz#1,#2,#3[#4,#5]{{\sl Yad.\ Fiz.\/ }{\bf #1} (19#2) #3%
\ [{\sl Sov.\ J.\ Nucl.\ Phys.\/ }{\bf #4} (19#2) #5]}
\def\zh#1,#2,#3[#4,#5]{{\sl Zh.\ Exp.\ Theor.\ Fiz.\/ }{\bf #1} (19#2) #3%
\ [{\sl Sov.\ Phys.\ JETP\/ }{\bf #4} (19#2) #5]}

\def\beq{\begin{equation}}
\def\eeq{\end{equation}}
\def\beqar{\begin{eqnarray}}
\def\eeqar{\end{eqnarray}}

\newcommand{\be}{\begin{equation}}
\newcommand{\ee}{\end{equation}}
\newcommand{\bea}{\begin{eqnarray}}
\newcommand{\eea}{\end{eqnarray}}

\def\nfrac#1#2{{\displaystyle{\vphantom1\smash{\lower.5ex\hbox{\small$#1$}}%
        \over\vphantom1\smash{\raise.25ex\hbox{\small$#2$}}}}}

\def\n#1{\mskip-#1mu}

\def\to{\rightarrow}

\def\lae{\mathrel{\mathop{\smash{\lower .5 ex \hbox{$\stackrel<\sim$}}}}}
\def\lae{\mathrel{\mathop{\smash{\lower .5 ex \hbox{$\stackrel>\sim$}}}}}

\def\Tr{{\rm Tr}}
\def\l:{\mathopen{:}\,}
\def\r:{\,\mathclose{:}}

\catcode`\@=11
\def\theequation{\arabic{equation}}
\catcode`\@=12

\nblack
\bcites

\nblack

\catcode`\@=11
\def\theequation{\thesection.\arabic{equation}}
\@addtoreset{equation}{section}
\@addtoreset{footnote}{section}
\@addtoreset{footnote}{subsection}
\catcode`\@=12

\newcommand{\beqn}{\begin{equation}}
\newcommand{\eeqn}{\end{equation}}
\newcommand{\beqnarray}{\begin{eqnarray}}
\newcommand{\eeqnarray}{\end{eqnarray}}
\newcommand {\bear} [1] {\begin {array} {#1}}
\newcommand {\ear} {\end {array}}

\newcommand{\CP}{{\bf C}{\rm P}}

\newcommand {\beqarn} {\begin{eqnarray*}}
\newcommand {\eeqarn} {\end{eqnarray*}}

\begin{document}

\begin{titlepage}

\begin{center}
\today
\hfill HUTP-00/A005\\
\hfill                  hep-th/0002222

\vskip 1.5 cm
{\large \bf Mirror Symmetry}
\vskip 1 cm 
{Kentaro Hori and Cumrun Vafa}\\
\vskip 0.5cm
{\sl Jefferson Physical Laboratory,
Harvard University\\
Cambridge, MA 02138, U.S.A.\\}

\end{center}

\vskip 0.5 cm
\begin{abstract}
We prove mirror symmetry for supersymmetric sigma models on Kahler
manifolds in $1+1$ dimensions. The proof involves establishing
the equivalence of the gauged linear sigma model, embedded in a theory
with an enlarged gauge symmetry, with a Landau-Ginzburg theory of Toda type.
Standard $R\rightarrow 1/R$ duality and dynamical generation of
superpotential by vortices are crucial in the derivation. This provides
not only a proof of mirror symmetry in the case of (local and global)
Calabi-Yau manifolds, but also for sigma models on manifolds with
positive first Chern class, including deformations of the action by
holomorphic isometries.

\end{abstract}

\end{titlepage}

\section{Introduction}

One of the most beautiful symmetries in string theory is
the radius inversion symmetry $R\rightarrow 1/R$  of a
circle, known as T-duality \cite{orig}.
This is the symmetry which exchanges winding modes on a circle
with momentum modes on the dual circle.
This symmetry has been the underlying
motivation for many of the subsequent dualities discovered
in string theory and in quantum field theories.
In particular, over a decade ago,
 it was conjectured in \cite{LVW,Dixon}
that a similar duality might exist in the context of
string propagation on Calabi-Yau manifolds, where the role
of the complex deformations on one manifold get exchanged
with the Kahler deformations on the dual manifold. The
pairs of manifolds satisfying this symmetry are known
as mirror pairs, and this duality is also called
mirror symmetry.

There has been a lot of progress since the original formulation
of this conjecture, in its support.  In particular many examples of this
phenomenon were found \cite{GP,candd}.
The intermediate step in the derivation
for this class of examples involved the construction of conformal
field theory for certain Calabi-Yau's \cite{Gepner}
and their identification with certain Landau-Ginzburg models
\cite{VW,Martinec,GVW}.
This connection was further elucidated in \cite{phases}
where it was shown that the linear
sigma model is a powerful tool in the study of strings
propagating on a Kahler manifold.

It was shown in \cite{candet}
how mirror symmetry can be used very effectively 
to gain insight into non-perturbative
effects involving worldsheet instantons.  Roughly speaking this
amounts to counting the number of holomorphic curves in
a Calabi-Yau
manifold.
This made the subject also interesting for algebraic geometers
in the context of enumerative geometry.  Motivated
by the existing examples some general class of mirror pairs
were formulated by mathematicians using
toric geometry \cite{Batyrev}.
Moreover a program to prove the rational curve counting formula,
predicted by mirror symmetry,
from the view point of localization and virtual fundamental cycles
was initiated in \cite{Kontsevich,ES,LT}
and was pushed to completion in \cite{G,yaucol,Bini,Pan}.
For reviews of various
aspects of mirror symmetry see 
\cite{yaubook};
for mathematical aspects of mirror symmetry see the excellent book
\cite{katcox}.

The question of a proof of mirror symmetry
and its relation with T-duality, which was its original motivation,
was further pursued in \cite{VWit} where it was shown
that for certain toroidal orbifold models mirror symmetry reduces to 
T-duality.  More generally, by following the prediction of
the map of D-branes under mirror symmetry, it was argued in
\cite{SYZ} that mirror symmetry should reduce to T-duality in a more
general context.  Furthermore,
it was shown in \cite{conv}, how the general
suggestion for construction
of mirror pairs proposed using toric geometry \cite{Batyrev} 
can be intuitively related to T-duality.

Mirror symmetry has also been extended from the case of Calabi-Yau
sigma models to more general cases.  On the one hand there
are proposals as to what the mirror theories are in the case
of certain sigma models with positive first Chern class
\cite{FI,CV,G2,EHY,EHX,G}. On the other hand
there are proposals for what the mirror of non-compact
Calabi-Yau manifolds are \cite{kkv,kmv,klyz}.

The aim of this paper is to present a proof of mirror
symmetry for all cases proposed thus far. The proof
depends crucially on establishing a dual description of
$(2,2)$ supersymmetric gauge theories in 1+1 dimensions.  The
dual theory is found using the idea analogous to Polyakov's
model of confinement in
 quantum electrodymanics in $2+1$ dimensions \cite{Polyakov}.
He considered a $U(1)$ gauge theory which includes magnetic monopoles
playing the role of instantons.
$U(1)$ Maxwell theory of gauge coupling constant $e$ in $2+1$ dimensions
is dual to
the theory of a periodic scalar field $\sigma\equiv\sigma +2\pi$
with the Lagrangian $e^2|\dd\sigma|^2$.
The gas of instantons and anti-instantons
with a long range interaction between them generates a potential term
\beq
U(\sigma)=\mu^3\,\cos(\sigma)
\label{poly}
\eeq
in the effective Lagrangian in terms of the dual variable $\sigma$,
where $\mu$ is the mass scale determined by $e$
and the monopole size.
One sees from this that a mass gap is generated and that
an electric flux is confined into a thin tube.
We note that the description in terms of the dual variable $\sigma$ was
essential in this argument.

In supersymmetric field theories,
instanton computation can be used to obtain exact results
for some important physical quantities.
For some of the striking examples, see
\cite{AHW,ADS,NSVZ,Amati,SW,SW3d}.
Among these, \cite{AHW} and \cite{SW3d} treat supersymmetric 
gauge theories in $2+1$ dimensions and
the effective theory is described
in terms of the dual variable as in \cite{Polyakov}.
Also, in \cite{SW}, duality between
vector and vector in $3+1$ dimensions
was used to solve the problem in an essential way.

We apply an analogous idea to study the long
distance behaviour of $(2,2)$ gauge theories, making
use of instantons which are vortices \cite{NO} in this case.
We dualize the phase of the charged fields in the sense of
$R\rightarrow 1/R$ duality and 
 describe the low energy effective theory
in terms of the dual variables. 
We will see that a superpotential is dynamically generated
by the instanton effect, as in \cite{Polyakov,AHW,ADS}, 
and we can exactly determine the (twisted) F-term
part of the effective Lagrangian.
To be specific, let us consider a $(2,2)$ supersymmetric
$U(1)$ gauge theory with
$N$ chiral multiplets of charge $Q_i$ ($i=1,\ldots,N$).
In addition to the gauge coupling, the theory has two parameters:
Fayet-Iliopoulos and Theta parameters. They are combined into a single
complex parameter $t$ and appear in the twisted superpotential
as $-t\Sigma$ where $\Sigma$ is
the twisted chiral field which is the field strength
of the gauge multiplet (and includes the scalar, the gaugino
and the field strength).
Each charged chiral field is sent by the duality on its phase
to a twisted chiral field $Y_i$ which
is a neutral periodic variable $Y_i\equiv Y_i+2\pi i$ that couples to
the field strength as a dynamical Theta angle $Q_iY_i\Sigma$.
The exact twisted superpotential we will find is given by
\beq
\widetilde{W}=\Sigma\left(\sum_{i=1}^NQ_iY_i-t\right)
\,+\,\mu\,\sum_{i=1}^N\e^{-Y_i},
\label{result}
\eeq
where $\mu$ is a scale parameter.
The term proportional to $\Sigma$ is the one that
appears already at the dualization process. The exponentials of
$Y_i$'s are the ones that are generated by instanton effect.
When $\sum_iQ_i\ne 0$, $\mu$ is a scale required to renormalize
the FI parameter $t$. In this case, a combination
of $t$ and $\mu$ is a fake and only one dimensionful parameter
$\Lambda=\mu\e^{-t/\sum_iQ_i}$
is the real parameter of the theory. This is
the standard dimensional transmutation.
In the case where $\sum_iQ_i=0$,
$t$ is the dimensionless parameter of the theory and 
$\mu$ is a fake
as it can be absorbed by a field redefinition.

Using the connection between $U(1)$ gauge theories
with matter and sigma models on Kahler manifolds \cite{phases}
we then relate the above result to the statement of
mirror symmetry\footnote{The idea to use the gauged linear
sigma model to derive mirror symmetry was also considered
in \cite{MP2}.}.  In particular we find that the mirror
to a sigma model is a Landau-Ginzburg model.
In the case of Calabi-Yau manifolds this can also
be related to the sigma model on another Calabi-Yau
manifold by the equivalence
of sigma models and Landau-Ginzburg models.
In the case of manifolds with non-zero first Chern class,
however, this is not possible
(we consider only manifolds with non-negative first Chern class
since otherwise the sigma model would not be well-defined):
the axial $U(1)$ R-symmetry is borken by an anomaly and
therefore the vector $U(1)$ R-symmetry of the mirror
theory must be broken by an inhomogenious superpotential.
Likewise, since the vector $U(1)$ R-symmetry of the original
non-linear sigma model
is an exact symmetry, the mirror manifold (on which
the Landau-Ginzburge superpotential is defined) must always
be Calabi-Yau so that the axial R-symmetry is unbroken.

A typical example of manifolds of positive first Chern class
is $\CP^1$.
It has been observed that
the supersymmetric $\CP^1$ sigma model is mirror to the $N=2$
sine-Gordon theory
which is a sigma model on a cylinder $\C^{\times}$
with a sine-Gordon superpotential \cite{FI,CV,G2,EHY,EHX}.
The two theories have $U(1)\times \Z_4$
vector-axial (or axial-vector) R-symmetries.
Both have two massive vacua which spontaneously
breaks $\Z_4$ to $\Z_2$.
Moreover, soliton spectrum and the scattering matrix have been observed
to agree \cite{FI}.
Actually, this mirror symmetry is the first non-trivial
one that can be derived by our method.
The linear sigma model for $\CP^1$ is a $U(1)$ gauge theory with
two chiral multiplets of charge $Q_1=Q_2=1$.
By integrating out the $\Sigma$ field from (\ref{result}),
we obtain the constraint $Y_1+Y_2=t$ which can be solved by
$Y_1=Y+t/2, Y_2=-Y+t/2$.
Then, the superpotential (\ref{result})
takes the form
\beq
\widetilde{W}(Y)=2\Lambda\, \cosh(Y),
\eeq
which is nothing but the sine-Gordon potential!
In fact, the mirror symmetry of other models,
including the
conformal field theories based on Calabi-Yau sigma models,
can be derived in a uniform way as
a natural generalization of this example. 
Furthermore,
the mirror theory provides an effective way to classify
the vacua of the theory
and to identify where they flow to in the infra-red limit.
 For example,
for a degree $d$ hypersurface on $\CP^{N-1}$ of size $t$,
we find the mirror
to be the orbifold of the Landau-Ginzburg model with
the superpotential
\beq
W=X_1^d+\cdots +X_N^d+e^{t/d} X_1\cdots X_N
\label{LGG}
\eeq
by the group $(\Z_d)^{N-1}$
which acts on the indivisual
fields $X_i$ by multiplication by $d$-th roots of unity, 
in such a way that $X_1\cdots X_N$ is invariant.
Note that the special case $d=N$ gives one the
proposed mirror
of sigma models on Calabi-Yau hypersurfaces
(Greene-Plesser construction \cite{GP}).
In the case with $d<N$, the hypersurface has a positive first Chern
class and the sigma model is asymptotic free \cite{AF}.
The mirror theory given above shows that
there are $N-d$ massive vacua at non-zero $X_i$'s
and another vacuum at $X_i=0$.
For $d>2$, the vacuum at $X_i=0$ flows to a non-trivial
fixed point described by the LG orbifold with the same group
$(\Z_d)^{N-1}$ and the
superpotential (\ref{LGG}) with the last term being dropped
as it is irrelevant at low energies.
In this example, the original linear sigma model actually leads to
the description of the low energy theory in terms of another
LG orbifold \cite{phases,QFTM},
with the same superpotential but with a different group
--- a single $\Z_d$ acting on $X_i$'s uniformly.
In fact, our result reproduces the mirror symmetry of
LG orbifolds \cite{LGO}.
In general, however, as the $\CP^1$ example shows,
our method provide information which
is hardly available in the original model.

The organization of this paper is as follows:
In section 2 we review certain aspects of mirror symmetry with
emphasis on its interplay with supersymmetry.  
We present the $N=2$ supersymmetric version of T-duality
which plays a crucial role for us later in the paper.  
In section 3
we consider dynamical aspects of $N=2$ gauge theories and establish
their equivalence with the above mentioned LG theories.
In section 4 we review aspects of linear sigma model
(i.e. gauge theory/sigma model connection).  In section 5
we use the results in section 3 and present a proof
of mirror symmetry.  Also in
this section we elaborate on what we mean by ``proving'' mirror
symmetry.  In section 6 we discuss
some aspects of D-branes in the
context of LG theories.  This
elucidates the relation between mirror symmetry for
local (non-compact)
and global (i.e. compact) sigma models which is discussed
in section 7.  Also the mirror of
complete intersections in toric varieties
are discussed there. In section 8 we discuss
some possible directions for future work.
In appendix A we present a conjectured generalization of our
results to the case of complete intersections in Grassmannians
and flag varieties.  In appendix B some aspects of supersymmetry
transformations needed in this paper are summarized.

\section{Mirror Symmetry Of $N=2$ Theories In Two Dimensions}

\newcommand{\bQ}{\overline{Q}}
\newcommand{\bD}{\overline{D}}
\newcommand{\bchi}{\overline{\chi}}
\newcommand{\bpsi}{\overline{\psi}}
\newcommand{\btheta}{\overline{\theta}}
\newcommand{\blambda}{\overline{\lambda}}
\newcommand{\bsigma}{\overline{\sigma}}
\newcommand{\bSigma}{\overline{\Sigma}}
\newcommand{\bPhi}{\overline{\Phi}}
\newcommand{\bj}{{\bar\jmath}}
\newcommand{\bi}{{\bar\imath}}
\newcommand{\bTheta}{\overline{\Theta}}

In this section, we review some basic facts and fix notations on
$(2,2)$ supersymmetric field theories in $1+1$ dimensions.
We also define the notion of mirror symmetry and present some examples.
In particular, we describe in detail the standard $R\to 1/R$ duality
and show how it can be viewed as
mirror symmetry in the case of complex torus.

\subsection{Supersymmetry}

\subsection*{\it $(2,2)$ Supersymmetry Algebra}

$(2,2)$ supersymmetry algebra is generated by
four supercharges
$Q_{\pm},\bQ_{\pm}$, space-time translations $P$, $H$
and rotation $M$, and
the generators $F_V$ and $F_A$ of two R-symmetries
$U(1)_V$ and $U(1)_A$. These obey the following
(anti-)commutation relations:
\beqa
&&~~~~~Q_+^2=Q_-^2=\bQ_+^2=\bQ_-^2=0,
\label{susy1}\\
&&~~~~~~~\{Q_{\pm},\bQ_{\pm}\}=2(H\mp P),
\label{susy2}\\
&&\{\bQ_+,\bQ_-\}=2Z,~~
\{Q_+,Q_-\}=2Z^*,
\label{susy3}\\
&&\{Q_-,\bQ_+\}=2\widetilde{Z},~~
\{Q_+,\bQ_-\}=2\widetilde{Z}^*,
\label{susy4}\\
&&~{[}M,Q_{\pm}]=\mp Q_{\pm},~~[M,\bQ_{\pm}]=\mp\bQ_{\pm},
\label{susy5}\\
&&~{[}F_V,Q_{\pm}]=-Q_{\pm},~~[F_V,\bQ_{\pm}]=\bQ_{\pm},
\label{susy6}\\
&&~{[}F_A,Q_{\pm}]=\mp Q_{\pm},~~[F_A,\bQ_{\pm}]=\pm \bQ_{\pm}.
\label{susy7}
\eeqa
The hermiticity of the generators is dictated by
\beq
Q_{\pm}^{\dag}=\bQ_{\pm}.
\label{herm}
\eeq
In the above expressions, $Z$ and $\widetilde{Z}$ are central charges.
The algebra with $Z=\widetilde{Z}=0$
can be obtained by dimensional reduction of the four-dimensional
$N=1$ supersymmetry algebra. $U(1)_V$ comes from the R-symmetry in four
dimensions and $U(1)_A$ is the rotation along the reduced directions.

It is not always the case that the two $U(1)$ R-symmetries are
the symmetry
of a given theory, except in super{\it conformal} field
theories where they both {\it must} be the symmetry.
In the class of theories we will consider in this paper,
at least one of them is a symmetry and the other may or may
not be
 broken to a discrete subgroup.
 
A central charge can be non-zero
if there is a soliton that interpolates different vacua
and/or if the theory has a continuous abelian symmetry. 
As the name suggests,
it must commute with the R-symmetry generators as well.
In particular, $Z$ ($\widetilde{Z}$) must always be zero
in a theory where $U(1)_V$ ($U(1)_A$) is unbroken.
In superconformal field theory, both must be vanishing.

\subsection*{\it Superfields and Supersymmetric Lagrangians}

Fields in a supermultiplet
can be combined into a single function, a {\it superfield},
of the superspace coordinates $x^0,x^1,\theta^{\pm},\btheta^{\pm}$.
Supersymmetry generators
$Q_{\pm}$, $\bQ_{\pm}$ act on the superfields as the derivatives
\beq
Q_{\pm}
={\partial\over\partial\theta^{\pm}}
+i\btheta^{\pm}\left({\partial\over\partial x^0}
\pm{\partial\over\partial x^1}\right),~~
\bQ_{\pm}
=-{\partial\over\partial\btheta^{\pm}}
-i\theta^{\pm}\left({\partial\over\partial x^0}
\pm{\partial\over\partial x^1}\right).
\eeq
These commute with another set of derivatives
\beq
D_{\pm}
={\partial\over\partial\theta^{\pm}}
-i\btheta^{\pm}\left({\partial\over\partial x^0}
\pm{\partial\over\partial x^1}\right),~~
\bD_{\pm}
=-{\partial\over\partial\btheta^{\pm}}
+i\theta^{\pm}\left({\partial\over\partial x^0}
\pm{\partial\over\partial x^1}\right).
\eeq
R-symmetries
act on a superfield ${\cal F}(x,\theta^{\pm},\btheta^{\pm})$
as
\beqa
&&\e^{i\alpha F_V}{\cal F}(x,\theta^{\pm},\btheta^{\pm})
=\e^{iq_V\alpha}
{\cal F}(x,\e^{-i\alpha}\theta^{\pm},\e^{i\alpha}\btheta^{\pm}),
\\
&&\e^{i\alpha F_A}{\cal F}(x,\theta^{\pm},\btheta^{\pm})
=\e^{iq_A\alpha}
{\cal F}(x,\e^{\mp i\alpha}\theta^{\pm},
\e^{\pm i\alpha}\btheta^{\pm}),
\eeqa
where $q_V$ and $q_A$ are the vector and axial R-charges of ${\cal F}$.

The basic representations of the supersymmetry algebra are
chiral and twisted chiral multiplets which
both consist of a complex scalar and a Dirac fermion.
These are represented by chiral (or $cc$) superfield and twisted chiral
(or $ac$) superfield respectively.
A chiral superfield $\Phi$ satisfies
\beq
\bD_{\pm}\Phi=0,
\eeq
and can be expanded
as
\beq
\Phi=\phi+\sqrt{2}\theta^+\psi_++\sqrt{2}\theta^-\psi_-
+2\theta^+\theta^-F+\cdots,
\eeq
where $F$ is a complex auxiliary field and
 $+\cdots$ involves only the derivatives of $\phi,\psi_{\pm}$.
The hermitian conjugate of $\Phi$ is
an anti-chiral (or $aa$) superfield
$D_{\pm}\overline{\Phi}=0$.
A twisted chiral superfield $Y$ satisfies \cite{Gates}
\beq
\bD_+Y=D_-Y=0,
\eeq
and can be expanded as
\beq
Y=y+\sqrt{2}\theta^+\bchi_+
+\sqrt{2}\btheta^-\chi_-
+2\theta^+\btheta^-G+\cdots.
\eeq
where $G$ is a complex auxiliary field and
$+\cdots$ involves only the derivatives of
the component fields.
The hermitian conjugate of $Y$
is a twisted anti-chiral (or $ca$)
superfield $D_+\overline{Y}=\bD_-\overline{Y}=0$.

We also introduce a vector multiplet.
It consists of a vector field $v_{\mu}$, Dirac fermions
$\lambda_{\pm}$, $\blambda_{\pm}$ which are conjugate to each other,
and a complex scalar $\sigma$ in the adjoint representation of the gauge
group.
It is represented in a vector superfield $V$ which is expanded
(in the Wess-Zumino gauge) as
\beqa
V&=&
\theta^-\btheta^-(v_0-v_1)+\theta^+\btheta^+(v_0+v_1)
-\theta^-\btheta^+\sigma
-\theta^+\btheta^-\bsigma
\label{Vect}\\
&&+\sqrt{2}i\theta^-\theta^+(\btheta^-\blambda_-+
\btheta^+\blambda_+)
+\sqrt{2}i\btheta^+\btheta^-(\theta^-\lambda_-+
\theta^+\lambda_+)+2\theta^-\theta^+\btheta^+\btheta^-D
\nonumber
\eeqa
where $D$ is a real auxiliary field. Using the gauge covariant derivatives
${\cal D}_{\pm}=\e^{-V}D_{\pm}\e^V$,
$\overline{\cal D}_{\pm}=\e^{V}\bD_{\pm}\e^{-V}$,
we can define the field strength as
\beqa
\Sigma&=&{1\over 2}\{\overline{\cal D}_+,{\cal D}_-\}\\
&=&\sigma+i\sqrt{2}\theta^+\blambda_+
-i\sqrt{2}\btheta^-\lambda_-+2\theta^+\btheta^-(D-iF_{01})
+\cdots,\nonumber
\eeqa
where $F_{01}$ is the curvature of $v_{\mu}$. This is a twisted
chiral (covariant) superfield $\overline{\cal D}_+\Sigma
={\cal D}_-\Sigma=0$.
The supersymmetry transformation of the component fields in this gauge
(based on \cite{WB,grassm,phases})
is recorded for convenience in Appendix B.

Supersymmetric Lagrangian can be obtained from
integrations over suitable fermionic coordinates.
There are
D-term, F-term, and twisted F-term.
D-term is for arbitrary superfields ${\cal F}_i$
and is given by
\beq
\int\dd^4\theta \,K({\cal F},\overline{\cal F})
={1\over 4}\int\dd\theta^+\dd\theta^-\dd\btheta^-\dd\btheta^+
K({\cal F},\overline{\cal F}),
\eeq
where $K({\cal F},\overline{\cal F})$ is an arbitrary
real function of ${\cal F}_i$'s.
This is classically R-invariant under any assignment of
R-charges.
F-term is for chiral superfields $\Phi_i$ and is given by
\beq
\int\dd^2\theta \,W(\Phi)+c.c.
={1\over 2}\int\dd\theta^-\dd\theta^+ W(\Phi)|_{\btheta^{\pm}=0}
+{1\over 2}\int\dd\btheta^+
\dd\btheta^- \overline{W}(\overline{\Phi})|_{\theta^{\pm}=0}
\label{Fterm}
\eeq
where $W(\Phi)$ is a holomorphic function of 
$\Phi_i$'s and is called a superpotential.
This is invariant under vector and axial
R-symmetries only when it is possible to
assign R-charges to $\Phi_i$'s so that $W(\Phi)$ has
vector and axial charge 2 and 0 respectively.
Twisted F-term is for twisted chiral superfields $Y_i$ and is given by
\beq
\int\dd^2\widetilde{\theta} \,\widetilde{W}(Y)+c.c.
={1\over 2}\int\dd\btheta^-\dd\theta^+
\widetilde{W}(Y)|_{\btheta^+=\theta^-=0}
+{1\over 2}\int\dd\btheta^+\dd\theta^-
\overline{\widetilde{W}}(\overline{Y})|_{\theta^+=\btheta^-=0}
\label{twFterm}
\eeq
where $\widetilde{W}(Y)$ is a holomorphic function of 
$Y_i$'s and is called a twisted superpotential.
For R-invariance, it is required that
R-charges can be assigned to $Y_i$'s so that $\widetilde{W}(Y_i)$ has
vector and axial charge 0 and 2 respectively.

\subsection*{\it Non-linear Sigma Model and Potentials}

Supersymmetric non-linear sigma model on a Kahler manifold $X$
is one of the main subjects of the present paper.
It is described by a set of chiral superfields
$\Phi^i$ $(i=1,\ldots,n)$ representing complex coordinates
of $X$.
Let
$g_{i\bj}=\partial^2K/\partial\phi^i\partial\bar\phi^{\bj}$
be the Kahler metric of $X$ where $K(\phi,\bar\phi)$
is a Kahler potential.
The Lagrangian can be given by the D-term
$\int\dd^4\theta\,K(\Phi,\bar\Phi)$ which is expressed
in terms of the component fields as
\beqa
L_K&=&-g_{i\bj}\partial^{\mu}\phi^i\partial_{\mu}\bar\phi^{\bj}
+ig_{i\bj}\,\bpsi_-^{\bj}(D_0+D_1)\psi^i_-
+ig_{i\bj}\,\bpsi_+^{\bj}(D_0-D_1)\psi^i_+
\nonumber\\
&&
+R_{i\bar kj\bar l}\,\psi_+^i\psi_-^j\bpsi_-^{\bar k}\bpsi_+^{\bar l},
\label{Kterm}
\eeqa
after the auxiliary fields are eliminated.
Here,
$R_{i\bar kj\bar l}$
is the curvature tensor with respect to the Levi-Civita connection
$\Gamma^i_{lj}=g^{i\bar k}\partial_lg_{j\bar k}$ of $X$.
Also,
\beq
D_{\mu}\psi_{\pm}^i
=\partial_{\mu}\psi_{\pm}^i+\partial_{\mu}
\phi^l\Gamma^i_{lj}\psi_{\pm}^j
\label{cov}
\eeq
is the covariant derivative with respect to the connection induced
on the worldsheet.

If $X$ has a holomorphic function $W(\phi)$,
the non-linear sigma model can be deformed by the F-term
with superpotential $W(\Phi)$,
\beq
L=\int\dd^4\theta\,K(\Phi,\bPhi)
+{1\over 2}\left(\int\dd^2\theta\,W(\Phi)\,+\,c.c.\right).
\eeq
Eliminating the auxiliary fields,
we obtain the Lagrangian $L_K+L_W$ where
$L_K$ is given in (\ref{Kterm}) and
the deformation term is
\beq
L_W=-{1\over 4}g^{\bj i}\partial_{\bj}\overline{W}\partial_i W
-{1\over 2}(D_i\partial_jW)\psi_+^i\psi_-^j
-{1\over 2}(D_{\bi}\partial_{\bj}\overline{W})\bpsi_-^{\bi}\bpsi_+^{\bj},
\eeq
in which
$D_i\partial_jW=\partial_i\partial_jW-\Gamma^l_{ij}\partial_lW$.
Note that a non-trivial holomorphic function $W(\phi)$ exists
only when $X$ is non-compact.
In some cases, the deformation discretize the energy spectrum
which would be continuous without $L_W$
because of the non-compactness.
The character of the theory therefore depends largely 
on the asymptotic behaviour of the superpotential.

If $X$ has a holomorphic isometry generated by a holomorphic
vector field $V$, the sigma model can
be deformed by another kind of potential term \cite{AGF}.
This is obtained by first gauging the isometry as in \cite{BagW},
taking the weak coupling limit,
and freezing the vector multiplet fields
at $\sigma=\widetilde{m}$, $v_{\mu}=0$, and
$\lambda_{\pm}=\blambda_{\pm}=0$.
A description in $(2,2)$ superspace was considered in \cite{gates2}.
The deformation term is
\beq
L_V=-g_{i\bj}|\widetilde{m}|^2V^i\overline{V}^{\bj}
-{i\over 2}\left(g_{i\bi}\partial_jV^i
-g_{j\bj}\partial_{\bi}\overline{V}^{\bj}\right)
\left(\widetilde{m}\,\bpsi_-^{\bi}\psi_+^j
+\overline{\widetilde{m}}\,\bpsi_+^{\bi}\psi_-^j\right).
\label{LV}
\eeq
The deformed Lagrangian
is invariant under the modified
$(2,2)$ supersymmetry
where the modification is given by
${\mit\Delta}Q_-\bpsi_+^{\bi}
=-\sqrt{2}i\widetilde{m}\overline{V}^{\bi}$,
${\mit\Delta}\bQ_+\psi_-^i
=-\sqrt{2}i\widetilde{m}V^i$, and
their complex conjugates (the action on bosonic fields is not modified).
The central charge of the modified supersymmetry is
non-vanishing and is given by
\beq
\widetilde{Z}\,=\,i\,\widetilde{m}{\cal L}_V
\label{Ztilm}
\eeq
where ${\cal L}_V$ acts on the fields as
${\cal L}_V\phi^i=V^i$ and
${\cal L}_V\psi_{\pm}^i=\partial_jV^i\psi_{\pm}^j$.
If the superpotential $W(\phi)$ is invariant under the isometry
$V^i\partial_iW=0$, then, the sigma model can be deformed
by $L_W$ and $L_V$ at the same time without breaking $(2,2)$
supersymmetry.
One can also consider the deformation by
a set of commuting holomorphic isometries
$V_1,\cdots,V_n$; simply replace
$\widetilde{m}V^i\to \sum_{a=1}^n\widetilde{m}_aV_a^i$,
$\overline{\widetilde{m}}V^i\to
\sum_{a=1}^n\overline{\widetilde{m}}_aV_a^i$
(the bosonic term $-|\widetilde{m}|^2|V|^2$
in (\ref{LV}) is replaced by
$-{1\over 2}|\sum_{a=1}^n\widetilde{m}_aV_a|^2
-{1\over 2}|\sum_{a=1}^n\overline{\widetilde{m}}_aV_a|^2$).

\subsection*{\it R-Symmetry}

The R-symmetries $U(1)_V$ and $U(1)_A$
are not always symmetries of the theory
(except their $\Z_2$ subgroups which acts as the sign flip
of spinors --- $2\pi$-rotation).
It can be broken at the classical level by potential terms
or at the quantum level by anomaly.
However, superconformal field theories always possess both symmetries.
We illustrate this
in the class of theories introduced above.

We first consider the non-linear sigma model on $X$
without potentials.
Both $U(1)$'s are classically unbroken. $U(1)_V$ remains a
symmetry of the quantum theory.
$U(1)_A$ is subject to the chiral anomaly
which is proportional to the trace of the
curvature of the connection
(\ref{cov}) of the tangent bundle of $X$.
Thus,
 $U(1)_A$ is anomalous if and only if the
first Chern class of $X$ is non-vanishing; $c_1(X)\ne 0$.
In particular, both $U(1)$'s are unbroken in
the sigma model on a CY manifold $X$, which is expected to
flow to a superconformal field theory of
central charge $c/3=\dim X$.
If $\int\phi^*c_1(X)$ is always an integer multiple of some $p$,
$U(1)_A$ is broken to its discrete subgroup $\Z_{2p}$
which can be further
broken spontaneously.
If the theory flows to a non-trivial fixed point,
$U(1)_A$ must be recovered there.

We next consider a theory with a superpotential $W(\phi)$.
$U(1)_A$ is classically unbroken but is subject to
the chiral anamaly.
$U(1)_V$ is unbroken if and only if
$W(\phi)$ is {\it scale invariant} in the sense that
it is possible to assign the vector R-charges to $\Phi^i$
so that $W(\Phi)$ has vector charge $2$.
The theory has a mass gap if at all the critical points of $W$
the Hessian
is non-degenerate, $\det\partial_i\partial_j W\ne 0$.
At a degenerate critical point,
the theory can flow to a non-trivial fixed point where $U(1)_V$
is recovered.
An example where both $U(1)$'s are unbroken is
the LG model on $\C^N$ with a quasi-homogeneous superpotential;
$W(\lambda^{2q_{i}}\phi_i)=\lambda^2 W(\phi_i)$
where vector R-charge $2q_i$ is assigned to $\phi_i$. 
It is believed that such a model
flows in the IR to an $N=2$ superconformal field theory
with central charge $c/3=\sum_i(1-2q_i)$ \cite{VW,Martinec}.

Finally, if the sigma model is perturbed by a holomorphic isometry,
$U(1)_A$ is explicitly broken by the fermion mass term in
(\ref{LV}). This is consistent with the non-vanishing of the
susy central charge
(\ref{Ztilm}).
This is also related to the fact that the scalar component of
the vector multiplet has a canonical axial charge $2$.

\subsection*{\it Supersymmetric Ground States}

Let us examine the ground states of the theory.
We compactify the spacial direction on $S^1$
and put a periodic boundary condition on all fields.
We also assume $Z=\widetilde{Z}=0$.
As in any supersymmetric field theory,
a state annihilated by all of $Q_{\pm},\bQ_{\pm}$
is a zero energy ground state, and vice versa.
Let $Q$ be one of $Q_-+\bQ_+$ and $\bQ_++\bQ_-$ or their hermitian
conjugates. Then, it follows
from the algebra (\ref{susy1})-(\ref{susy4}) that
\beq
\{Q,Q^{\dag}\}=2H.
\label{QQH}
\eeq
It also follows from (\ref{susy1})-(\ref{susy4}) that
\beq
Q^2=0,
\label{nilp}
\eeq
and we can consider the cohomology of states using $Q$
as the coboundary operator.
In the theory where $H$ has a discrete spectrum,
(\ref{QQH}) and (\ref{nilp}) imply
that
the supersymmetric ground states
are in one to one
correspondence with
the $Q$-cohomology classes.
The index of the operator $Q$ is the Witten index
$\Tr(-1)^F$ which is invariant under perturbation of the
theory.

If the central charge in the supersymmetry
algebra is non-vanishing because of
a continuous abelian global symmetry group $T$, say, as
$\widetilde{Z}=\sum_{a=1}^n\lambda_a S_a$ where $S_a$
are the generators of $T$,
(\ref{QQH}) still holds but (\ref{nilp}) is
modified as $Q^2=2\sum_{a=1}^n\lambda_a S_a$.
However, when restricted to $T$-invariant states,
$Q$ is still nilpotent and we can consider $Q$-cohomology.
This is the $T$-equivariant cohomology.
Since a continuous symmetry cannot be broken in
$1+1$ dimensions, the supersymmetric ground states
are still in one to one
correspondence with
the $T$-equivariant $Q$-cohomology classes.

For a sigma model on a compact Kahler manifold $X$,
$Q=Q_-+\bQ_+$ reduces in the zero momentum sector
to the exterior derivative $\dd=\partial+\bar\partial$ 
acting on differential forms on $X$.
In fact, the zero momentum approximation is exact as far as
vacuum counting is concerned \cite{JDG} and the supersymmetric vacua
and harmonic forms are in one to one correspondence.
In particular $\Tr(-1)^F=\chi(X)$.
The R-charge of a vacuum corresponding to
a harmonic $(p,q)$-form is $q_V=-p+q$ and $q_A=p+q-\dim_{\C}X$,
where the shift by $\dim_{\C}X$ is for the invariance of the spectrum
under conjugation $q_A\leftrightarrow -q_A$.
Of course, if $c_1(X)$ is non-zero and $U(1)_A$ is anomalous,
the axial charge $q_A$ does not make sense.
However, if $\Z_{2p}\subset U(1)_A$ is non anomalous,
there is a $\Z_{2p}$ grading in the space of vacua.

If $X$ is non-compact and has a superpotential $W(\Phi)$,
the spectrum is discrete at sufficiently low energies if 
$g^{\bj i}\partial_{\bj}\overline{W}\partial_iW\geq c>0$ at all infinity
in the field space.
When $W(\Phi)$ has only non-degenerate critical points,
the supersymmetric vacua are in one to one correspondence
with the critical points.
When $W(\Phi)$ can be perturbed to such a situation,
the index is the number of critical points.

If the compact sigma model
is deformed by a holomorphic isometry $V$,
the susy central charge is non-vanishing
and the nilpotency of $Q=Q_-+\bQ_+$ is modified as
$Q^2=2i\widetilde{m}{\cal L}_V$.
Since this is a small perturbation, the index remains the same
as the $\widetilde{m}=0$ case, ${\rm Tr}(-1)^F=\chi(X)$.
In the zero momentum sector,
$Q$ is proportional to
$\dd_{\widetilde{m}}=\dd-\sqrt{2}i\widetilde{m}\,i_V$.
If $V$ has only non-degenerate zeroes,
the supersymmetric vacua are in one to one correspondence with 
the zeroes of $V$. In particular, the index is the number of zeroes
(the Hopf index theorem).
See \cite{JDG} for more details.

\subsection*{\it Chiral Ring}

We can also consider cohomology of local operators
with respect to
$Q=\bQ_++\bQ_-$ or $Q=Q_-+\bQ_+$ (when the central charge is zero).
We shall call a local operator commuting with
$Q_{cc}=\bQ_++\bQ_-$ (resp. $Q_{ac}=\bQ_++Q_-$)
a chiral or $cc$ operator
(resp. a twisted chiral or $ac$ operator).
The lowest component of a (twisted) chiral superfield is
a (twisted) chiral operator.
It follows from the supersymmetry algebra that
the space-time translation of a (twisted) chiral operator
is $Q$-exact and does not change the
cohomology class of the operator.

A product of two (twisted) chiral operators is annihilated by $Q$.
They commute with each other up to $Q$-exact operators since
one can make them space-like separated.
Therefore, the $Q$-cohomology group of local operators form
a commutative ring \cite{LVW}. This is called chiral or $cc$ ring
for $Q_{cc}$-cohomology and twisted chiral or $ac$ ring
for $Q_{ac}$-cohomology.
In general $cc$ and $ac$ rings in a given theory are different from each
other.

If the susy central charge is non-zero because of an abelian
global symmetry, we can define equivariant
chiral ring in an obviuos way.

\subsection*{\it Twisting to Topological Field Theory}

It is often useful to twist $N=2$ theories
to topological field theories \cite{TFT}.
This is possible when the quantum theory
possesses at least either one of $U(1)_V$ or $U(1)_A$
R-symmetries (which we call here $U(1)_R$, generator $R$)
under which the R-charges are all integral.
It is standard to call it A-twist for $R=F_V$
and B-twist for $R=F_A$.

We start with the Euclidean version of the theory (obtained
by Wick rotation $x^0=-ix^2$ from the Minkowski theory).
It has the supersymmetry with the same algebra
(\ref{susy1})-(\ref{susy7}) and the same hermiticity condition
(\ref{herm}).
Twisting is to replace the group
$U(1)_E$ of space-time rotation
generated by $M$ by the diagonal subgroup of
$U(1)_E\times U(1)_R$, considering $M^{\prime}=M+R$
as the new generator of the rotation group.
In particular, the twisted theory on a curved worldsheet
is obtained by gauging the diagonal subgroup of
$U(1)_E\times U(1)_R$ (instead of $U(1)_E$) by the spin connection.
The energy momentum tensor
on the flat worldsheet is thus modified \cite{topS,EY,mirWit} as
$T^{\rm twisted}_{\mu\nu}
=T_{\mu\nu}+(1/4)(\epsilon_{\mu}^{\,\,\lambda}
\partial_{\lambda}J^R_{\nu}
+\epsilon_{\nu}^{\,\,\lambda}
\partial_{\lambda}J^R_{\mu})$
where $J^R_{\mu}$ is the $U(1)_R$ current,
and the spin of fields and conserved currents
are modified.

An important aspect of the twisted theory is that
some of the supercharges have spin zero and make sense
without reference to the coordinates. These
are $Q_-$ and $\bQ_+$ for A-twist while
$\bQ_{\pm}$ for B-twist (see
(\ref{susy5})-(\ref{susy7})).
As we have seen,
$Q_{ac}=Q_-+\bQ_+$ and $Q_{cc}=\bQ_++\bQ_-$
are nilpotent if the susy central charges are zero.
Then, we can consider $Q=Q_{ac}$ or $Q_{cc}$ as
a BRST operator that selects operators
in the A-twisted or B-twisted model respectively.
In particular, $ac$ ring elements are the
physical operators in the A-model
and $cc$ ring elements are the physical operators in the B-model.
$T^{\rm twisted}_{\mu\nu}$ is $Q$-exact
in the class of theories we consider in this paper,
and the correlation functions
of $Q$-invariant operators are independent of the choice of the
worldsheet metric.
In this sense the twisted theory
is a topological field theory.
Even if the susy central charge is non-zero because of
an abelian global symmetry, the twisted theory can still be considered
as topological field theory, physical operators selected by
equivariant cohomology.


For the non-linear sigma model possibly with a superpotential,
A-twist is possible when $W(\Phi)$ is scale invariant
while B-twist is possible when $c_1(X)=0$.
A-twist of a model with $W\equiv 0$ yields topological
sigma model \cite{topS}
while B-model on a flat manifold $X$ is called
topological LG \cite{topLG}.
The fermion path-integral of a B-model is
a chiral determinant and is made well-defined
using the anomaly cancellation condition $c_1(X)=0$.
The definition involves the choice of a multipicative factor which
can be translated to the choice of a nowhere vanishing
holomorphic $n$-form where
$n$ is the complex dimension of $X$. Indeed,
$c_1(X)=0$ assures the existence of such an $n$-form.

\subsection*{\it Spectral Flow}

Let us return to the untwisted $N=2$ theories (where we assume
$Z=\widetilde{Z}=0$ for now).
We have seen that supersymmetric ground states
of the theory on a periodic circle
and the (twisted) chiral ring are both
characterized as the cohomology with
respect to the nilpotent supercharge
$Q$. There is in fact an intimate relation between them:
{\it In a theory which can be A-twisted (resp. B-twisted),
there is a one-to-one correspondence
between supersymmetric ground states and $ac$ ring elements
(resp. $cc$ ring elements).}
In a theory where both A and B-twist are possible,
the space of supersymmetric ground states, $ac$ ring and $cc$ ring
are all the same as a vector space
(but of course $ac$ and $cc$ rings are different as a ring in general).

This can be seen conveniently using the twisted theory.
We consider a theory which is A-twistable.
Let us insert an $ac$ ring element ${\cal O}$
at the tip of a long cigar-like hemi-sphere
in the A-twisted theory.
The twisted theory is equivalent to the untwisted $N=2$
theory on the flat cylinder region,
and we obtain a state of the untwisted theory at the boundary circle.
Because of the twisting in the curved region,
the fermions are periodic on the boundary circle.
Now, the supersymmetric ground state corresponding to
${\cal O}$ is the state at the boundary circle
in the limit where the cylinder region becomes infinitely long.
The state is indeed a ground state because the infinitely long cylinder
plays the role of projection to zero energy states.
Various aspects of this relation and the geometry
of vacuum states and its relation to the chiral rings have been
studied in \cite{CV1}.  In particular this geometry
is captured by what is called the $tt^*$ equations.

\subsection{The Mirror Symmetry}

The $(2,2)$ supersymmetry algebra (\ref{susy1})-(\ref{susy7})
is invariant under the outer automorphism given 
by the exchange of the generators
\beq
Q_-\leftrightarrow \bQ_-,~~
F_V\leftrightarrow F_A,~~
Z\leftrightarrow \widetilde{Z}.
\label{exchange}
\eeq
{\it Mirror symmetry} is an equivalence
of two $(2,2)$ supersymmetric field theories under
which the generators of supersymmetry algebra
are exchanged according to (\ref{exchange}).
A chiral multiplet of one theory is mapped to
a twisted chiral multiplet of the mirror, and vice versa.
It is of course a matter of convention which to call
$Q_-$ or $\bQ_-$.
Here,
we are assuming the {\it standard} convention where,
in the sigma model on a Kahler manifold
(possibly with a superpotential), the complex coordinates
are the lowest components of chiral superfields.
If we flip the convention of one of a mirror pair,
the two theories
are equivalent {\it without} the exchange of (\ref{exchange}).

\subsubsection{Mirror Symmetry between Tori}

~~~~In the case of supersymmetric sigma model on a flat torus,
it has been known that mirror symmetry reduces to
the $R\to 1/R$ duality performed on a middle dimensional torus.
Below,
we review this in the simplest case of sigma model into
the algebraic torus $\C^{\times}=\R\times S^1$.
We start with recalling the bosonic $R\to 1/R$ duality.

\subsection*{\it $R\to 1/R$ Duality}

Let us consider the following action
for a periodic scalar field
$\varphi$ of period $2\pi$ 
\beq
S_{\varphi}={1\over4\pi}\int_W R^2h^{\mu\nu}
\partial_{\mu}\varphi\partial_{\nu}\varphi\,\sqrt{h}\,\dd^2x,
\label{sR}
\eeq
where $h_{\mu\nu}$ is the metric on
the worldsheet $W$ which we choose to be of Euclidean signature.
This is the action for a sigma model into a circle $S^1$ of
radius $R$.
This action can be obtained also from the following action for
$\varphi$ and a one-form field $B_{\mu}$
\beq
S^{\prime}={1\over 2\pi}\int_W
{1\over 2R^2}
h^{\mu\nu}
B_{\mu}B_{\nu}\sqrt{h}\dd^2x
+{i\over 2\pi}\int_W B\wedge\dd\varphi.
\label{Spr}
\eeq
Completing the square with respect to $B_{\mu}$ which is solved
by
\beq
B=iR^2*\dd\varphi,
\label{Bvphi}
\eeq
and integrating it out,
we obtain the action (\ref{sR}) for the sigma model.

If,
changing the order of integration,
we first integrate over the
periodic scalar $\varphi$,
we obtain a constraint $\dd B=0$.
If the worldsheet $W$ is a genus $g$ surface,
there is a $2g$-dimensional space of
closed one-forms modulo exact forms\footnote{This can
be easily extended to the case of worldsheets with boundaries.}. One can choose a basis
$\omega^i$ ($i=1,\ldots,2g$) such that
each element has integral periods on one-cycles on $W$ and
that $\int_W\omega^i\wedge\omega^j=J^{ij}$
is a non-degenerate matrix of integral entry.
Then, a general solution to $\dd B=0$ is
\beq
B=\dd\vartheta_0+\sum_{i=1}^{2g} a_i\omega^i,
\label{Btha}
\eeq
where $\vartheta_0$ is a real scalar field and
$a_i$'s are real numbers.
Integration over
$\varphi$ actually yields constraints on $a_j$'s as well.
Recall that $\varphi$ is a periodic variable of period $2\pi$.
This means that $\varphi$ does not have to come back to its original value
when circling along a nontrivial one-cycles in $W$, but comes back to
itself up to $2\pi$ shifts. For such a topologically
nontrivial configuration, $\dd\varphi$ has an expansion like
(\ref{Btha}) with non-zero coefficient
$a_i$ for $\omega^i$ which is dual to the one-cycle.
That the shift is only allowed to take integer multiples of $2\pi$ means
that such $a_i$ is constrained to be $2\pi n_i$ where $n_i$ is an integer.
Thus, for a general configuration of $\varphi$ we have
\beq
\dd\varphi=\dd\varphi_0+\sum_{i=1}^{2g} 2\pi n_i\omega^i,
\label{varpom}
\eeq
where $\varphi_0$ is a single valued function on $W$.
Now, integration over $\varphi$ means integration over the function
$\varphi_0$ and summation over the integers $n_i$'s.
Integration over $\varphi_0$ yields the constraint $\dd B=0$
which is solved by (\ref{Btha}).
What about the summation over $n_i$'s?
To see this we substitute in $\int B\wedge \dd\varphi$
for $B$ from (\ref{Btha});
\beq
\int_W B\wedge\dd\varphi=2\pi \sum_{i,j}a_iJ^{ij}n_j.
\eeq
Now, noting that $J^{ij}$ is a non-degenerate integral matrix and
using the fact that $\sum_n\e^{ian}=2\pi\sum_m\delta(a-2\pi m)$,
we see that
summation over $n_i$ constrains $a_i$'s to be an integer multiples
of $2\pi$;
\beq
a_i=2\pi m_i,~~~m_i\in\Z.
\eeq
Inserting this into (\ref{Btha}), we see that $B$ can be written as
\beq
B=\dd\vartheta,
\label{Btheta}
\eeq
where now $\vartheta$ is a periodic variable of period $2\pi$.
Now, inserting this to the original action we obtain
\beq
S_{\vartheta}={1\over 4\pi}\int_W{1\over R^2}
h^{\mu\nu}\partial_{\mu}\vartheta\partial_{\nu}\vartheta
\sqrt{h}\dd^2x
\eeq
which is an action for a sigma model into $S^1$ of
radius $1/R$.

Thus, we have shown that the sigma model into $S^1$ of radius $R$ is
equivalent to the model with radius $1/R$.
This is the $R\to 1/R$
duality (which is called {\it target space duality} or {\it T-duality}
in string theory).

Comparing (\ref{Bvphi}) and (\ref{Btheta}), we obtain the relation
\beq
R\,\dd\varphi\,=\,i\,{1\over R}*\dd\vartheta.
\label{momwind}
\eeq
Since $R\dd\varphi$ and $iR*\dd\varphi$ are the conserved currents
in the original system
that count momentum and winding number respectively,
the relation (\ref{momwind})
means that {\it momentum and winding number are exchanged
under the $R\to 1/R$ duality}.
In particular, the vertex operator
\beq
\exp(i\vartheta)
\eeq
that creates a unit momentum in the dual theory
must be equivalent to an operator that creates a unit winding
number in the original theory.
This can be confirmed by the following
path integral manipulation.
Let us consider the insertion of
\beq
\exp\left(-i\int_p^q B\right)
\eeq
in the system with the action (\ref{Spr}),
where the integration is along a path $\tau$ emanating from $p$
and ending on $q$.
Then, using (\ref{Btheta}) we see that
\beq
\exp\left(-i\int_p^q B\right)=\e^{-i\vartheta(q)}\,
\e^{i\vartheta(p)}.
\eeq
On the other hand, the insertion of
$\e^{-i\int_p^qB}$
changes the $B$-linear term in (\ref{Spr}).
We note that $\int_p^qB$ can be expressed as
$\int_WB\wedge\omega$, where 
$\omega$ is a one-form with delta function support along
the path $\tau$.
This $\omega$ can be written as $\omega=\dd \theta_{\tau}$
where $\theta_{\tau}$ is a multi-valued function on $W$
that jumps by one when crossing the path $\tau$.
Now, the modification of the action (\ref{Spr}) can be written as
\beq
{i\over 2\pi}\int_WB\wedge\dd\varphi\longrightarrow
{i\over 2\pi}\int_WB\wedge\dd\varphi+i\int_p^qB=
{i\over 2\pi}\int_WB\wedge\dd(\varphi+2\pi\theta_{\tau}).
\eeq
Integrating out $B_{\mu}$, we obtain the action (\ref{sR})
with $\varphi$ replaced by
$\varphi^{\prime}=\varphi+2\pi\theta_{\tau}$.
Note that $\varphi^{\prime}$ jumps by $2\pi$
when crossing the path $\tau$ which starts and ends on $p$ and $q$.
In particular, it has winding number
$1$ and $-1$ around $p$ and $q$ respectively.
Thus, the insertion of $\e^{i\vartheta}$
creates the unit winding number in the original system.

\subsection*{\it Mirror Symmetry as $R\to 1/R$ Duality}

We now proceed to a supersymmetric sigma model
on the algebraic torus, or the cylinder  $\C^{\times}=\R\times S^1$.
We show that $R\to 1/R$ duality performed on the $S^1$ factor
is indeed a mirror symmetry.
We work now in Minkowski signature.

We denote the complex coordinate of the cylinder $\R\times S^1$
as
\beq
\phi=\varrho+i\varphi
\eeq
where $\varrho$ is the coordinates of $\R$ and $\varphi$
is the periodic coordinate of $S^1$ of period $2\pi$.
The Lagrangian of the system is
\beq
L=\int\dd^4\theta\,{R^2\over 2}|\Phi|^2
={R^2\over 2}\left(\,
-\eta^{\mu\nu}\partial_{\mu}\overline{\phi}\partial_{\nu}\phi
+i\bpsi_-(\partial_0+\partial_1)\psi_-
+i\bpsi_+(\partial_0-\partial_1)\psi_+\,\right),
\label{LagR}
\eeq
where $\Phi$ is the chiral superfield whose lowest component is $\phi$.
The Kahler metric for $\phi$ is
$\dd s^2=R^2|\dd\phi|^2
=R^2(\dd\varrho^2+\dd\varphi^2)$ so that $S^1$
has radius $R$.\footnote{
In this paper, we take the convention
$S={1\over 2\pi}\int\dd^2 x L$ as the relation of
the action and the Lagrangian.
Thus, the weight factor
in Path-Integral is $\exp({i\over 2\pi}\int \dd^2x L)$
(in Minkowski signature).}

We perform the duality transformation on $\varphi$.
As we have seen, this yields another periodic variable
$\vartheta$ of period $2\pi$ with the Kinetic term
$-(1/2R^2)\eta^{\mu\nu}\partial_{\mu}\vartheta\partial_{\nu}\vartheta$.
Thus, the dual theory is also a sigma model into
a cylinder, but with a metric
\beq
\dd\widetilde{s}^2=\dd\varrho^2+{1\over R^2}\dd\vartheta^2
={1\over R^2}\left(R^4\dd\varrho^2+\dd\vartheta^2\right).
\eeq
Thus, either $R^2\varrho+i\vartheta$ or $R^2\varrho-i\vartheta$
is the complex coordinates of the new cylinder.
What is the superpartner of this (anti-)holomorphic variable?
We note the supersymmetry transformations
$\delta\psi_{\pm}=-i\sqrt{2}(\partial_0\pm\partial_1)\phi
{\bar\epsilon}^{\pm}$ and
$\delta\bpsi_{\pm}=i\sqrt{2}(\partial_0\pm\partial_1)\phi
\epsilon^{\pm}$.
{}From (\ref{momwind}), we see (after continuation back to Minkowski
signature by $x^2=ix^0$) that
$R^2(\partial_0\pm\partial_1)\varphi=
\mp(\partial_0\pm\partial_1)\vartheta$
and therefore
$R^2(\partial_0+\partial_1)\phi=(\partial_0+\partial_1)\eta$
and
$R^2(\partial_0-\partial_1)\phi
=(\partial_0-\partial_1)\overline{\eta}$
where
\beq
\eta=R^2\varrho-i\vartheta.
\eeq
Thus, the supersymmetry transformation is expressed as
\beqa
&&R^2\delta\psi_+=-i\sqrt{2}(\partial_0+\partial_1)\eta
\overline{\epsilon}^+,
~~~
R^2\delta\bpsi_+=i\sqrt{2}(\partial_0+\partial_1)\overline{\eta}
\epsilon^+,\\
&&R^2\delta\psi_-=-i\sqrt{2}(\partial_0-\partial_1)
\overline{\eta}\overline{\epsilon}^+,
~~~
R^2\delta\bpsi_-=i\sqrt{2}(\partial_0+\partial_1)\eta
\epsilon^-.
\eeqa
This is not a supersymmetry transformation for a chiral multiplet,
but that for a twisted chiral multiplet.
Indeed, renaming the fermions as
\beq
R^2 \psi_{\pm}=\pm\bchi_{\pm},~~
R^2\bpsi_{\pm}=\pm\chi_{\pm},
\label{Rpsichi}
\eeq
the Lagrangian for the dual theory becomes
$\int\dd^4\theta (-{1\over 2R^2}|\Theta|^2)$
for a twisted chiral superfield
$\Theta=\eta+\sqrt{2}(\theta^+\bchi_++\theta^-\chi_-)+\cdots$.
Thus, we have seen that $R\to 1/R$ duality on $S^1$
transforms a theory of a chiral multiplet
to another theory of a twisted chiral multiplet.
Thus, this is a mirror symmetry.

The above manipulation can be simplified
by performing the dualization in superspace.
We follow the procedure developed in \cite{RV}.
We start with the following
Lagrangian for a real superfield $B$ and a twisted chiral
superfield $\Theta$.
\beq
L^{\prime}=\int\dd^4\theta 
\left({R^2\over 4}B^2-{1\over 2}(\Theta+\bTheta)B\right)
\eeq
We first integrate over the twisted chiral field $\Theta$, $\bTheta$.
This yields the following constraint on $B$
\beq
\bD_+D_-B=D_+\bD_-B=0,
\eeq
which is solved by
\beq
B=\Phi+\bPhi,
\label{BPhiPhi}
\eeq
where $\Phi$ is a chiral superfield.
Now, inserting this into the original Lagrangian we obtain the
Lagrangian (\ref{LagR})
\beq
L=\int\dd^4\theta \,{R^2\over 4}(\Phi+\bPhi)^2=
\int\dd^4\theta ~{R^2\over 2}\,\bPhi\Phi.
\eeq
for the sigma model into the cylinder with radius $R$ on $S^1$.
Now, reversing the order of integration, we consider integrating out $B$
first. Then, $B$ is solved by
\beq
B={1\over R^2}(\Theta+\bTheta).
\label{BThTh}
\eeq
Inserting this into $L^{\prime}$ we obtain
\beq
\widetilde{L}=\int\dd^4\theta \,
\left(\,-{1\over 2R^2}\bTheta\Theta\right),
\eeq
which is again
the Lagrangian for supersymmetric sigma model on the cylinder.
This time, the radius of $S^1$ is $1/R$ and
the complex coordinate is described by the twisted chiral superfield
$\Theta$.
{}From (\ref{BPhiPhi}) and (\ref{BThTh}), we
obtain $R^2(\Phi+\bPhi)=\Theta+\overline{\Theta}$ which
reproduces the relation between the component fields
obtained above (e.g. (\ref{Rpsichi})).

\subsubsection{Examples}

~~~~Here we present three classes of
examples of (conjectural) mirror symmetry.
They are mirror symmetry between LG model and LG model,
sigma model and sigma model, and sigma model and LG model.

\subsection*{\it Minimal Models and Orbifolds}

$N=2$ minimal models are the simplest class of exactly
solvable $(2,2)$ SCFTs.
It has been argued that
the $(d-2)$-th minimal model arises as the IR fixed point
of the LG model of chiral superfield $X$ with the superpotential
\cite{Martinec,VW}
\beq
W=X^d.
\label{min}
\eeq
If we orbifold the model by the $\Z_d$ symmetry generated by
$X\to \e^{2\pi i/d}X$,
we obtain again the $(d-2)$-th minimal model \cite{LGO}
\beq
\widetilde{W}=\widetilde{X}^d.
\label{twmin}
\eeq
This time, however, $\widetilde{X}$ is a twisted chiral superfield
and $\widetilde{W}$ is a twisted superpotential.
This means that the minimal model and its 
$\Z_d$-orbifold can be considered as mirror to each other.

If a CFT ${\cal C}$
has a discrete abelian symmetry group $\Gamma$, the orbifold CFT
${\cal C}^{\prime}={\cal C}/\Gamma$ has a symmetry group
$\Gamma^{\prime}$ isomorphic to $\Gamma$ and the orbifold
${\cal C}^{\prime}/\Gamma^{\prime}$ is identical to the original
CFT ${\cal C}$. We apply this general fact to the sum of $N$ copies of
minimal models
\beq
W=X_1^d+\cdots +X_N^d,
\eeq
modulo its $\Z_d$ symmetry group
generated by $X_i\mapsto X_i\to\e^{2\pi i/d} X_i$ for all $i$.
This LG orbifold ${\cal C}=\Bigl(W=\sum_iX_i^d\Bigr)/\Z_d$
has a symmetry group $\Gamma\cong (\Z_d)^{N-1}$
generated by $X_i\mapsto \e^{2\pi i\alpha_i/d}X_i$.
Orbifold of ${\cal C}$ by this $\Gamma$ is
${\cal C}^{\prime}=
\Bigl(W=\sum_iX_i^d\Bigr)/(\Z_d)^N=\oplus_i\Bigl(W=X_i^d\Bigr)/\Z_d$.
By the mirror symmetry of (\ref{min}) mod $\Z_d$ and (\ref{twmin}),
this is identical to
the sum of $N$ copies of the minimal model given by the twisted chiral
superpotential
\beq
\widetilde{W}=\widetilde{X}_1^d+\cdots +\widetilde{X}_N^d.
\eeq
This indeed has a symmetry $\Gamma^{\prime}$ isomorphic to
$(\Z_d)^{N-1}$ generated by
$\widetilde{X_i}\mapsto\e^{2\pi i\alpha_i/d}X_i$ with $\sum_i\alpha_i=0$
(mod $d$).
By the general fact on the orbifold, we see that 
the orbifold
$\Bigl(\widetilde{W}=\sum_i\widetilde{X}_i^d\Bigr)/(\Z_d)^{N-1}$
is identical to the original SCFT $\Bigl(W=\sum_iX_i^d\Bigr)/\Z_d$.
In other words, the $\Z_d$ orbifold and the
$(\Z_d)^{N-1}$ orbifold of the sum of $N$ copies of the
minimal model are mirror to each other.

\subsection*{\it Quintic}

Applying the above construction of mirror pair to the orbifold
minimal models \cite{Gepner}
corresponding to a CY sigma model (at the special
point of its moduli space), Greene and Plesser constructed 
pairs of CY manifold whose sigma models are mirror to each other
\cite{GP}.  The connection between Landau-Ginzburg models and Calabi-Yau
sigma models was first discussed in \cite{GVW,Martinec}.  
The derivation of this connection in
\cite{GVW} involved a change of variables in field space, as if one
were dealing with an ordinary integral.  This heuristic
derivation of the relation between LG models
and Calabi-Yau sigma models was made precise by Cecotti
 \cite{cecoin} who showed the arguments are precise in the
 context of computing periods associated to special geometry
 of the LG model, and that the derivation of \cite{GVW} 
can be viewed as showing the equivalence of periods and special
geometry of the Calabi-Yau with an LG model
(in modern terminology as far
as the middle dimensional D-brane masses are concerned).
In fact this identification of LG models and special
geometry associated to the vacuum geometry
of the sigma model is crucial in our derivation of mirror symmetry
for the case of complete intersections in toric varieties.

The connection between LG models and Calabi-Yau
sigma models was further elucidated in \cite{phases} using the
linear sigma model description, which we will review in section 4 in this
paper.

\subsection*{\it $\CP^{N-1}$ and Affine Toda Thoery}

Less known class of mirror symmetry is between
non-liner sigma models on manifolds of positive first Chern class
and LG models without scale invariance.
The typical example is the mirror symmetry of the
$\CP^{N-1}$ sigma model and supersymmetric $A_{N-1}$ affine Toda
theory.
The $\CP^{N-1}$ model is asymptotic free and generates
a dynamical scale $\Lambda$. It has $N$ vacua with mass gap.
$U(1)_V$ is unbroken but $U(1)_A$ is anomalously broken to
$\Z_{2N}$ which is spontaneously broken to $\Z_2$.
The $A_{N-1}$ affine Toda theory is an LG model of $N-1$
periodic variables $X_i$ having superpotential
\beq
W=\Lambda
\left(\e^{X_1}+\cdots+\e^{X_{N-1}}+\prod_{i=1}^{N-1}\e^{-X_i}\right).
\eeq
This theory has the same properties of the $\CP^{N-1}$ model
mentioned above, except that
$U(1)_V$ and $U(1)_A$ are exchanged, the mass scale is explicitly
introduced and $U(1)_V\to\Z_{2N}$ is an explicit breaking.
This duality is in many ways an $N=2$ generalization of
the duality of the
(bosonic) sine-Gordon theory and the massive Thirrling model
\cite{sineGTh}.
In particular, solitons of the affine Toda theory are mapped to
the fundamental fields in the $\CP^{N-1}$ model
(in the linear sigma model realization) \cite{Witten79}.

The equivalence of the two theories
has been observed from various points of view.
The agreement of BPS soliton spectrum and their scattering matrix
\cite{FI,CV},
of correlation functions of topologically twisted theories 
(coupled to topological gravity)
\cite{G2,EHY,EHX}.
This example is extended to a more general class of
manifolds in \cite{CV,G2,G,EHX}.

\section{The Dynamics of $N=2$ Gauge Thoeries in Two Dimensions}

In this section, we study the dynamics of $(2,2)$
supersymmetric gauge theories in $1+1$ dimensions.
We consider $U(1)$ gauge theory with charged chiral
multiplets and assume that there is no superpotential
for the charged fields.
The theory is described by vector superfield $V$ with field strength
$\Sigma$ which is a twisted chiral superfield.
We consider $N$ chiral superfields $\Phi_i$ of charge $Q_i$.
For earlier studies of this class of theories, see
\cite{Witten79,DDL,phases}.

The classical theory is parametrized by the
gauge coupling $e$, Fayet-Iliopoulos (FI) parameter $r$
and Theta angle $\theta$ where $e$ has dimension of mass but $r$ and
$\theta$ are dimensionless.
The Lagrangian of the theory is given by
\beq
L=\int\dd^4\theta
\left(\,
\sum_{i=1}^N\bPhi_i\,\e^{2Q_iV}\Phi_i
-{1\over 2e^2}\bSigma\Sigma
\,\right)
+{1\over 2}\Biggl(
-\int\dd^2\widetilde{\theta}~t\,\Sigma
\,+\, c.c.\,
\Biggr),
\label{lag}
\eeq
where $t$ is the complex combination
\beq
t=r-i\theta.
\eeq

The theory is super-renormalizable with respect to
the gauge coupling $e$. However, the FI parameter $r$ is
renormalized to cancell a one-loop
divergence unless $\sum_iQ_i=0$.
The dependence of the bare parameter $r_0$
on the cut-off $\Lambda_{UV}$ is
\beq
r_0=\sum_{i=1}^NQ_i\log\left({\Lambda_{UV}\over \Lambda}\right)
\label{renor}
\eeq
where $\Lambda$ is a scale parameter that replaces
$r$ as the parameter of the theory if $\sum_iQ_i\ne 0$.

The classical theory is invariant under $U(1)_V\times U(1)_A$
R-symmetry where $\Sigma$ is assigned an axial charge $2$
and zero vector charge.
$U(1)_V$ is an exact symmetry of the theory
but $U(1)_A$ is subject to the chiral anomaly.
The axial rotation by $\e^{i\alpha}$
shifts the theta angle by
\beq
\theta\to\theta-2\sum_{i=1}^NQ_i\alpha.
\label{anomaly}
\eeq
Thus $U(1)_A$ is unbroken if $\sum_iQ_i=0$ but otherwise is
broken to the discrete subgroup
$\Z_{2p}$ with $p=\sum_iQ_i$.

In addition, the theory has other global symmetries.
There are at least $N-1$ $U(1)$ symmetries which are the
phase rotation of the $N$ chiral superfield modulo
$U(1)$ gauge transformations.
This will be important in our study.
Of course there could be larger symmetry if some of the $U(1)$
charges $Q_i$ coincide. These global symmetries are non-anomalous
and are the symmetries of the quantum theory.

In what follows, we study the dynamics of this gauge theory.
We first dualize each of the charged chiral
fields $\Phi_i$ using the phase rotation symmetry.
We then study the effective theory
described in terms of the dual variables.
The goal of this section is to show that the twisted
superpotential as mentioned in the introduction of the paper
is dynamically generated.

\subsection{Abelian Duality}

Let us consider a complex scalar field $\phi$
which is minimally coupled to a gauge field
$A_{\mu}$. 
In terms of the polar variables $(\rho,\varphi)$ defined by
$\phi=\rho\,\e^{i\varphi}$,
the kinetic term $-\eta^{\mu\nu}D_{\mu}\phi^{\dag}D_{\nu}\phi$
is written as the sum of
$-(\partial_{\mu} \rho)^2$
and
\beq
L_{\varphi}=-\rho^2(\partial_{\mu}\varphi+QA_{\mu})^2,
\label{psiA}
\eeq
where $Q$ is the charge of $\phi$.
This Lagrangian is invariant under the shift
$\varphi\to \varphi+$constant,
and therefore we can consider dualizing $\varphi$
as we have done when we discussed $R\to 1/R$ duality.
What is new here is that we have a gauge field coupled to
$\varphi$ as in (\ref{psiA}).
The dualization procedure
start with the following Lagrangian for
the vector field $B_{\mu}$,
the angle variable $\varphi$, plus the gauge field $A_{\mu}$:
\beq
L^{\prime}=-{1\over 4\rho^2}(B_{\mu})^2
+\epsilon^{\mu\nu}B_{\mu}(\partial_{\nu}\varphi+QA_{\nu}).
\eeq
Integration over $B$ yields the Lagrangian (\ref{psiA}).
If, instead, we first integrate over $\varphi$, we obtain the constraint
$B_{\mu}=\partial_{\mu}\vartheta$
where $\vartheta$ is an angle variable of period
$2\pi$. Plugging this into $L^{\prime}$,
we obtain the new Lagrangian
\beq
L_{\vartheta}
=-{1\over 4\rho^2}(\partial_{\mu}\vartheta)^2
+Q\epsilon^{\mu\nu}\partial_{\mu}\vartheta A_{\nu}
=
-{1\over 4\rho^2}(\partial_{\mu}\vartheta)^2
-Q\,\vartheta\,\epsilon^{\mu\nu}{1\over 2}F_{\mu\nu},
\label{Ltheta}
\eeq
where a partial integration is used.
Thus, the dual variable $\vartheta$ is coupled to the
gauge field $A_{\mu}$ as a dynamical Theta angle.

\subsection*{\it Supersymmetric Case}

It is straightforward to repeat this dualization in the supersymmetric
theory with a chiral superfield $\Phi$ of charge $Q$.
In fact, we only have to dualize the phase of
$\Phi$ and suitably rename other fields.
As we have seen above, the dual variable couples to the gauge field
as a dynamical Theta angle.
Such a variable must
be in a twisted chiral multiplet $Y$ that couples to the field strength
$\Sigma$ in the twisted superpotential as
\beq
Q\,Y\,\Sigma.
\eeq
One can see this explicitly by performing 
the duality transformation in the superspace, as we now show.

We start with the following
Lagrangian for a vector superfield $V$,
a real superfield $B$ and a twisted chiral
superfield $Y$ whose imaginary part is periodic with period $2\pi$.
\beq
L^{\prime}=\int\dd^4\theta 
\left(\e^{2QV+B}-{1\over 2}(Y+\overline{Y})B\right),
\eeq
where $Q$ is an integer.
We first integrate over $Y$.
This yields the constraint
$\bD_+D_-B=D_+\bD_-B=0$ on $B$
which is solved by
\beq
B=\Psi+\overline{\Psi},
\label{BPsi}
\eeq
where $\Psi$ is a chiral superfield.
Since the imaginary part of $Y$ is an
angular variable of period $2\pi$, so is the imaginary part of $\Psi$.
Now, inserting this into the original Lagrangian we obtain
\beq
L=\int\dd^4\theta \,\,\e^{2QV+\Psi+\overline{\Psi}}
\eeq
which is nothing but the Lagrangian for the chiral superfield
$\Phi=\e^{\Psi}$ of charge $Q$.
Now, reversing the order of integration, we consider integrating out $B$
first. Then, $B$ is solved by
\beq
B=-2QV+\log\left({Y+\overline{Y}\over 2}\right).
\label{BVY}
\eeq
Inserting this into $L^{\prime}$ we obtain
\beq
\widetilde{L}=
\int\dd^4\theta
\left(\,QV(Y+\overline{Y})
-{1\over 2}(Y+\overline{Y})\log(Y+\overline{Y})\,\right)
\eeq
Using the fact that $Y$ is a twisted chiral superfield,
$\bD_+Y=D_-Y=0$,
the term proportional to $V$ can be written as
\beqa
\int\dd^4\theta \,VY
&=&-{1\over 4}\int\dd\theta^+\dd\btheta^-\bD_+D_-V\,Y\\
&=&
{1\over 2}\int\dd^2\widetilde{\theta}\,\Sigma Y
\eeqa
where we have used $\Sigma=\bD_+D_-V$
which holds for abelian gauge group.
Together with the gauge kinetic term and
the classical FI-Theta terms, we obtain
the following Lagrangian \footnote{In the usual T-duality
the dilaton shifts, proportional to the volume of the
space. In the case at hand, since translations
in the dualizing
circle is gauged this shift in dilaton does not arise.}:
\beq
\widetilde{L}=\int\dd^4\theta
\left\{\,
-{1\over 2e^2}\overline{\Sigma}\Sigma
-{1\over 2}(Y+\overline{Y})\log(Y+\overline{Y})\,\right\}
+{1\over 2}\left(\,
\int\dd^2\widetilde{\theta}\,\,
\Sigma(\,QY-t\,)
+\,c.c.\,
\right)
\label{tilLdual}
\eeq
We indeed see that the charged chiral superfield $\Phi$
has turned into a neutral twisted chiral superfield $Y$
which couples to the field strength $\Sigma$ as the 
dynamical Theta angle.

It follows from
(\ref{BPsi}) and (\ref{BVY}) that
the original chiral field $\Phi$
and the twisted chiral field $Y$ are related by
\beq
Y+\overline{Y}=2\,\bPhi\,\e^{2QV}\Phi.
\label{YYPhi}
\eeq
We see that the dual field $Y$ is a gauge invariant
composite of the original field $\Phi$. 
Using the expression (\ref{Vect}) of $V$ in the Wess-Zumino gauge,
we can write down the relation between
the components fields
$y=\varrho-i\vartheta,\bchi_+,\chi_-$ of $Y$
and those $\phi=\rho\,\e^{i\varphi},\psi_+,\psi_-$ of $\Phi$:
\beqa
\varrho&=&\rho^2,
\label{yy}\\
\partial_{\pm}\vartheta&=&
\pm 2\left(-\rho^2(\partial_{\pm}\varphi+QA_{\pm})
+\overline{\psi}_{\pm}\psi_{\pm}\right)
\label{thpsi}
\eeqa
where $\partial_{\pm}=\partial_0\pm\partial_1$ etc, and
\beqa
\bchi_+=2\phi^{\dag}\psi_+,&&\chi_-=-2\bpsi_-\phi,
\label{chipsi1}\\
\chi_+=2\bpsi_+\phi,&&\bchi_-=-2\phi^{\dag}\psi_-.
\label{chipsi2}
\eeqa
We note that the term $\pm 2\overline{\psi}_{\pm}\psi_{\pm}$
in (\ref{thpsi}) reflects the fact that we are dualizing
on the phase of the whole superfield $\Phi$.
The Kahler metric of the field $Y$ is given by
\beq
\dd s^2={|\dd y|^2\over 2(y+\bar y)}
={1\over 4\varrho}(\dd\varrho^2+\dd\vartheta^2)
=\dd\rho^2+{1\over4 \rho^2}\dd\vartheta^2,
\label{lastone}
\eeq
as can also be seen from the bosonic treatment (e.g. (\ref{Ltheta})).
We note that
the relation (\ref{yy}) implies a condition that the real part
of $y$ is allowed to take only non-negative values,
${\rm Re}(y)\geq 0$.
The boundary ${\rm Re}(y)= 0$ corresponds to $|\Phi |=0$
where the circle on which we are dualizing shrinks to zero
size.  With respect to the metric (\ref{lastone}), the boundary
is at finite distance from any point with finite ${\rm Re}(y)$.
Naively we expect a lot of singularities coming from such end points.
However,
it will be argued below that the physically relevant region is
infinitely far away from such a boundary region, once the renormalization
is taken into account.

\subsection*{\it Renormalization}

We recall that we had to renormalize the FI parameter of the theory
as (\ref{renor}).
In order for the coupling $\Sigma(QY-t)$ to be finite,
we have to renormalize also the field $Y$.
This can be done by letting the bare dual field
$Y_0$ to depend on the UV cut-off as
\beq
Y_0=\log(\Lambda_{UV}/\mu)+Y,
\eeq
where $\mu$ is the scale at which the field is renormalized.

Note that this resolves the issue of the bound
${\rm Re}(y)\geq 0$.
In fact, the correct condition is ${\rm Re}(y_0)\geq 0$
and therefore it means
\beq
{\rm Re}(y)\geq -\log(\Lambda_{UV}/\mu)
\eeq
for the renormalized variable.
In particular, in the continuum limit
$\Lambda_{UV}/\mu\to\infty$, there is no bound on the renormalized field.
Also, we note that the Kahler metric for the renormalized field
is
\beq
\dd s^2={|\dd y|^2\over
2(2\log(\Lambda_{UV}/\mu)+y+\overline{y})}
\simeq
{|\dd y|^2\over
4\log(\Lambda_{UV}/\mu)}
\label{ymet}
\eeq
which becomes flat in the continuum limit.

\subsection*{\it R-Symmetry}

We would like to know the transformation property of the new field $Y$
under the vector and axial R-symmetries.
It appears from the relation (\ref{YYPhi}) that the superfield
$Y$ transforms as a charge zero field
under both $U(1)_V$ and $U(1)_A$.
However, it cannot be directly seen from (\ref{YYPhi})
how the imaginary part $\vartheta$
of $Y$ transforms.
Nevertheless, one can read the transformation of $\vartheta$
from (\ref{YYPhi}) or (\ref{thpsi}) in a way similar to \cite{AHW}.
Let us note that the conserved currents of the vector and axial
R-symmetries are given by
\beq
J^V_{\pm}=\overline{\psi}_{\pm}\psi_{\pm}+\cdots,~~~
J^A_{\pm}=\pm \overline{\psi}_{\pm}\psi_{\pm}+\cdots,
\eeq
where $+\cdots$ are contributions from the vector multiplet fields.
The operator product of these with $\partial_{\pm}\vartheta$
expressed as (\ref{thpsi}) has the following singularity:
\beqa
&&J^V_{\pm}(x)\partial_{\pm}\vartheta(y)\sim
{\!\pm 2\,\,\,\over (x^{\pm}-y^{\pm})^2},\\
&&J^A_{\pm}(x)\partial_{\pm}\vartheta(y)\sim
{2\over (x^{\pm}-y^{\pm})^2}.
\eeqa
These show that $\vartheta$ is invariant under vector R-symmetry
but is shifted by $2$ by the axial
R-symmetry. Therefore the superfield $Y$
transforms under $U(1)_V\times U(1)_A$ as
\beqa
&&
\e^{i\alpha F_V}Y(\theta^{\pm},\btheta^{\pm})\e^{-i\alpha F_V}
=Y_i(\e^{-i\alpha}\theta^{\pm},
\e^{i\alpha}\btheta^{\pm}),
\label{VY}\\
&&
\e^{i\alpha F_A}Y(\theta^{\pm},\btheta^{\pm})\e^{-i\alpha F_A}
=Y(\e^{\mp i\alpha}\theta^{\pm},
\e^{\pm i\alpha}\btheta^{\pm})-2i\alpha.
\label{AY}
\eeqa
Indeed, the Lagrangian (\ref{tilLdual}) exhibits the $U(1)_V$ invariance
and the correct $U(1)_A$ anomaly under this action.

\subsection*{\it Multi-Flavor Case}

It is straightforward to extend the
dualization considered above to the case where there are several
charged chiral fields. Dualizing each of the chiral fields $\Phi_i$,
we obtain the following twisted superpotential for the
twisted chiral fields $Y_{i0}$
\beq
\widetilde{W}
=\Sigma\,\left(\,\sum_{i=1}^NQ_iY_{i0}-t_0\,\right).
\label{multi0}
\eeq
The relation between $\Phi_i$ and $Y_{i0}$ are the same as
in (\ref{YYPhi})-(\ref{chipsi2}).
We renormalize the fields as
\beq
Y_{i0}=\log(\Lambda_{UV}/\mu)+Y_i.
\label{renoYi}
\eeq
so that (\ref{multi0}) is finite in the continuum limit
$\Lambda_{UV}\to\infty$ in the case $\sum_iQ_i\ne 0$.
In the case $\sum_iQ_i=0$, (\ref{multi0}) is invariant under
an $i$-independent shift of $Y_{i0}$'s and we also do this
field redefinition (\ref{renoYi}).
In any case, the bound ${\rm Re}(y_{i0})\geq 0$ is eliminated from
$y_i$'s.
With respect to the renormalized fields the twisted superpotential
can be written as
\beq
\widetilde{W}
=\Sigma\,\left(\,\sum_{i=1}^NQ_iY_{i}-t(\mu)\,\right).
\label{multi}
\eeq
where $t(\mu)$ is the effective FI-Theta parameter
\beq
t(\mu)
=\left\{
\begin{array}{ll}
\sum_{i=1}^NQ_i\log(\mu/\Lambda)-i\theta&
\mbox{if $\sum_{i=1}^NQ_i\ne 0$},\\[0.2cm]
r-i\theta&\mbox{if $\sum_{i=1}^NQ_i=0$}.
\end{array}
\right.
\label{teff}
\eeq
As in the single flavor case, one can see that the
R-symmetry group $U(1)_V\times U(1)_A$ acts on the fields $Y_i$ as
\beqa
&&
\e^{i\alpha F_V}Y_i(\theta^{\pm},\btheta^{\pm})\e^{-i\alpha F_V}
=Y_i(\e^{-i\alpha}\theta^{\pm},
\e^{i\alpha}\btheta^{\pm}),
\label{VYi}\\
&&
\e^{i\alpha F_A}Y_i(\theta^{\pm},\btheta^{\pm})\e^{-i\alpha F_A}
=Y_i(\e^{\mp i\alpha}\theta^{\pm},
\e^{\pm i\alpha}\btheta^{\pm})-2i\alpha.
\label{AYi}
\eeqa
We see that the superpotential (\ref{multi})
is invariant under $U(1)_V$.
Note also
that (\ref{multi}) exhibits the axial anomaly (\ref{anomaly})
for $p=\sum_iQ_i\ne 0$
and the breaking of $U(1)_A$ down to $\Z_{2p}$.
It may appear that there are extra symmetries
$Y_i\to Y_i+c_i$ with $c_i\ne c_j$ and
 $\sum_{i=1}^NQ_ic_i=0$ that makes the transformation 
(\ref{AYi}) ambiguous.
However, there is no
corresponding symmetry in the original system;
$Y_i$'s can be shifted only by axial symmetries and
the only axial symmetry in the system is the
$U(1)_A$ with current
$J^A_{\pm}=\pm\sum_{i=1}^N\bpsi_{i\pm}\psi_{i\pm}+\cdots$
as long as $Q_i$'s are all non-zero.
Thus, there is no room for ambiguity in the R-transformation
(\ref{AYi}).
In fact, new terms in $\widetilde{W}$ that violates the
invariance under the extra shifts with $c_i\ne c_j$
is generated,
as we will show next.

\subsection{Dynamical Generation Of Superpotential}

The twisted superpotential (\ref{multi})
obtained from dualization is an exact expression in
perturbation theory with resepect to $1/r$;
it is simply impossible to write down a perturbative correction
that respects the R-symmetry and/or anomaly, holomorphy in
$t$, and periodicity of Theta angle. The D-term is of course
subject to perturbative corrections.

However, the twisted superpotential is
possibly corrected by non-perturbative effects.
A typical non-perturbative effect in quantum field theory is
by the presence of instantons.
The bosonic part of our theory is an abelian Higgs model
which can have an instanton configuration ---
vortex.
It has been known that in an abelian Higgs model
a Theta dependent vacuum energy density is generated by
the effect of the gas of vortices and anti-vortices \cite{CDG}.
As in that case, and also
as in Polyakov's model of confinement where a bosonic potential
for the dual field is
generated from the gas of monopoles and anti-monopoles,
we expect that a superpotential for $Y_i$'s
can be generated by the gas of vortices and anti-vortices.

Around the vortex for a charged scalar $\phi_i$,
the phase of $\phi_i$ has winding number one.
As we have seen in $R\to 1/R$ duality,
a winding configuration is dual to the insertion of the vertex operator
$\e^{i\vartheta_i}$.
The supersymmetric completion of this operator is
the twisted chiral superfield
\beq
\e^{-Y_i}.
\label{expo}
\eeq
These exponentials 
have vector R-charge $0$ and axial R-charge $2$, as can be seen from
(\ref{VYi}) and (\ref{AYi}).
Thus, we can add these to the twisted superpotential without
violating the $U(1)_V$ R-symmetry, and maintaining the correct anomaly
of $U(1)_A$.

In what follows we shall show that
a correction of the form (\ref{expo}) is indeed generated.
In fact we will show that the correction is simply the sum of them
and the exact superpotential is given by
\beq
\widetilde{W}
=\Sigma\left(\sum_{i=1}^NQ_iY_i-t(\mu)\right)
\,+\,\mu\,\sum_{i=1}^N\e^{-Y_i}.
\label{Weff}
\eeq
This is one of the main results of this paper.
Note that the change of the renormalization scale $\mu$
can be absorbed by the shift of $Y_i$'s dictated by
(\ref{renoYi}).  The actual parameter
 of the theory
 is still the dynamical scale $\Lambda$ for $\sum_iQ_i\ne 0$
and the FI-Theta parameter $t$ for $\sum_iQ_i=0$.

Before embarking on the computation to show that $\e^{-Y_i}$'s are
indeed generated,
we make a simple consistency check of (\ref{Weff}).
Let us consider
integrating out $Y_i$'s for a fixed configuration of
$\Sigma$ whose
lowest component of $\Sigma$ is large and slowly varying.
The variation with respect to $Y_i$'s yields the relation
$Q_i\Sigma-\mu\,\e^{-Y_i}=0$ or $Y_i=-\log(Q_i\Sigma/\mu)$.
Inserting this into (\ref{Weff})
we obtain the effective superpotential for the $\Sigma$ field
\beq
\widetilde{W}_{\it eff}(\Sigma)
=\Sigma\,\left(\,
-\sum_{i=1}^NQ_i\Bigl(\log\left(Q_i\Sigma/\mu\right)-1\Bigr)
-t(\mu)\,\right).
\label{Weffsig}
\eeq
This is nothing but what we would obtain when we integrate out
the chiral superfield $\Phi_i$'s in the original gauge theory
for a fixed configuration of
$\Sigma$ \cite{dadda,CV,phases,MP}.

\subsubsection{Localization}

~~~~We first establish that the validity of
(\ref{Weff}) for general $N$ and $Q_i$
is a consequence of the case with $N=1$.
This can be seen by considering a theory with a larger gauge symmetry
where the $U(1)^{N-1}$ global symmetries are gauged and recovering
the original theory in the weak coupling limit of the extra
gauge interactions.
  Sometimes
we will refer to this procedure as {\it localization}.

We start with the $U(1)^N$ gauge theory with a single charged
matter for each $U(1)$ which is described by
the following classical Lagrangian
\beq
L=
-\sum_{i,j}
\int\dd^4\theta\,{1\over 2e^2_{\bi j}}
\overline{\Sigma}_i\Sigma_j
+
\sum_{i=1}^N
\left\{\,\int\dd^4\theta\,
\overline{\Phi}_i\e^{2Q_iV_i}\Phi_i
+{1\over 2}\Biggl(-\int\dd^2\widetilde{\theta}\,
t_i\,\Sigma_i+c.c.\,\Biggr)\,
\right\}.
\eeq
If the gauge coupling matrix is diagonal
\beq
{1\over e^2_{\bi j}}
=\delta_{i,j}{1\over e^2_i},
\label{diag}
\eeq
the $N$ single flavor theories are decoupled from each other.
In such a case,
the exact twisted superpotential is given by
the sum of those for single flavor theories.
If we assume that (\ref{Weff}) holds for the single
flavor cases, it is
\beq
\widetilde{W}
=\sum_{i=1}^N
\left(\,\Sigma_i\Bigl(Q_iY_i-t_i(\mu)\Bigr)+\mu\,\e^{-Y_i}\,\right),
\label{weffsum}
\eeq
where we have chosen the scale $\mu$ common to all $i$, and
$t_i(\mu)$ is the effective FI-Theta parameter at $\mu$
\beq
t_i(\mu)=Q_i\log(\Lambda_i/\mu)-i\,\theta_i,
\eeq
for the $i$-th theory.

Now the essential point here
is that the (twisted) superpotential is
independent of variation of the D-term.
Thus, (\ref{weffsum}) is valid for all values of
$1/e^2_{\bi j}$ as long as there is no singularity in going
from the diagonal one (\ref{diag}).
In particular, let us consider the coupling matrix of the following form
\beq
\sum_{i,j}
{1\over 2e^2_{\bi j}}
\overline{\Sigma}_i\Sigma_j
={1\over 2e^2}\left|{1\over N}\sum_{i=1}^N\Sigma_i\right|^2
+{1\over \epsilon e^2}
\sum_{i=1}^{N-1}|\Sigma_i-\Sigma_{i+1}|^2.
\eeq
In the limit
\beq
\epsilon \rightarrow 0,
\label{decoup}
\eeq
the only dynamical gauge symmetry is the diagonal
$U(1)$ subgroup of $U(1)^N$
with the field strength given by
\beq
\Sigma_1+\Sigma_2+\cdots +\Sigma_N=:N\Sigma,
\eeq
and we recover the $U(1)$ gauge theory with $N$ matter
fields $\Phi_i$ with 
charge $Q_i$.
To be precise, the limit (\ref{decoup}) itself does not completely fix
$\Sigma_i-\Sigma_j$, but rather sets it to be a constant.  In other 
words we have the choice
$\Sigma_i=\Sigma+{\mit\Delta}_i$ with ${\mit\Delta}_i$ being a constant.
However, such a shift ${\mit\Delta}_i$
would yield a twisted superpotential whose perturbative part
does not match with what we have (\ref{multi}) for the single
$U(1)$ gauge theory. 
Therefore we must have ${\mit\Delta}_i=0$ in the present theory.
 Nevertheless as we will discuss later
a non-zero ${\mit\Delta}_i$ is indeed allowed
when we consider a perturbation of the theory with
``twisted masses''.  Therefore we will in general
take into account the above possible deformation of the $U(1)$
gauge theory.

The bare FI and Theta parameter of the $U(1)$ gauge theory is
related to those of the $U(1)^N$ theories by
\beq
t_0=\sum_{i=1}^N (Q_i\log(\Lambda_{UV}/\Lambda_i)-i\,\theta_i).
\eeq
In particular, for the theory
with $\sum_{i=1}^NQ_i\ne 0$, the dynamical scale $\Lambda$ is given by
$\prod_i\Lambda^{Q_i}=\prod_i\Lambda_i^{Q_i}$, while
the FI parameter of the theory with $\sum_{i=1}^NQ_i=0$
is $r=-\sum_{i=1}^NQ_i\log(\Lambda_i)$. 
Then, the effective coupling $t(\mu)$ at energy $\mu$
is simply the sum $\sum_{i=1}^Nt_i(\mu)$.
Thus, the twisted superpotential (\ref{weffsum})
becomes (\ref{Weff}) in the limit (\ref{decoup}).
This shows that (\ref{Weff}) for general $N$ and $Q_i$ follows from
the $N=1$ case.

Thus, to show (\ref{Weff}) we only have to show the single flavor case.
We note also that in the single flavor case
we only have to show it in the case of unit charge $Q=1$.
Other cases just follow from that case by a redefinition of the
gauge field and the FI-Theta parameter;
$Q\Sigma\to \Sigma$, $Qt\to t$.

\subsubsection{The Generation Of Superpotential}

~~~~Now, let us consider the single flavor case with $Q=1$.
We first determine the
possible form of the non-perturbative correction
${\mit\Delta}\widetilde{W}$
 from the general requirements \cite{Seiberg}
--- holomorphy in $t$, periodicity in Theta angle, R-symmetry,
and asymptotic behaviour.
Let $\Lambda=\mu\e^{-t}$ be the dynamical scale and
let us put $\widetilde{Y}=Y-t$.
Since $t$ and $Y$ are periodic with period $2\pi i$,
${\mit\Delta}\widetilde{W}$ must be a holomorphic function of
$\Sigma$, $\Lambda$ and $\e^{-\widetilde{Y}}$.
The anomaly (\ref{anomaly})
of the axial R-symmetry is absorbed by the shift of the
Theta angle $\theta\to\theta+2\alpha$, or $t\to t-2i\alpha$.
This modified $U(1)_A$ symmetry transforms the variables as
\beq
\Sigma\to\e^{2i\alpha}\Sigma,~~
\Lambda\to\e^{2i\alpha}\Lambda,~~
\e^{-\widetilde{Y}}\to\e^{-\widetilde{Y}}.
\eeq
Since the twisted superpotential must have
charge $2$ under this transformation,
${\mit\Delta}\widetilde{W}$ must be of the form
$\Sigma f(\Lambda/\Sigma,\e^{-\widetilde{Y}})$
which is expanded in a Laurent series as
\beq
{\mit\Delta}\widetilde{W}
=\Sigma\sum_{n,m}c_{n,m}(\Lambda/\Sigma)^n\e^{-m\widetilde{Y}}
=\sum_{n,m}c_{n,m}\Sigma^{1-n}\mu^n\e^{-(n-m)t-mY}.
\label{laurent}
\eeq
Now, we recall that the field
$Y$ was introduced by dualizing the circle
of radius $|\phi|$.
Therefore, in the description in terms of $Y$ and $\Sigma$,
we are looking at the region $\phi\ne 0$ of the field space
where the gauge symmetry is broken and
$\Sigma$ is therefore massive.
Thus,
the twisted superpotential must be analytic at $\Sigma=0$.
This means that only terms of $1-n\geq 0$ is non-vanishing
in (\ref{laurent}).
On the other hand,
in the semi-classical limit where $r$ is very large,
 the correction must be small compared to
the perturbative term $\Sigma(Y-t)$.
This requires $n\geq m$.
Finally, since $Y$ is unbounded in real positive direction
(but is bounded from below as ${\rm Re}Y\sim |\Phi|^2\geq 0$
in the semi-classical description), $m\geq 0$ is also required
for the correction to be small.
To summarize, only $1\geq n\geq m\geq 0$ is allowed.
$n=m=0$ is of the same order as the perturbative term and is
excluded. $n=1, m=0$ is just a constant term. The final candidate
$n=m=1$ is a non-trivial term which is $\e^{-Y}$.
Thus, the twisted superpotential must be of the form
\beq
\widetilde{W}=\Sigma(Y-t(\mu))+c\mu\,\e^{-Y},
\label{Weffsingle}
\eeq
where $c$ is a dimensionless constant.

The question is therefore whether the coefficient $c$
is zero or not.
We have already observed an
evidence that supports non-vanishing of $c$;
Integration over $Y$ yields
the correct effective superpotential for $\Sigma$,
$\widetilde{W}_{\it eff}(\Sigma)=\Sigma\log(\Sigma/\Lambda)$,
if $c$ is non-zero.
In what follows we show
that the term $\mu\,\e^{-Y}$ is indeed generated by an instanton effect.

\subsection*{\it The Vortex}

We continue our gauge theory to
the Euclidean signature by Wick rotation
$x^0=-ix^2$. We choose the orientation so that $z=x^1+ix^2$
is the complex coordinates.
This leads to $F_{01}\to -iF_{12}$, $D_0+D_1\to 2D_{\bar z}$
and $D_0-D_1\to -2D_z$.
After solving for the auxiliary field
as $D=-e^2(|\phi|^2-r_0)$ and $F=0$,
the Euclidean action is given by
\beqa
S_E\,=\,{1\over 2\pi}\int\dd^2x&&\!\!\!\!\!\!\!\!
\Biggl(~
|D_{\mu}\phi|^2+|\sigma\phi|^2
+{1\over 2e^2}|\partial_{\mu}\sigma|^2+{1\over 2e^2}(F_{12}^2+D^2)
+i\theta F_{12}
\nonumber\\
&&\!\!
-2i\bpsi_-D_{\bar z}\psi_-+2i\bpsi_+D_z\psi_+
+\bpsi_-\sigma\psi_++\bpsi_+\overline{\sigma}\psi_-
\nonumber\\
&&\!\!
+{1\over e^2}\left(\,-i\blambda_-\partial_{\bar z}\lambda_-
+i\blambda_+\partial_z\lambda_+\,\right)
\nonumber\\
&&\!\!
+i\left(\,
\phi^{\dag}\lambda_-\psi_+-\phi^{\dag}\lambda_+\psi_-
-\bpsi_+\blambda_-\phi+\bpsi_-\blambda_+\phi\,\right)~\Biggr).
\label{SE}
\eeqa
An instanton is a topologically non-trivial configuration
that minimizes the bosonic
part of this action.

An instanton can contribute to the
(twisted) superpotential only when it carries two
fermionic zero modes of the right kind.
Since a twisted F-term is obtained by the integration over
two fermionic coordinates other than $\theta^-$ and $\btheta^+$, 
a relevant configuration must be invariant under the supercharges
$Q_-$ and $\overline{Q}_+$.
The invariance of the fermions under these supercharges
requires (see Appendix B)
\beqa
&&\sigma=0,
\label{vort0}\\
&&D_{\bar z}\phi=0,
\label{vort1}\\
&&F_{12}=e^2(|\phi|^2-r_0).
\label{vort2}
\eeqa
The bosonic part of the action
is ${1\over 2\pi}\int
\dd^2x({1\over 2e^2}|\partial_{\mu}\sigma|^2+|\sigma\phi|^2)$
plus
\beqa
\lefteqn{{1\over 2\pi}\int\dd^2 x\left(\,
|D_{\mu}\phi|^2+{1\over 2e^2}(F_{12}^2+D^2)
+i\theta F_{12}\,\right)}
\nonumber\\
&&
={1\over 2\pi}\int\dd^2x\left(\,
|2D_{\bar z}\phi|^2-F_{12}|\phi|^2
+{1\over 2e^2}(F_{12}+D)^2-{1\over e^2}DF_{12}+i\theta F_{12}\,
\right)
\nonumber\\
&&
={1\over 2\pi}\int\dd^2x\left(
|2D_{\bar z}\phi|^2+{1\over 2e^2}(F_{12}+D)^2\,\right)
-{t_0\over 2\pi}\int F_{12}\,\dd^2x,
\eeqa
where $D=-e^2(|\phi|^2-r_0)$ and $t_0=r_0-i\theta$.
For a given topological number
\beq
k=-{1\over 2\pi}\int F_{12}\,\dd^2x,
\eeq
the real part of the
action is bounded by $kr_0$,
and the minimun is indeed attained by
a solution to the equations (\ref{vort0})-(\ref{vort2}).
The value of the action for such an instanton is
\beq
S_E=kt_0.
\eeq
Under the axial rotation by $\e^{i\alpha}$,
the path-integral measure in this topological sector
changes by the phase $\e^{2ik\alpha}$. Since the twisted superpotential
has axial R-charge 2, we see that the relevant configurations are
those with $k=1$.

A solution to (\ref{vort1}) and (\ref{vort2})
is the vortex.
For each vortex with $k=1$, $\phi$ has a single simple zero.
The moduli space of gauge equivalence classes of
$k=1$ vortices is complex one-dimensional and is
parametrized by the location of the zero of $\phi$.
To see this, we note the following well-known fact.
The orbit of a solution to (\ref{vort1}) under
the complexified gauge transformations
contains vortex solutions in one gauge equivalence class,
and conversely, any gauge equivalence class of vortex solutions is
contained in one such orbit.
Here, a complexified gauge transformation is a rotation
$(iA_{\bar z},\phi)\to(iA_{\bar z}+h\partial_{\bar z}h^{-1},h\phi)$
by a function $h$ with values in $\C^{\times}$.
Thus, we only
have to find solutions to the equation $D_{\bar z}\phi=0$
modulo the complexified gauge transformations.
In other words, we only have to find pairs of
a holomorphic line bundle with a holomorphic section.
Here,
it is convenient to compactify our Euclidean 2-plane to a Riemann
sphere.
Then, there is a unique holomorphic line bundle of $k=1$.
Such a bundle has two-dimensional
space of holomorphic sections, where each section has a single simple zero.
The residual complexified gauge symmetry is a multiplication by
a constant and it does not change the location of the zero
of a section.
Thus, the space of
equivalence classes is complex one-dimensional and is parametrized
by the zero locus of $\phi$.

Let us examine in more detail
the behaviour of a vortex solution.
We consider the vortex at $z=0$.
For the finiteness of the action,
$|\phi|$ must approach the vacuum value $|\phi|^2=r_0$ at infinity.
By rescaling $\phi=\sqrt{r_0}\widehat{\phi}$
where $|\widehat{\phi}|\to 1$ at infinity,
it becomes clear that the equations
depend only on one length parameter $1/e\sqrt{r_0}$.
This characterizes the size of the vortex.
Thus, we expect that
the gauge field is nearly flat on $|z|\gg1/e\sqrt{r_0}$
and $\widehat{\phi}$ is nearly covariantly constant there.
Since $(-1/2\pi)\int F_{12}=1$,
we have $A_{\mu}=-\partial_{\mu}\arg(z)$ and
$\phi=\sqrt{r_0}z/|z|$ at infinity.
The exact solution takes the form
\beqa
iA&=&-{1-f\over 2}\left(\,{\dd z\over z}-{\dd \bar z\over \bar z}\,
\right),\\
\phi&=&\sqrt{r_0}
\exp\left(-\int^{\infty}_{|z|^2}
{\dd w f(w)\over 2w}\right)\,
{z\over |z|},
\eeqa
where $f$ is a function of $w=|z|^2$
which satisfies the equation
$wf^{\prime\prime}={e^2r_0\over 2}f+ff^{\prime}$
and the boundary condition $f(0)=1,f(+\infty)=0$.
The asymptotic behaviour of the function $f(|z|^2)$
at $|z|\gg 1/e\sqrt{r_0}$ is
\beq
f={\it const}\,\sqrt{m|z|}\,\e^{-m|z|}+\cdots,
\eeq
where {\it const} is a numerical constant and 
\beq
m=e\sqrt{2r_0}.
\eeq
The vortex at $z=z_0$
is obtained from the above
solution
simply by the replacement $z\to z-z_0$.

\subsection*{\it Fermionic Zero Modes}

Now let us examine the fermionic zero modes in the instanton background.
Since $\sigma=0$ in this background,
the fermionic part of the action (\ref{SE})
decomposes into two parts as
\beq
-i(\bpsi_-,\lambda_+)
\left(\begin{array}{cc}
2D_{\bar z}&\!\!\!-\phi\\
\phi^{\dag}&\!\!\!-{1\over e^2}\partial_z
\end{array}\right)
{\psi_-\choose\blambda_+}
+i(\bpsi_+,\lambda_-)
\left(\begin{array}{cc}
2D_{z}&\!\!\!-\phi\\
\phi^{\dag}&\!\!\!-{1\over e^2}\partial_{\bar z}
\end{array}\right)
{\psi_+\choose \blambda_-}
\label{zeroeqn}
\eeq
Using the index theorem of \cite{EWeinberg}, one can see
that the operators of the first and the second terms have
index $1$ and $-1$ respectively in our background.
Furthermore, there is no normalizable zero modes for
$(\bpsi_-,\lambda_+)$ nor $(\psi_+,\blambda_-)$.
To see this we note that
\beqa
\lefteqn{
\int\dd^2z\left(\Bigl|2D_z\psi_+-\phi\blambda_-\Bigr|^2
+2e^2\Bigl|\phi^{\dag}\psi_+
-{1\over e^2}\partial_{\bar z}\blambda_-\Bigr|^2\right)}\\
&=&
\int \dd^2z\left(|2D_z\psi_+|^2+2e^2|\phi|^2|\psi_+|^2
+{2\over e^2}|\partial_{\bar z}\blambda_-|^2
+|\phi|^2|\blambda_-|^2\right)
\eeqa
in the vortex background,
where $\psi_+$ and $\blambda_-$ are considered here
as commuting spinors.
The left hand side vanishes for a zero mode of
$(\psi_+,\blambda_-)$,
but then vanishing of the right hand side requires
$\psi_+=\blambda_-=0$.
The argument for $(\bpsi_-,\lambda_+)$ is the same.

Thus, each of $(\psi_-,\blambda_+)$ and
$(\bpsi_+,\lambda_-)$
has exactly one zero mode.
Actually, the expression for the zero mode in terms of
the vortex solution is available;
\beq
\left(\begin{array}{c}
\!\psi_-^{(0)}\!\!\\[0.1cm]
\!\blambda_-^{(0)}\!\!
\end{array}\right)
=\left(\begin{array}{c}
\!D_z\phi\!\\[0.1cm]
\!F_{12}\!
\end{array}\right),~~
\left(\begin{array}{c}
\!\bpsi_+^{(0)}\!\!\\[0.1cm]
\!\lambda_+^{(0)}\!\!
\end{array}
\right)=
\left(\begin{array}{c}
\!D_{\bar z}\phi^{\dag}\!\\[0.1cm]
\!F_{12}\!
\end{array}
\right).
\eeq
It is easy to verify using the vortex equations (\ref{vort1})
and (\ref{vort2}) that these are indeed the zero modes.
These come from the supersymmetry transformation under
the broken supercharges $\bQ_-$ and $Q_+$.
The asymptotic behaviour of $\phi^{\dag}\psi_-^{(0)}$
and
$\bpsi_+^{(0)}\phi$
are
\beqa
\phi^{\dag}\psi_-^{(0)}\!&=&\!\partial_z|\phi|^2
={\it const}\times r_0\,{\bar z\over |z|}\,\sqrt{m\over |z|}\e^{-m|z|}
+\cdots\\
\bpsi_+^{(0)}\phi\!&=&\!\partial_{\bar z}|\phi|^2
={\it const}\times r_0\,{z\over |z|}\,\sqrt{m\over |z|}\e^{-m|z|}
+\cdots,
\eeqa
for the vortex at $z=0$.
These agree with the asymptotic behaviour of the propagators
for a Dirac fermion of mass $m=e\sqrt{2r_0}$;
\beq
\phi^{\dag}\psi_-^{(0)}\propto r_0\,S_{--}^F(z),~~~
\bpsi_+^{(0)}\phi\propto - r_0\,S_{++}^F(z).
\label{zeroprop}
\eeq
This is not an accident;
After a suitable similarity transformation,
the operators in (\ref{zeroeqn})
near infinity (where $\phi=\sqrt{r}z/|z|$
and $A_{\mu}=-\partial_{\mu}\arg(z)$) become nothing but the Dirac
operator for a free fermion of mass $m=e\sqrt{2r}$.

\subsection*{\it The Computation}

In order to prove the generation of $\e^{-Y}$ term in the twisted
superpotential, we would like
to compute a quantity that is non-vanishing
only when such a term is generated.
To find out what is the appropriate object to look at,
let us return to our theory of $\Sigma $ and $Y$.
We recall that the Lagrangian is
\beq
\widetilde{L}=\int\dd^4\theta\left(\,
-{1\over 2e^2}\overline{\Sigma}\Sigma
-{1\over 4r_0}\overline{Y}Y\,\right)
+{1\over 2}\Biggl(\,\int\dd^2\widetilde{\theta}\,
\left(\,\Sigma(Y-t)+c\mu\,\e^{-Y}\right)\,+\, c.c.\,\Biggr),
\eeq
and we were asking whether $c$ is zero or not.
Here we have approximated the Kahler potential for $Y$ by the
one in the continuum limit 
(\ref{ymet}).

This theory involves a $U(1)$ gauge field.
However, there is no charged field and
the only appearance of the gauge field is in the kinetic term
and in the Theta term (with the Theta angle being
$\vartheta-\theta$ where $\vartheta=-{\rm Im}(y)$).
As is well known \cite{Coleman},
the effect of the gauge field is to generate a mass term
for $\vartheta$:
\beq
U={e^2\over 2}\left(\widetilde{\vartheta-\theta}\right)^2,
\eeq
where $(\widetilde{\alpha})^2=
{\rm min}\{(\alpha+2\pi n)^2|n\in \Z\}$.
Thus, we can treat $\Sigma$ as an ordinary twisted chiral superfield
which has a complex auxiliary field.
In particular, the theory without $\e^{-Y}$ term
is a free theory of two twisted chiral multiplets.
It is easy to diagonalize the $\Sigma Y$ mixing and it turns out
that the combinations
$X^{(\pm)}=\pm\Sigma/(2e)+(Y-t)/(2\sqrt{2r_0})$
are superfields of mass
\beq
\pm m=\pm e\sqrt{2r_0}.
\eeq
Now, it is easy to see that
the fermionic components $\chi_+$ and $\bchi_-$ of $\overline{Y}$
has vanishing two point function in the free theory;
\beq
\langle\chi_+(x)\,\bchi_-(y)\rangle=0, ~~~\mbox{if ~$c=0$}.
\eeq
However, if $\e^{-Y}$ is generated, the twisted F-term
would contain a term $-\e^{-y}\bchi_+\chi_-$ in the Lagrangian.
This would contribute to the two point function as
\beq
\langle\chi_+(x)\,\bchi_-(y)\rangle=c\mu r_0^2\int\dd^2z\,
\e^{-t}\,S_{++}^F(x-z)S_{--}^F(z-y)
\label{effxx}
\eeq
where $S_{\alpha\beta}^F(x-y)$ is the Dirac propagator for the fermions of
mass $m=e\sqrt{2r_0}$.

Now, let us compute the two point function
$\langle\chi_+(x)\,\bchi_-(y)\rangle$ in the original gauge theory.
We recall from (\ref{chipsi2}) that
\beq
\chi_+=2\bpsi_+\phi,~~\bchi_-=-2\phi^{\dag}\psi_-.
\eeq
Since the product of
these carries an axial R-charge 2,
only the vortex backgrounds with
$k=1$ can contribute to the two point function.
The contribution is expressed as an integration
over the location $z$ of the vortex
\beq
\langle\chi_+(x)\,\bchi_-(y)\rangle=-4 \Lambda_{UV}\int\dd^2z\,
\e^{-t_0}\Bigl(\bpsi_{+}^{(0)}\phi_{(z)}\Bigr)(x)
\left(\phi_{(z)}^{\dag}\psi_-^{(0)}\right)(y),
\eeq
where $\phi_{(z)}$ is the vortex solution at $z$.
The factor of $\Lambda_{UV}$ comes from the measure of
bosonic zero modes ($\propto \Lambda_{UV}^2$)
and that for the fermionic zero modes
($\propto \Lambda_{UV}^{-1}$).
As we have seen, the fermionic zero mode multiplied by $\phi$
is proportional to the Dirac propagator (\ref{zeroprop}).
Thus, we obtain
\beq
\langle\chi_+(x)\,\bchi_-(y)\rangle={\it const}\times
r_0^2 \Lambda_{UV}\int\dd^2z\,
\e^{-t_0}\,S_{++}^F(x-z)S_{--}^F(z-y),
\eeq
which agrees with (\ref{effxx}) considering the relation
$\Lambda_{UV}\e^{-t_0}=\mu\,\e^{-t}$.
Thus, we have shown that our dual theory
correctly reproduces the gauge theory result if and only if $c\ne 0$.

\subsubsection{Solitons and Dualization}
In the $R\rightarrow 1/R$ duality, as reviewed before,
the momentum and winding modes get exchanged.  This view
provides us with another way to interpret the generation
of superpotential (\ref{Weffsingle}).
Let us turn off the gauge interaction and consider $\Sigma$ as a
non-dynamical parameter.
Before dualization, we have a field $\Phi$ of mass
$\Sigma$.
In the dual description $\Phi$ should arise as a winding mode.
Indeed if
we consider the superpotential
\beq
W=\Sigma(Y-t)+e^{-Y}.
\eeq
Viewing $\Sigma$ as non-dynamical, the
vacua are labeled by $\partial_YW=0$ and we obtain
\beq
\partial_YW=0\rightarrow e^{-Y}=\Sigma
\label{critw}
\eeq
and the critical points are given by
\beq
Y_n=Y_0+2n \pi i
\eeq
where $Y_0$ is a special solution of (\ref{critw}).
The vacua are indexed by an integer $|n\rangle $, corresponding
to winding number in the $Y$ plane.  This should correspond
to momentum mode in the $\Phi$ variable.  In other words
the $\Phi$ field should have $Y$-winding number charge $+1$
and acts on vacua
\beq
\Phi: |n\rangle \rightarrow |n+1\rangle
\eeq
In other words $\Phi$ should be identified with the soliton
interpolating between the vacua.  The BPS mass in this sector is
given by \cite{mathuretal}
\beq
|m|={1\over 2\pi}|W(Y_n)-W(Y_{n-1})|= |\Sigma|,
\eeq
in agreement with the expected mass of $\Phi$.
This reasoning provides another view point on the superpotential
in the actual gauge system.
Related ideas were recently discussed in \cite{oogv}.

\subsection{A Few Generalizations}

It is straightforward to extend the description
using the dual fields to more general gauge theories.
We consider here two generalizations;
the case with $U(1)^k$ gauge group and the case with twisted masses.

\subsection*{\it Many $U(1)$ Gauge Groups}

The first example is $U(1)^k$ gauge group with $N$ matter fields.
We denote the field strength superfield for the $a$-th gauge group
by $\Sigma_a$ ($a=1,\ldots,k$), and the chiral superfield for the
$i$-th charged matter by $\Phi_i$ ($i=1,\ldots,N$).
We denote by $Q_{ia}$ the charge of the $i$-th matter
under the $a$-th gauge group.

The exact twisted superpotential can be obtained
by the localization argument which reduces the problem to
the sum of copies of the single flavor case.
This time, instead of keeping only one gauge coupling
we keep $k$ of them.
We start with the sum of
$N$ copies of $U(1)$ theory with charge $1$ matter,
and take the weak coupling limit except for the
$U(1)^k$ gauge group embedded in $U(1)^N$ according to the charge matrix
$Q_{ia}$.
This constrains the $N$ gauge fields as
\beq
\Sigma_i=\sum_{a=1}^kQ_{ia}\Sigma_a.
\label{freeze}
\eeq
The exact superpotential for $\Sigma_a$'s and the dual
$Y_i$ of $\Phi_i$ is then given by
\beq
\widetilde{W}=
\sum_{a=1}^k\Sigma_a\left(\sum_{i=1}^NQ_{ia}Y_i-t_a\right)
+\mu\sum_{i=1}^N\,\e^{-Y_i}.
\label{multig}
\eeq
Integrating over $Y_i$'s, we obtain the following effective
twisted superpotential for $\Sigma_a$'s:
\beq
\widetilde{W}_{\it eff}(\Sigma_a)=
-\sum_{a=1}^k\Sigma_a\left(
\,\sum_{i=1}^NQ_{ia}\left(\log\Bigl(
\sum_{b=1}^kQ_{ib}\Sigma_b/\mu\Bigr)-1\right)
+t_a\,\right).
\label{multiWeffsig}
\eeq
This is the effective superpotential
that we would obtain if we integrate out $\Phi_i$'s
in the original gauge theory \cite{MP}.

\subsection*{\it Twisted Masses}

The next example is $U(1)^k$ gauge theory with $N$ charged matter fields
as above,
but the twisted masses $\widetilde{m}_i$ for the matter fields are
turned on.
The twisted mass can be considered as the lowest components of the
field strength superfields of the $U(1)^N/U(1)^k$
flavor symmetry group.
This is a non-trivial deformation of our gauge theory.
The (anomalous) axial R-symmetry is explicitly
broken by this perturbation
but is restored if $\widetilde{m}_i$ are rotated as
$\widetilde{m}_i\mapsto\e^{2i\alpha}\widetilde{m}_i$.

As noted before, the dualization for this case follows
from extending the $U(1)^k$ gauge group to $U(1)^N$ and
taking the suitable decoupling limit.
With the twisted masses, the constraint on the field strengths
(\ref{freeze}) is shifted as
\beq
\Sigma_i=\sum_{a=1}^kQ_{ia}\Sigma_a-\widetilde{m}_i.
\eeq
The exact twisted superpotential is thus
\beq
\widetilde{W}=\sum_{a=1}^k\Sigma_a\left(
\sum_{i=1}^NQ_{ia}Y_i-t_a\right)
+\mu\sum_{i=1}^N\,\e^{-Y_i}-\sum_{i=1}^N\widetilde{m}_iY_i.
\label{twism}
\eeq
Note that the net number of
deformation parameters is $(N-k)$ from the flavor group
$U(1)^N/U(1)^k$;
$\delta\widetilde{m}_i=\sum_{a=1}^kQ_{ia}c_a$ is absorbed by
a shift of the origin of $\Sigma_a$'s.
Integration over $Y_i$'s yields an
effective superpotential for $\Sigma$
that we would obtain if we integrate out $\Phi_i$'s in the original
gauge theory (as is done in
\cite{HH} for $k=1$ case with $Q_i=\pm 1$).

\section{Sigma Models From Gauge Theories}

The gauge theory studied in the previous section
reduces at low enough energies to the non-linear sigma model
on a certain manifold.
To see this, we examine
the space of classical vacua of the theory.
This
can be read by looking at the potential energy
for the scalar fields $\sigma,\phi_i$
\beq
U={e^2\over 2}
\left(\,\sum_{i=1}^NQ_i|\phi_i|^2-r_0\,\right)^{\!2}
+\sum_{i=1}^NQ_i^2|\sigma|^2|\phi_i|^2.
\label{potU}
\eeq
We notice two branches of solutions to the vacuum equation
$U=0$;
Higgs branch where $\sigma=0$ and $\sum_{i=1}^NQ_i|\phi_i|^2=0$,
and Coulomb branch where $\sigma$ is free and all $\phi_i=0$.
If $\pm\sum_{i=1}^NQ_i> 0$, $\pm r_0$ is bound to be large
and there is only a Higgs branch.
If $\sum_{i=1}^NQ_i=0$,
there is again only a Higgs branch except $r_0\sim 0$
where a Coulomb branch develops.
The degrees of freedom transverse to the
Higgs branch have masses of order $e\sqrt{|r_0|}$, as can be seen from
the Lagrangian $\sum_i|D\phi_i|^2+(1/2e^2)|\dd\sigma|^2+U$.
Thus, for a generic value of the parameter $r_0$,
the theory at energies much smaller than $e\sqrt{|r_0|}$
describes the non-linear sigma model on the Higgs branch.
This in particular means that all aspects of
the sigma model at finite energies (including the
BPS soliton spectra for massive sigma models) 
can be seen by studying the corresponding
gauge theory.

To be more precise, the Higgs branch $X$ is the space of solutions
to
\beq
\sum_{i=1}^NQ_i|\phi_i|^2=r_0
\label{moment}
\eeq
modulo $U(1)$ gauge transformations
$\phi_i\mapsto\e^{iQ_i\gamma}\phi_i$.
This space has complex dimension $N-1$ and
inherits a structure of Kahler manifold from
that of flat $\C^N$ of $\phi_i$'s. 
It is a standard fact that this is equivalent as a complex manifold
to the quotient of $\C^N-{\cal P}$ by the $\C^{\times}$
action $\phi_i\mapsto \lambda^{Q_i}\phi_i$, where
${\cal P}$ is some subset of codimension $\geq 1$.
In particular, complex coordinates of $X$ are represented
in the sigma model by the lowest components of chiral superfields.
The parameter
$t_0=r_0-i\theta$ is identified as the complexified Kahler class.
The first Chern class of this space is proportional to
$|\sum_{i=1}^NQ_i|$ and the cut-off dependence (\ref{renor})
of $r_0$ corresponds to
the renormalization of the sigma model metric \cite{AF}.
The sigma model limit is thus
\beq
e\gg\Lambda 
\eeq
for $\sum_{i=1}^NQ_i\ne 0$.

The space $X=(\C^N-{\cal P})/\C^{\times}$ has
an algebraic torus $(\C^{\times})^{N-1}$ as a group of holomorphic
automorphisms; the group $(\C^{\times})^N$ acting on $\C^N$ in a standard
way modulo the comlpexified gauge group $\C^{\times}$.
There is an open subset $\{\phi_i\ne 0,\forall i\}$ of $X$ on which
$(\C^{\times})^{N-1}$ acts freely and transitively.
Such a space $X$ is called a {\it toric variety}, or {\it toric manifold}
if it is smooth.
As is clear from the equation (\ref{moment}), if
$Q_i$ are all positive (or all negative), the manifold $X$ is compact
but if there is a mixture of positive and negative $Q_i$'s
$X$ is non-compact.

For $\sum_{i=1}^NQ_i=0$, the FI parametr $r=r_0$ does not run and
$t=r-i\theta$ is a free parameter of the theory.
At $r=0$, $X$ contains a singular point $\phi_i=0$ where the
$U(1)$ gauge group is unbroken.
A new flat direction (Coulomb branch) develops there
and the sigma model on $X$ becomes singular. 
The actual singularity is determined
by the vanishing of the quantum effective potential
at large values of $\sigma$.
Since there is an effective superpotential 
$\widetilde{W}_{\it eff}(\sigma)$ (\ref{Weffsig}) which
is valid at large $\sigma$, this is equivalent to the
condition that $\partial_{\sigma}\widetilde{W}_{\it eff}(\sigma)=0$
at large values of $\sigma$.
Thus, singularity of the quantum theory is located at
\beq
t=-\sum_{i=1}^NQ_i\log(Q_i).
\eeq
Note that the theory is singular only for
a particular value of $\theta$ (either $0$ or $\pi$) and $r$.
In particular, the theories with $r\gg0$ and $r\ll 0$
are smoothly connected to each other \cite{phases},
even though the space $X$ at $r\gg 0$ differs from
$X$ at $r\ll 0$.

One can also consider abelian gauge theory
with gauge group $U(1)^k$ and $N$ matter fields with $N\times k$
charge matrix $Q_{ia}$.
For a generic value of $t_a$ in a certain range,
the vacuum manifold is a Kahler manifold $X$ of dimension $N-k$.
$X$ as a complex manifold is of the type
$(\C^N-{\cal P})/(\C^{\times})^k$ where
${\cal P}$ is a union of certain planes in $\C^N$.
The algebraic torus $(\C^{\times})^{N-k}=(\C^{\times})^N/(\C^{\times})^k$
acts on $X$ in an obvious way as holomorphic automorphisms
and $X$ contains an open subset on which $(\C^{\times})^{N-k}$
acts freely and transitively.
Thus, $X$ is a toric variety. (In fact, any normal toric variety is
obtained this way). We refer the physics reader to
\cite{MP} for more precise definition
of a toric variety and its general properties.
See also \cite{Oda,Fulton} for more detail.
If $\sum_{i=1}^NQ_{ia}=0$ for all $a$,
the theory is parametrized by $k$ dimensionless
parameters $t_a=r_a-i\theta_a$.
Otherwise, there is a single scale parameter
and $k-1$ dimensionless parameters.
Unlike in the single $U(1)$ case,
the theory can possibly become singular
at some locus even if $\sum_{i=1}^NQ_{ia}\ne 0$.
Singular locus in the parameter space is
determined by finding a flat direction in the $\sigma_a$ space;
$\partial_{\sigma_a}\widetilde{W}_{\it eff}=0$ where
$\widetilde{W}_{\it eff}$ is given in (\ref{multiWeffsig})
(more precisely, we must consider all possible ``mixed branches''
and do the same computation in the reduced theory).
This was studied in detail in \cite{MP}.
However, as in the single $U(1)$ case,
two generic points in the parameter space can be smoothly
connected to each other without meeting a singularity.

The first Chern class $c_1(X)$ of $X$ is not necessarily
positive semi-definite.
In the present paper, we only consider $X$ with $c_1(X)\geq 0$.
In such a case, the running of the FI parameter of the gauge theory
matches the running of the sigma model coupling
at one-loop level, and our gauge theory
indeed describes the non-linear sigma model on $X$ at energies well below
the gauge coupling constants.
To be more precise,
the precise relation of the parameters is possibly complicated
when $c_1(X)$ is close to zero
and the one-loop running is not dominant.
This can also be understood by comparing the size of the moduli spaces of
gauge theory instantons and instantons of the non-linear sigma model
(as shown in \cite{phases} in the case of projective hypersurfaces).
It was also noted in \cite{LNS} that the formal parameters
corresponding to
irrelevant deformations are in complicated relation
even when $c_1$ is large.
The precise relation can be determined by finding the so called
flat coordinates with the expansion point at infinity. That would
lead to the natural coordinates used in the large volume expansion of
the non-linear sigma model \cite{BCOV}.
(This in particular applies to the mirror QFTs that we will obtain later
in this paper in finding the map 
between the parameters of the non-linear sigma model and the mirror.)
If there is a negative component in $c_1(X)$,
the sigma model is not asymptotic free and is not well-defined.
In such a case, our gauge theory has little to
do with the manifold $X$.
If $c_1(X)$ is negative definite, it is infra-red free and
the sigma model (defined as a cut-off theory)
flows to a free theory of $c/3=\dim X$.

\subsubsection{Examples of Toric Varieties}

Let us consider some examples of toric varieties.
If we consider a $U(1)$ gauge theory with $N$ fields with charges
$+1$, this gives a linear sigma model realization of $\CP^{N-1}$.
More generally, if the charges of the matter
fields are positive but not necessarily equal, it gives
a realization of weighted projective space, with weights
determined by the charges. 

If we consider a $U(1)$ gauge theory with $N$ fields
with charge $+1$ and one with charge $-d$, this gives a realization of
the total space of the ${\cal O}(-d)$ line bundle over $\CP^{N-1}$.

Hirzebruch surface $F_a$ ($a=0,1,\ldots$) is a toric manifold of
dimension two
which is realized as the vacuum manifold of the
$U(1)\times U(1)$ gauge theory with four chiral fields with charges
$(1,0)$, $(1,0)$, $(0,1)$, and $(a,1)$.
$F_0$ is the product $\CP^1\times \CP^1$ and
$F_1$ is the blow up of $\CP^2$ at one point.
The first Chern class is positive
for these two cases and
it is positive-semi-definite for $F_2$.
In all these cases, the gauge theory
describes the non-linear sigma model.
For $a\geq 3$, however, $c_1(F_a)$ is not even positive semi-definite
and the sigma model is not well-defined.

\subsubsection{Holomorphic Vector Fields and Deformations of Linear
Sigma Model}

The unitary subgroup $U(1)^{N-k}$ of the holomorphic
automorphism group $(\C^{\times})^{N-k}$ is actually an isometry group
of the Kahler manifold $X$.
Since it is an abelian group,
as explained in section 2,
we can use this to deform the sigma model on $X$ by a potential
term like (\ref{LV}).
One may be interested in whether the deformation can be
realized in the gauge theory.
In fact, this $U(1)^{N-k}$ is a commutative subgroup of
the flavor symmetry group of the gauge theory, and we can consider the
deformation by twisted masses.
By construction, the potential deformation of the
sigma model is naturally identified as
the twisted mass deformation of the gauge group.
One can also confirm this by computing the bosonic potential
in gauge theory and by checking that it agrees with
the potential ${1\over 2}|\widetilde{m}_AV_A|^2
+{1\over 2}|\overline{\widetilde{m}}_AV_A|^2$
from the holomorphic isometry (where $V_A$ are the generators of
$U(1)^{N-k}$). We exhibit this in the simplest case
$X=\CP^1$ and leave the general case as an exercise.
The $\CP^1$ sigma model is realized by a $U(1)$ gauge theory
with two fields $\Phi_1$ and $\Phi_2$ of unit charge.
The potential of the gauge theory with the twisted mass is given by
\beq
U={e^2\over 2}(|\phi_1|^2+|\phi_2|^2-r)^2
+|\sigma-\widetilde{m}|^2|\phi_1|^2+|\sigma|^2|\phi_2|^2.
\label{Um}
\eeq
We now fix $\phi_1$ and $\phi_2$ to lie in the $\CP^1$ before
perturbation ($|\phi_1|^2+|\phi_2|^2=r$) and then extremize
the potential with respect to $\sigma$.
Plugging the result back into (\ref{Um}), we obtain the potential
\beq
U_{\widetilde{m}}={|\widetilde{m}|^2\over r}|\phi_1|^2|\phi_2|^2.
\label{Umex}
\eeq
On the other hand, the $U(1)$ isometry generates a holomorphic vector
field with $V^z=iz\partial/\partial z$ where $z$ is the coordinate of
$\CP^1$ given by
$z=\phi_1/\phi_2$. The metric of $\CP^1$ determined by the
quotient is the standard Fubini-Study metric
$\dd s^2=r|\dd z|^2/(1+|z|^2)^2$. Measuring $V$ by this,
we obtain
the potential
\beq
|\widetilde{m}|^2\parallel V\parallel^2=
|\widetilde{m}|^2{r|z|^2\over(1+|z|^2)^2}
=|\widetilde{m}|^2{|\phi_1|^2|\phi_2|^2\over r}
\eeq
which is nothing but (\ref{Umex}).

\subsection{Hypersurfaces and Complete Intersections}

The non-linear sigma models of
hypersurface or complete intersections in a compact
toric manifold
can also be realized as a gauge theory \cite{phases}.
Let $X$ be the compact toric manifold realized as the vacuum manifold
of $U(1)^k$ gauge theory with $N$ matter fields $\Phi_i$
of charge matrix $Q_{ia}$.
We shall consider the submanifold $M$ of $X$ defined by
the equations
\beq
G_{\beta}=0,~~~~\beta=1,\ldots,l,
\eeq
where $G_{\beta}$ are polynomials of $\Phi_i$ of charge
$d_{\beta a}$ for the $a$-th $U(1)$ gauge group.
Let us add $l$ matter fields
$P_{\beta}$
of charge matrix $-d_{\beta a}$ to the $U(1)^k$ gauge theory.
This theory by itself realizes
a non-linear sigma model on a non-compact toric manifold
$V$. $V$ is the total space of the sum of $l$ line bundles on $X$;
The new coordinates $p_{\beta}$ parametrize the fibre directions
and $X$ is embedded in $V$ as the zero section $p_{\beta}=0$.
Now let us consider the gauge theory,
with the same gauge group and the same matter content,
but having a gauge invariant superpotential of the form
\beq
W=\sum_{\beta=1}^lP_{\beta}G_{\beta}(\Phi).
\label{supM}
\eeq
Then, the vacuum equation requires
$G_{\beta}=0$ and $\sum_{\beta}p_{\beta}\partial_iG_{\beta}=0$.
If $M$ is a smooth complete intersection in $X$,
this means that $p_{\beta}=0$ for all $\beta$ and
the vacuum manifold is $M$ itself.
Thus, the gauge theory with the superpotential (\ref{supM})
realizes the non-linear sigma model on the complete intersection $M$
in $X$.

  Thus the sigma model on the compact manifold $M$
is closely related to the sigma model on the
non-compact toric manifold $V$.  
In fact they both have the same gauge field and matter content.
The only difference between them
is that, in the compact theory, there is an F-term
involving a gauge invariant superpotential $W$
which yields $cc$ ring deformation.
In this sense the compact theory can be
embedded in the non-compact theory.
The non-compact theory is obtained by considering $W\rightarrow 
\epsilon W$ in the limit of setting $\epsilon \rightarrow 0$.
Of course this limit changes drastically the behavior of the theory and
in particular the theory has $2l$ more complex dimensions in the UV.

However, we can ask if there are any quantities which are unaffected
by this deformation.
The answer is that if we are considering quantities (such as $ac$ ring)
which are sensitive only to the twisted F-terms such as
Kahler parameters (the FI terms of the gauge theory),
then they should not depend on the F-term deformations.
Thus for those questions it should be irrelevant whether we are
considering the compact theory or the non-compact theory.  
All that we have to do is to find out how the states and the operators
of the compact theory are embedded in that
of the non-compact theory and compute the protected quantities
that way.  This is in fact analogous to embedding
questions of vacuum geometry from one theory
in a theory of higher central charge, discussed in \cite{CV}.
It is also similar to the computation of the elliptic
genus of minimal models using a free theory with higher
central charge \cite{Wittenel}.

In the present case this issue has been studied in 
\cite{MP} where they also obtain
the embedding of the chiral field of the sigma model on $M$ theory
with that of $V$.
This result is stated as follows.
Let $\delta_{\beta}$ ($\beta=1,\ldots,l$) be 
the $ac$ ring element of the theory on $V$ defined by
\beq
\delta_{\beta}=\sum_{a=1}^kd_{\beta a}\Sigma_a,
\eeq
where $\Sigma_a$ is the field strength of the $a$-th gauge group.
Then, the correlation functions $\langle ...\rangle_M$
of the A-twisted model on
$M$ are obtained from those $\langle ...\rangle_V$ on $V$ by
\beq
\langle{\cal O}_1\cdots {\cal O}_s\rangle_M
=\langle{\cal O}_1\cdots {\cal O}_s
(-\delta_1^2)\cdots(-\delta_l^2)\rangle_V.
\label{MPrelation}
\eeq
An intuitive understanding of this relation is available
in the quantum mechanics obtained by dimensional reduction.
Let $L_a$ be a line bundle over $V$
defined by the equivalence relation
$(p_{\beta},\phi_i;c)\equiv
(\prod_a\lambda_a^{-d_{\beta a}}p_{\beta},
\prod_a\lambda_a^{Q_{ia}}\phi_i;\lambda_a c)$
where $(\lambda_a)$ is the element of the complexified gauge group
$(\C^{\times})^k$ and
 $c$ is the fibre coordinate.
Then, one can show that the $ac$ ring element $\Sigma_a$
corresponds to the first Chern class
$c_1(L_a)$ of $L_a$.
Now, let us consider the tensor product
\beq
L_{\beta}~=~
\bigotimes_{a=1}^kL_a^{\otimes d_{\beta a}},
\eeq
whose first Chern class corresponds to the $ac$ ring element
$\delta_{\beta}$.
The line bundle $L_{\beta}^{-1}$ has a section proportional to
$p_{\beta}$ and thus
$-\delta_{\beta}$ represents the divisor class of $p_{\beta}=0$ in $V$.
In particular, the product $(-\delta_1)\cdots(-\delta_l)$
represents the class $X$ in $V$.
On the other hand, the bundle $L_{\beta}$ has a section
proportional to $G_{\beta}$ and thus
$\delta_{\beta}$ represents the divisor class of $G_{\beta}=0$.
Therefore, $(-\delta_1^2)\cdots(-\delta_l^2)$ represents a delta function
supported on $M$.
What is shown in \cite{MP} is basically that this quantum mechanical
interpretation remains true 
for the full 2d QFT as long as topological correlators are
concerned.  It turns out that we need a stronger version of the
relation (\ref{MPrelation}).   We thus proceed to a physical 
derivation of this relation, which uses the fact that the
compact and non-compact theories are embedded in the same underlying
physical theory, and yields the stronger version that we
need.

The basic idea is that with $\epsilon \not=0$ we have turned
on a superpotential and we can follow the states from
the non-compact theory to the compact theory.  In this sense
the non-compact theory flows in the IR limit (for non-vanishing
$\epsilon$) to the compact theory.  We can thus follow the
states in the non-compact theory and ask which ones survive
in the IR limit.  By the nature of the RG flow, we will be losing
some states as we take the IR limit.  However, if we concentrate
on the ground states in the Ramond sector which correspond
to normalizable states in the non-compact theory, then in the compact theory
they are bound to survive as a ground state (the same
cannot be said of the non-normalizable ground states in
the non-compact theory, which might disappear from the spectrum
of the compact theory).

In the sigma model on the non-compact manifold $V$,
the ground state corresponding to
the operator $\delta=\delta_1\cdots\delta_l$ is a normalizable
state which is the product of Kahler forms that control
the sizes of the compact part of the geometry.
We denote this state by
\beq
|\delta\rangle_{V}.
\eeq
In the large volume limit, this state has the axial R-charge
$-\dim_{\C}V+2l=-\dim_{\C} M$
which is the lowest among normalizable ground states.
Now, let us turn on the superpotential $\epsilon W$
where $W$ is the one given in (\ref{supM}).  The
state $|\delta\rangle_V$ which is the unique normalizable ground
state in the Ramond sector with axial charge $-\dim_{\C} V+2l$ will be
deformed to a state which we denote by $|\delta
\rangle_{\epsilon}$. 
By standard supersymmetry arguments,
this state is a normalizable ground state
of the Ramond sector.
For large $\epsilon$  (or for any
finite $\epsilon$ in the IR limit) the theory corresponds to the
sigma model on the compact manifold $M$ for which there
is a unique ground state with axial charge $-\dim_{\C} M$, which
is also sometimes denoted by $|1\rangle_M$.  Thus
we can identify $|\delta\rangle_{\epsilon}=|1\rangle_M$
as long as axial charge is conserved. 
Thus, we have the following
correspondence of states as we increase $\epsilon$;
\beq
|\delta\rangle_V\stackrel{\epsilon\ne 0}{\longrightarrow}
|\delta\rangle_{\epsilon}\stackrel{\epsilon\to\infty}{\longrightarrow}
|1\rangle_M.
\label{stateq}
\eeq
Here we have used the axial R-charge to identify the
state to which $|\delta \rangle_V$ is deformed. 
For the more general case,
where the axial R-sysmmetry is broken by an anomaly, the result
still remains true, as we will now argue.  
To see this, note that at large Kahler
parameters, i.e. as $t_a\rightarrow \infty$ axial R-charge
is a good symmetry. So at least to leading order it is
correct. To show it is true for all $t_a$ we proceed as follows:
According to \cite{CV1} the topologically twisted theory
picks a section of the vacuum bundle which is holomorphic
in the sense defined by the topological twisting.  In particular
we can choose the pertubed state $|\delta \rangle_{\epsilon}$
so that
\beq
{\partial \over \partial {\overline{t}_a}}|\delta\rangle_{\epsilon} =0.
\eeq
For infinitely large $t_a$, since the
axial charge is conserved the state $|\delta \rangle_{V} $ is
deformed for finite $\epsilon$
to $|1\rangle_{M}$.  For finite but large $t_a$
the fact that we can choose a holomorphic
section of the vacuum bundle shows that the difference between
$|\delta \rangle_{\epsilon}$
and $|1\rangle_M$
can only be given by states with coefficients involving
some powers of $q_a=e^{-t_a}$.
Here we note that the axial R-charge can be made conserved by
shifting the Theta angle to cancell (\ref{anomaly}).
This in particular means that we
assign {\it non-negative} R-charges
to $q_a$ as long as the first Chern class of $M$ is
positive-semi-definite.
It thus follows that $q_a$'s
would be accompanied by states which will have too
small an R-charge to correspond to a normalizable state, since
$|\delta \rangle_V$ is the unique normalizable ground state
with minimum  R-charge.  This 
establishes the relation (\ref{stateq}) in general.

The relation (\ref{stateq}) can be used to yield
another derivation of the relation (\ref{MPrelation}):
The topological correlations for sigma model on $M$ can be written
as
\beq
\langle{\cal O}_1\cdots {\cal O}_s\rangle_M
=\langle 1|{\cal O}_1\cdots {\cal O}_s|1\rangle_M=
\langle \delta |{\cal O}_1\cdots {\cal O}_s|\delta \rangle_{V
}=\langle
(-\delta_1^2)\cdots (-\delta_l^2){\cal O}_1\cdots {\cal
O}_s\rangle_V.
\label{newder}
\eeq
where we have used the $\epsilon $ independence in the topologial
A-model computations, as $\epsilon W$ does not affect the $ac$
correlation functions
(note that the overall proportionality factor is not fixed, and depends
on the choice of normalization of the topologically twisted theory).

\subsubsection{Examples of Hypersurfaces}
Consider a $U(1)$ gauge theory
with $N$ fields $\Phi_i$ of charge $+1$ and one field $P$ with
charge $-d$.   So far this is the same as the linear sigma model
description of the total space of the line bundle ${\cal O}(-d)$
over $\CP^{N-1}$.  However, now 
consider adding to the action the F-term with superpotential
\beq
W=PG(\Phi_i)
\eeq
where $G$ is a polynomial of degree $d$ in $\Phi_i$.  This 
gauge theory describes, in the infrared, the sigma model
on a hypersurface of degree $d$ in $\CP^{N-1}$.
The hypersurface for $d=N$ is a Calabi-Yau manifold of
complex dimension $N-2$. For $d<n$ it is
a manifold of positive $c_1$ and the sigma model is
asymptotically free.
For $d>n$ it is a manifold with negative $c_1$ and
the sigma model is not asymptotically free.

The geometric regime corresponds to when the FI parameter
is very large $r\gg 0$
(which is the UV limit for $d<N$).
If we consider the limit
$r\rightarrow -\infty$ (which is the IR limit for $d<N$),
we end up with an LG theory \cite{phases}:  
$P$ picks up an expectation value and breaks the $U(1)$
gauge symmetry to 
a discrete subgroup $\Z_d$ and the effective theory
is given by the LG model
with $W=G(\Phi_i)$ divided by a $\Z_d$ action, which corresponds
to multiplying the fields $\Phi_i$ by a $d$-th root of unity.  If
we turn off the superpotential, i.e., back in the non-compact
toric model corresponding to the total space of the line bundle
${\cal O}(-d)$ over $\CP^{N-1}$ the limit $r\rightarrow -\infty$
would correspond to the orbifold
of the free theory of $N$ fields $\Phi_i$ modded
out by $\Z_d$, acting as the $d$-th root of unity on $\Phi_i$.

The generalization of the above to the case of hypersurfaces
of weighted projective space are straightforward.

\section{A Proof Of Mirror Symmetry}

In this section we show how the results of section 3 on
the dynamics of supersymmetric gauge theories
naturally lead to a proof of mirror symmetry.
Before embarking on the proof, we discuss what we mean
by the ``proof''.

\subsection{What We Mean by Proof of Mirror Symmetry}

As we have discussed before the action for a $(2,2)$ supersymmetric
field theory has three types of terms: D-terms, F-terms
and twisted F-terms.
In the case of a supersymmetric sigma model, these three
types of terms have the following interpretations:  The D-terms
 correspond to changing the metric on the target manifold
without changing the Kahler or complex parameters. For example
for $\CP^1$ we can consider a fixed total area, but deform
the metric from the round metric to any other metric with the
same total area by varying
the D-term.
Variations of F-terms and twisted F-terms correspond
to deformations of complex and Kahler parameters of the target
space.  As discussed before, F-terms and twisted F-terms
control many important aspects of the $(2,2)$
theories.  In particular the $cc$ and $ac$ rings depend
only on F-terms and twisted F-terms respectively and are independent
of D-terms.  These in particular encode the instanton corrections
of the sigma model to the $ac$ ring.
Also BPS structure of the massive $(2,2)$ theories are completely determined
by the F-terms and twisted F-terms \cite{CV1,CV,NSI}.  One can
 define the notion of D-branes for $(2,2)$
theories (as is familiar in the conformal case, and can
be easily extended to the non-conformal case as we will
discuss in section 6).  One can also show
that the overlap of the corresponding boundary states with
vacua will only depend on the F-terms (or twisted F-terms
depending on which combination of supercharge the D-brane
preserves).  

These indicate the importance of F-terms and
twisted F-terms for the $(2,2)$ theories.  In fact
they become even more prominent in the conformal case,
for example for the case of Calabi-Yau manifolds.
Namely in that case
the sigma model flows in the infrared limit to a conformal
field theory which is determined entirely
by the 
complex and Kahler parameters of the manifold.
In other words the metric on the Calabi-Yau manifold adjusts
itself to the unique form consistent with conformal invariance
for a given Kahler and complex structure.  In particular
the D-terms are entirely {\it determined} by the F-terms
and twisted F-terms in this case, as far as the IR behavior
is concerned.
As a consequence, a deformation of the conformal field theory 
corresponds to a deformation of the Kahler or complex
structure of the Calabi-Yau manifold, which in turn is realized
through variations of F-terms and twisted F-terms.
 From these facts, and from the fact that
the $cc$ and $ac$ rings are determined in terms
of F-terms and twisted F-terms one wonders
whether the opposite is true (i.e. whether the $cc$
and $ac$ rings determine the F-terms and twisted F-terms
and thus the full theory).  
This is, however, not completely true; 
theories with the same chiral ring can differ
in the ``integral structure'' \cite{Aspinwallmorrison}.
More precisely, in order to completely specify the theory
we need to know the structure of the allowed D-branes.  

We prove mirror symmetry in two different senses:
In the strong sense, we find a dual theory, for which
we prove it is equivalent to the original theory, up
to deformations involving D-terms.  In the weak sense,
we propose a dual theory for which we can only show
that the $cc$, $ac$, BPS structure of solitons and
the $D$-brane structure are the same.
The weak and strong senses become equivalent if we assume that
the $cc$ and $ac$ ring, together with the integral structure
determine the theory up to D-term variations.

We are now ready to discuss the proof for mirror
symmetry.  We divide the discussion
into several cases and indicate in each
case what is the sense of the proof (strong/weak)
that we shall present.

\subsection{The Proof}

The cases of interest naturally
divide into three different classes: 

(i) $X=\CP^{N-1}$ or a more general compact toric manifold with
$c_1(X)\geq 0$,

(ii) $X$ is a non-compact Calabi-Yau or a more general
non-compact toric manifold.

\noindent
And finally,

(iii) Sigma models on hypersurfaces or complete intersection
in $X$.

It turns out that the mirror for the cases (i) and (ii) are rather
straightforward to derive, using the tools we have developed
so far and we shall prove them in the strong sense.
For the derivation of the case (iii) we need an additional
tool, which we will discuss in section 6, and we postpone a complete
discussion of (iii) to section 7.  This will lead
to a proof of case (iii) in the weak sense.
  In this section after
deriving the general mirror for the cases (i) and (ii)
we briefly mention how it effectively gives an
answer of the case (iii) (to be more fully developed in section
7).   We present
some examples for cases (i) and (ii) and how it relates
to case (iii)
 to illustrate the meaning of the results.

We first consider the sigma model on
a toric manifold which is described by a gauge theory
with a single $U(1)$ gauge group with chiral fields of
charge $Q_i$ ($i=1,\ldots,N$).
We recall that the dual of the gauge theory is described by
a vector multiplet with field strength $\Sigma$ and $N$ periodic
variable $Y_i$ dual to the charged matter fields.
It has an approximate Kahler potential
$(1/2e^2)|\Sigma|^2+\sum_i|Y_i|^2/4r_0$
and an exact twisted superpotential
\beq
\widetilde{W}
=\Sigma\left(\sum_{i=1}^NQ_iY_i-t(\mu)\right)
\,+\,\mu\,\sum_{i=1}^N\e^{-Y_i}.
\eeq
The fields $\Sigma$ and $\sum_i Q_iY_i$ have
mass of order $e\sqrt{|r_0|}$
while the mass scale for the modes tangent to
$\sum_iQ_iY_i=t$ is of order $\mu\sqrt{|r_0|}$
where $\mu\sim\Lambda$ for $\sum_iQ_i\ne 0$.
In the sigma model limit $e\gg\mu$,
these mass scales are well-separated and
it becomes appropriate to integrate out $\Sigma$.
This yields the constraint
\beq
\sum_{i=1}^NQ_iY_i=t.
\label{cons}
\eeq
Clearly this can be solved by $N-1$ periodic variables.
Now, the dual theory becomes the theory of such $N-1$
variables solving (\ref{cons}) with the twisted superpotential
\beq
\widetilde{W}\,=\,\mu\,\sum_{i=1}^N\e^{-Y_i}.
\eeq
This can be interpreted as a sigma model on $(\C^{\times})^{N-1}$
with a twisted superpotential.
Since the complex coordinates are the lowest components
of twisted chiral superfields,
this theory can be identified as the mirror
of the non-linear sigma model on $X$.

It is straightforward to extend this to the more general case
where we start with $k$ $U(1)$ gauge groups
and $N$ matter fields of charge $Q_{ia}$ ($i=1,\ldots,N$, $a=1,\ldots,k$).
The target space $X$ of the sigma model is the quotient of
\beq
\sum_{i=1}^NQ_{ia}|\phi_i|^2=r_a
\eeq
by the $U(1)^k$ action $\phi_i\mapsto \e^{iQ_{ia}\gamma_a}\phi_i$,
which is a toric variety of dimension $N-k$.
The dual description
of the gauge theory is obtained in (\ref{multig}).
Integrating out the vector multiplet, we obtain
the algebraic torus
$(\C^{\times})^{N-k}$ as the solutions to
\beq
\sum_{i=1}^NQ_{ia}Y_i=t_a.
\label{hypc}
\eeq
The dual theory is a sigma model on this $(\C^{\times})^{N-k}$ 
with the twisted superpotential
\beq
\widetilde{W}\,=\,\mu\,\sum_{i=1}^N\e^{-Y_i}.
\label{toricW}
\eeq
This can be identified as
the mirror of the sigma model on $X$.
We thus have established the mirror symmetry of the non-linear sigma model
on a toric variety $X^{N-k}$
and the theory on the algebraic torus $(\C^{\times})^{N-k}$ 
with a superpotential.

We can also consider deforming the sigma model using
holomorphic vector fields, as discussed in section 2.
There are $(N-k)$ such parameters corresponding
to the $U(1)^N/U(1)^k$ holomorphic isometry
group for the above toric variety.  
We parameterize these deformations by
$N$ parameters $\widetilde{m}_i$ (as we will see
below $k$ of them are redundant).
The corresponding gauge theory is the one with
the twisted masses $\widetilde{m}_i$ and the twisted superpotential for the
dual theory is obtained in (\ref{twism}).
Adding the twisted masses does not affect the elimination of the
field strength $\Sigma_a$ and we obtain the same constraints
(\ref{hypc}).
All it does is to shift the superpotential (\ref{toricW}) as
\beq
\widetilde{W}\,=\,\mu\,\sum_{i=1}^N\e^{-Y_i}
-\sum_{i=1}^N\widetilde{m}_iY_i.
\label{twistmir}
\eeq
This is not a single valued function on $(\C^{\times})^{N-k}$,
but the multivaluedness is a constant shift and therefore the Lagrangian
itself is well-defined.
It is easy to see, using (\ref{hypc}) that $k$ of the parameters
$\widetilde{m}_i$ are redundant and do not affect the superpotential.

We will now illustrate these ideas in two concrete cases:
One is in the context of compact toric varieties which we
exemplify by using the case of $\CP^{N-1}$.  The second
one is for the case of non-compact toric varieties
which we exemplify by considering the total space
of ${\cal O}(-d)$ line bundle over $\CP^{N-1}$.  In the context
of the latter example we also explain briefly how
the mirror for the hypersurface case arises.  We complete the discussion
for the hypersurface case in section 7.

\subsection{Compact Toric Manifold}

\subsection*{\it The $\CP^{N-1}$ Model}

The linear sigma model for $X=\CP^{N-1}$ is the
$U(1)$ gauge theory with $N$
matter fields of charge $1$.
The constraint $\sum_{i=1}^NY_i=t$ is solved by
$Y_i=t/N-\Theta_i$ ($i=1,\ldots,N-1$) and
$Y_N=t/N+\sum_{i=1}^{N-1}\Theta_i$,
where $\Theta_i$ are periodic variables of period $2\pi i$
and can be considered as coordinates of $(\C^{\times})^{N-1}$.
The superpotential is
\beq
\widetilde{W}=
\Lambda\left(\,\e^{\Theta_1}+\cdots+\e^{\Theta_{N-1}}
+\prod_{i=1}^{N-1}\e^{-\Theta_i}\,\right),
\label{aToda}
\eeq
where $\Lambda=\mu\,\e^{-t/N}$ is the dynamical scale of the theory.
This is the superpotential for supersymmetric
affine $A_{N-1}$ Toda field theory.
Thus, we have derived the mirror symmetry
of $\CP^{N-1}$ model and affine Toda theory which was
observed in
\cite{FI,CV,G,EHY,EHX} from various points of view.

Having no F-term, the theory is invariant under $U(1)_V$ R-symmetry.
The twisted superpotential (\ref{aToda}) explicitly
breaks $U(1)_A$
but its $\Z_{2N}$ subgroup remains unbroken;
for
$\Theta_j\to\Theta_j+2\pi i/N$,
$\widetilde{W}\to \e^{2\pi i/N}\widetilde{W}$.
The vacua of the theory are given by the
critical points of $\widetilde{W}$,
$\partial_{\Theta_i}\widetilde{W}=0$. It is solved by
$\e^{\Theta_1}=\cdots=\e^{\Theta_{N-1}}=:X$ where $X=X^{-(N-1)}$.
Namely, there are $N$ vacua at $\e^{\Theta_j}=\e^{2\pi i\ell/N}$
(all $j$)
with the critical value
\beq
\widetilde{W}\,=N\Lambda\,\e^{2\pi i\ell/N},
\label{Wvac}
\eeq
($\ell=0,\ldots,N-1$).
Each vacuum is massive and
breaks spontaneously the axial R-symmetry $\Z_{2N}$ to
$\Z_2$. All these are indeed the properties which are
possessed by the $\CP^{N-1}$ model.

An advantage of LG type description is that the BPS soliton
spectrum can be exactly analyzed using the superpotential.
A BPS soliton corresponds to a trajectory connecting
two vacua which projects onto a straight
line in the $\widetilde{W}$ space. The mass is equal to the absolute value
of the susy central charge
which is given by the
difference of the two critical values of $\widetilde{W}$.
{}From (\ref{Wvac}) we see that the BPS soliton
connecting the $0$-th and the $\ell$-th vacua has central charge
\beq
\widetilde{Z}_{0\ell}=N\Lambda(\,\e^{2\pi i\ell/N}-1).
\eeq
One can also see that there are ${N\choose \ell}$
such solitons \cite{HIqV}.
 For each of them,
$\ell$ of $\e^{-Y_i}$ are equal to a trajectory $f_{\ell}$ in $\C^{\times}$
while the remaining $(N-\ell)$ of them to another $f_{N-\ell}$.
They both starts from $\e^{-t/N}$ and ends at $\e^{-t/N+2\pi i\ell/N}$
but $f_{\ell}$ has a relative winding number $(-1)$ compared to
$f_{N-\ell}$.
Since the winding in $Y_i$ is dual to the charge for phase rotation
of $\Phi_i$,
the soliton has the same quantum number as the product
$\Phi_{i_1}\cdots\Phi_{i_{\ell}}$.
Indeed, the soliton spectrum of the $\CP^{N-1}$ model
has been studied in \cite{Witten79} and it was found that
the $\ell=1$ solitons are the elementary electrons $\Phi_i$
and the higher-$\ell$ solitons are their
$\ell$-th antisymmetric products.  This spectrum
for the solitons was also recovered from the $tt^*$
geometry in \cite{CVexact}.

We can also consider deforming the $\CP^{N-1}$ sigma
model by the addition of a combination of the holomorphic
vector fields $U(1)^{N-1}$.  In such a case we obtain
the mirror (\ref{twistmir}):
\beq
\widetilde{W}=
\Lambda\left(\,\e^{\Theta_1}+\cdots+\e^{\Theta_{N-1}}
+\prod_{i=1}^{N-1}\e^{-\Theta_i}\right)+\sum_{i=1}^{N-1}
(\widetilde{m}_i-\widetilde{m}_N) \Theta_i,
\label{atwToda}
\eeq
As has been noted, $\widetilde{m}_i$'s changes
the central charge of the 
supersymmetry algebra by a
term proportional to the charges $S_i$ of the global abelian
symmetry $U(1)^{N-1}$,
\beq
\delta\widetilde{Z}=\sum_{i=1}^{N-1}
(\widetilde{m}_i-\widetilde{m}_N)S_i.
\eeq
The charges $S_i$ can be identified
with the weights of the $SU(N)$ global
symmetry. In this way we can recover not only the soliton
spectrum, but also their quantum numbers under the $SU(N)$ global
symmetry, and obtain the anticipated result noted above \cite{HIqV}.
Note also, that the above deformation, deforms the quantum
cohomology ring.  If we denote by
$x=-d\widetilde{W}/dt$
the generator of the
chiral ring (corresponding to the Kahler class of $\CP^{N-1}$),
from the above superpotential one obtains
\beq
\prod_{i=1}^N(x-\widetilde{m}_i)
=\Lambda^N.
\eeq
Note that the twisted masses
deforms the cohomologry ring from the simple form $x^N=\Lambda^N$
to an arbitrary polynomial of degree $N$.
The result for the case of $\CP^1$ was first derived through other
arguments in \cite{cecop}.
The general case was conjectured in \cite{HH}
from brane construction of the theory.

As noted before, we do not attempt to specify the
D-terms which are subject to perturbative corrections in general.
However it is expected that the sine-Gordon theory
($A_1$ affine Toda theory) is integrable
and the D-term is protected from quantum correction \cite{KU,FI}.
Thus, one may expect that a more detailed statement of the
mirror symmetry can be made.
The $\CP^1$ model realized as the linear sigma model
possesses the $SU(2)$ isometry group as the global
symmetry.
Recall that we have seen that
the Kahler potential for $Y_i$'s is approximately
$|Y_i|^2/4r_0$ at the classical level.
Recall also that $r_0\to \infty$ in the continuum limit.
This suggests that the equivalence of the $SU(2)$ invariant
$\CP^1$ model and the
sine-Gordon theory holds only in the limit where
the Kahler potential of the latter vanishes.
This is actually consistent with the observation \cite{KU}
that
the $N=2$ sine-Gordon theory possesses $SU(2)$ global symmetry,
rather than its q-deformation, in the limit of vanishing Kahler potential.
Also, it was observed in \cite{FI} that the scattering matrix of
the BPS solitons 
of the $\CP^1$ model and the sine-Gordon theory agree with each other
in such a limit.
It would be interesting to investigate the integrability and
the protection of the D-term
in more general cases.

\subsection*{\it More General Cases}

\newcommand{\A}{{\rm A}}

We now study a general aspects of the mirror theory for more general
toric manifold $X$.
Let us consider the equations
$\sum_{i=1}^Nv_iQ_{ia}=0$ for integers $v_i$.
The space of solutions form a lattice of rank $N-k$.
Let $v_i^{\A}$ be the integral basis of this
lattice;
\beq
\sum_{i=1}^Nv_i^{\A}Q_{ia}=0,~~~
\left\{\begin{array}{l}
\A=1,\ldots,N-k,\\
a=1,\ldots,k.
\end{array}\right.
\eeq
The constraints (\ref{hypc}) on $Y_i$ can be solved by
$N-k$ periodic variables $\Theta_{\A}$ as
\beq
Y_i=\sum_{\A=1}^{N-k}v_i^{\A}\Theta_{\A}+t_i
\eeq
where $(t_i)$ is a solution to $\sum_{i=1}^NQ_{ia}t_i=t_a$
(an arbitrary choice will do;
another choice is related by a shift of $\Theta_{\A}$'s).
Now, the superpotential (\ref{toricW}) of the mirror theory
can be expressed as
\beq
\widetilde{W}=\sum_{i=1}^N\exp\left(-t_i-\langle \Theta,v_i\rangle\right),
\label{Wtor}
\eeq
where $\langle \Theta,v_i\rangle$ is the short hand notation for
$\sum_{\A=1}^{N-k}v_i^{\A}\Theta_{\A}$.
Note that the expression (\ref{Wtor})
is the same as the function in \cite{Bat}
which determines the quantum cohomology of toric manifolds.
Namely, we have derived the result of
\cite{Bat} as a straightforward consequence of our dual description.

It is useful to consider $v_i=(v_i^{\A})$ as
(generically linearly dependent)
$N$ vectors in the lattice $\N=\Z^{N-k}$
and $\Theta_{\A}$ as the coordinates on $\M_{\C}=\M\otimes\C$
where $\M$ is the dual lattice of $\N$.
Then, $\langle\Theta,v_i\rangle$ that appears in (\ref{Wtor})
can be identified as the natural pairing.
It is well-known in toric geometry \cite{Oda,Fulton}
that $X$ is compact if and only if the cone generated by $v_i$
covers the whole $\N_{\R}=\N\otimes\R$ (i.e. any element of
$\N_{\R}$ is expressed as a linear combination of $v_i$'s with
non-negative coefficients).
This in particular means that there is no value of $\Theta\in \M_{\C}$
such that ${\rm Re}\langle\Theta,v_i\rangle\geq 0$ for all $i$.
In other words,
for any non-zero ${\rm Re}\Theta$,
${\rm Re}\langle\Theta,v_i\rangle< 0$ for some $i$.
Thus, there is no obvious run-away direction of the superpotential
(\ref{Wtor}) and we generically expect a discrete spectrum.
\footnote{There can be an accidental situation (which appears
in a sublocus of the parameter space) where
there is a run-away direction. This comes from the singularity
of the sigma model that is associated with the development of the Coulomb
or mixed branch, which is studied in \cite{MP}.}
This is of course 
an expected property for a mirror of a compact sigma model.

Since the Witten index of the sigma model is equal to the
Euler number
\beq
{\rm Tr}(-1)^F=\chi(X),
\eeq
the number of critical points of the
superpotential (\ref{Wtor})
must agree with $\chi(X)$.
To check this, we note another well-known fact in toric geometry:
Our algebraic torus $(\C^{\times})^{N-k}$ (with coordinates
$\e^{\Theta_{\A}}$) is embedded as an open subset of a
``dual'' toric variety $Y$ and
each term in (\ref{Wtor}) extends to a section of the anti-canonical bundle
$K_Y^{-1}$ of $Y$.\footnote{The ``dual'' toric variety is constructed
as follows. Take the convex hull of $v_i$'s in $\N_{\R}$;
this makes a convex polytope $\Delta$ in $\N_{\R}$. Consider the
dual $\Delta^{\circ}\subset \M_{\R}$ of $\Delta\subset \N_{\R}$
defined as the set of points in $\M_{\R}$ whose values at $\Delta$
are $\geq -1$. Then, the vertices $\{u_I\}$ of $\Delta^{\circ}$
determines the ``dual'' toric variety $Y$.}
Thus, each partial derivative
$\partial_{\A}\widetilde{W}=\partial\widetilde{W}/\partial\Theta_{\A}$
also extends to a section $s_{\A}$ of $K_Y^{-1}$.
Since a critical point is a common zero of all
$\partial_{\A}\widetilde{W}$'s,
the number of critical points is the number of intersection points of
the divisors $\partial_{\A}\widetilde{W}=0$ in $(\C^{\times})^{N-k}$.
Since there is no run-away behaviour of the potential
for generic values of $t_a$, we do not expect a common zero
of $s_{\A}$'s at infinity
$Y-(\C^{\times})^{N-k}$.
Thus, the number of critical points must be the same as
the topological intersection number of the $N-k$ divisors
$s_{\A}=0$
in $Y$ which is counted as
\beq
{\rm Tr}_{\rm mirror}(-1)^F
=\langle[Y],c_1(K_Y^{-1})^{N-k}\rangle.
\label{chernn}
\eeq
Therefore the number (\ref{chernn}) must agree with the Euler number of
$X$. It appears that this is not known in general. However,
this certainly holds in every example one can check
as long as $c_1(X)\geq 0$.

Since the sigma model is well-defined and the gauge theory agrees with it
only when the first Chern class of $X$ is positive semi-definite,
the above result makes sense only in such cases $c_1(X)\geq 0$.
To illustrate this, we consider the Hirzebruch surface $F_a$.
As noted before, the charge matrix for $F_a$ is
$Q_{i1}={}^t(1,0,1,a)$ and $Q_{i2}={}^t(0,1,0,1)$ and therefore
the vectors $v_i^{\A}$ are given by
$v_1=(1,0)$, $v_2=(0,1)$, $v_3=(-1,a)$ and $v_4=(0,-1)$.
Then the superpotential is given by
\beq
\widetilde{W}=
\e^{-\Theta_1}+\e^{-t_2}\e^{-\Theta_2}+
\e^{-t_1}\e^{\Theta_1-a\Theta_2}+\e^{\Theta_2}.
\eeq
It is easy to see that the number of critical points are
$4$ for $a=0,1$ and $2a$ for $a\geq 2$.
Since the Euler number of $F_a$ is always $4$, the result is ``correct''
only for $a=0,1,2$. This is consistent because $c_1(F_a)$
is positive semi-definite only for $a=0,1,2$.

\subsection{Non-Compact Case}

In the case where $X$ is non-compact there is an obvious run-away
direction of the superpotential $\widetilde{W}$.
To see this, we note that $X$ is non-compact if and only if
$v_i$'s generate a proper convex cone in $\N_{\R}$.
This means that there is a point $\Theta_0$ in $\M_{\R}$ 
such that $\langle \Theta_0,v_i\rangle\geq 0$ for all
$i$.
Now, consider the behaviour of
the superpotential $\widetilde{W}$ at $\Theta=t\Theta_0$
in the limit $t\to +\infty$.
In the case where $\langle\Theta_0,v_i\rangle> 0$ for all $i$,
each term of the superpotential vanishes in the limit
and this is the run-away direction.
If there is some $i$ such that $\langle\Theta_0,v_i\rangle=0$
the superpotential stays finite but can be extremized by choosing
an appropriate $\Theta_0$.
In any case, there is a run-away direction of the superpotential.
We thus expect a continuous spectrum, which is indeed a property of
the sigma model on a non-compact manifold.

Below, we present two basic examples of
non-compact toric manifolds.
The first one, the total space of
${\cal O}(-d)$ over $\CP^{N-1}$,
provides a starting point of the discussion of
the mirror for hypersurfaces in $\CP^{N-1}$ and more general
toric complete intersections.
The second one,
the total space of ${\cal O}(-1)\oplus{\cal O}(-1)$
over $\CP^1$, is important for the study of phase transition
in the sense of \cite{phases,AGM}.

\subsection*{\it ${\cal O}(-d)$ over $\CP^{N-1}$}

The linear sigma model for the non-compact
space given by the total space ${\cal O}(-d)$
over $\CP^{N-1}$ is given by a $U(1)$ gauge theory with
$N$ fields with charge $+1$ and one field with charge
$-d$.
  As discussed before we find that the mirror for this
theory is an LG theory given by an LG theory with superpotential
$$W=\sum_{i=1}^{N}\e^{-Y_i}+\e^{-Y_P}$$
where
\beq
-d Y_P+\sum_{i=1}^N Y_i =t
\label{NCO}
\eeq
and $Y_P$ is the dual field to the charge $-d$ matter
and $Y_i$ for $i=1,...,N$ are dual to the charge $+1$ matter fields.
It is natural to use (\ref{NCO}) to solve for $Y_P$. Let us define
\beq
X_i=\e^{-Y_i/d}
\label{NNO}
\eeq
for $i=1,...,N$. Then we have
\beq
\e^{-Y_P}=\e^{t/d}X_1X_2...X_N
\label{elim}
\eeq
and therefore 
\beq
W=X_1^d+X_2^d+...+X_N^d+\e^{t/d}X_1X_2...X_N
\label{lgh}
\eeq
However, we have to note that the field redefinition
(\ref{NNO}) is not single valued.  In fact as we shift
$Y_i\rightarrow Y_i+2\pi i$ we transform $X_i$ by a primitive
$d$-th root of unity
$$X_i \rightarrow X_i \e^{-2\pi i/d}.$$
However, from (\ref{elim}) we see that the product
of the $X_i$ is well defined. Thus the LG theory we get is given
by the above superpotential $W$ modded out by $G=(\Z_d)^{N-1}$
where $G$ is the group which acts on $X_i$ by all
$d$-th roots of unity which preserve the product $X_1X_2...X_N$.

As we discussed in section 4, the linear sigma model
for the above non-compact theory
and the one given by a hypersurface of degree
$d$ in $\CP^{N-1}$ differ by deformations
involving only the $cc$ ring elements, which thus do not affect
the $ac$ ring, which we are studying through mirror symmetry.
Thus we expect the same LG with superpotential (\ref{lgh})
(modded out by $G$) to describe
the mirror of hypersurface of degree $d$ in $\CP^{N-1}$.
Note that for the case of Calabi-Yau manifolds $d=N$, and
this gives the mirror construction of Greene and Plesser 
\cite{GP}.  However
clearly there should be a difference between non-compact
case and the compact theory.  For example their conformal central
charges (in $N=2$ units) differ by 2 in the Calabi-Yau case.
What we will find is that indeed there is a difference and
this is reflected in what is the appropriate field variable.
In the compact case (i.e. the hypersurface case) we will find
that the fundamental fields are $X_i$, whereas in the non-compact
case the fundamental fields are $Y_i$ which are related to $X_i$
as defined in (\ref{NNO}).  This will be discussed after our
discussion of the realization of D-branes in LG theories and
their associated ``BPS masses''.

\subsection*{\it ${\cal  O}(-1)+{\cal O}(-1)$ over $\CP^1$}

As a special case of interest, which has
been studied in \cite{phases} let us consider the mirror
for ${\cal  O}(-1)+{\cal O}(-1)$ over $\CP^1$. This
is given by the linear sigma model with a single $U(1)$
with four fields with charges $(-1,-1,+1,+1)$.  For
non-vanishing $t$ we can eliminate $\Sigma$ as before
and we obtain, in the dual formulation the superpotential
$$W=X_1+X_2+X_3+X_4$$
with $X_i=\e^{-Y_i}$ and 
$$X_1X_2=X_3X_4 \e^t.$$
We can eliminate, say $X_1$ and write
$$W=X_2+X_3+X_4+\e^tX_3X_4/X_2$$
Later in this paper we will return to this example and show how
the prepotential for this model (i.e., the famous tri-logarithmic structure
of \cite{candet,aspinmorri}) can be obtained using the mirror potential
we have found here.

\section{D-branes, BPS Mass and LG Models}

In the case of Calabi-Yau manifolds one can consider special
n-dimensional Lagrangian submanifolds, and they represent
a class in $H_n(M)$.  We can imagine a D-brane wrapping that class.  In
the sigma model we consider worldsheet with boundaries where
the boundary can end on these manifolds.  Such boundaries
preserve the A-model supercharges.
Given the fact that the neutral observables
of the $B$ model correspond to $n$-form (after
contraction with the holomorphic $n$-form) we have a natural pairing
between $B$-model chiral fields,
and $A$-model boundary states.
 The
pairing
is simply given by integrating the corresponding n-form
on the corresponding n-cycle.  Moreover, this integration
can also be interpreted as the inner product of the boundary
state defined by the D-brane and state defined (through
topological twisting) by the $B$-model observable \cite{OOY}.
In other words
$$\langle \gamma_i |\phi_\alpha \rangle=\int_{\gamma_i }\phi_\alpha$$
where $\gamma_i$ represents an n-dimensional cycle giving
rise to boundary state $\langle \gamma_i|$ and
$\phi_\alpha$ represents an element of the B-model
chiral field corresponding to an $n$-form.  There is
one distinguished element for the chiral ring,
the identity operator, which corresponds to the
holomorphic $n$-form $\Omega$ on the Calabi-Yau.
In particular we have
$$Z_i=\langle \gamma_i |1\rangle =\int_{\gamma_i} \Omega$$
For type IIB compactifications
on Calabi-Yau 3-folds, the $Z_i$ also represents central extension of
supersymmetry algebra, in a sector with D-brane wrapping
the $\gamma_i$ cycle.

In the context of LG models, the natural topological
theory corresponds to the B-model one \cite{topLG}.
 It is thus natural to ask the analog of this
pairing in the context of
the LG models.  This is important in our applications because
we have found an equivalence between the linear
sigma models and
LG models.

Consider a Landau Ginzburg theory with superpotential
$W(x_i)$ where $x_i$ are chiral fields, with $i=1,...,n$. Then it is
known
that the chiral ring for this theory is given by \cite{VW,Martinec}:
$${\cal R}={\bf C}[x_i]/dW$$
i.e.,  the ring generated by $x_i$ subject to setting $dW=0$.  It is
also
known \cite{arno} that there is a corresponding homology
group which is equal to the dimension of the ring, namely consider
the homology group
$$H_n({\bf C}^n, B)$$
where $B$ denotes the asymptotic region in ${\bf C}^n$ where $Re
W\rightarrow
+\infty$, i.e. n-cycles in the $x_i$ space with boundaries
ending on $B$. Then the dimension of this space is equal to
the dimension of the ring ${\cal R}$. In fact as we
will now discuss,
there is a natural pairing between them
and this identifies them as dual spaces. This is completely
parallel to the case of the Calabi-Yau case and the pairing
between mid-dimensional cycles and B-model observables
(with equal left/right $U(1)$ charge).

As is shown in \cite{HIqV}, in fact we can
construct analogs of Lagrangian submanifolds in the context of
LG models as well, representing a basis for $H_n({\bf C}^n,B)$
and preserving the A-combination of supercharges $Q_{ac}=Q_- +{\overline
Q}_+$.  Moreover the image of these cycles on the W-plane
is a straight line extending from the critical values of W
to infinity along the positive real axis.
Let $\gamma_i$ represent one such cycle.
We would like now to discuss the analog of the pairing
discussed above for the case of Calabi-Yau manifolds, between
the B-model observables and A-model boundary states.

Consider the pairing
$$\Pi_{i,a} =\langle {\gamma_i}|\phi_a \rangle$$
where $\phi^a$ denotes a B-model chiral field observable
and $\langle \gamma_i|$ denotes the boundary state corresponding
to $\gamma_i$.  It has been
shown in \cite{CV1} that the
$\Pi_b$ satisfy the flatness equation in the context of $tt^*$:
$$\nabla_a\Pi_b=(D_a+C_a)\Pi_b =0\qquad {\overline \nabla_a}
{\overline \Pi_b}=({\overline D_a}+{\overline C_a}){\overline \Pi_b}=0
$$
where $C_a$ denotes the multiplication by the chiral field corresponding
to $\phi_a$ on the chiral fields and $D_a$ is a covariant derivative
defined in \cite{CV1}.
This was formulated in \cite{CV1} by showing that the above quantity
can be computed by reducing to the case of quantum mechanics.  The same
result can also be derived more directly using the gymnastics leading to $tt^*$
equation.  In fact for the conformal case this has already been
shown in \cite{OOY} and the same argument also applies
to the massive cases (by considering the geometry of
semi-infinite cigar, as in the derivation of $tt^*$, with topological
B-twisting on the tip of the cigar, as will be discussed in \cite{HIqV}).

A special case of this overlap, is given by $\Pi_{i,1}$, i.e.
overlap with the state corresponding to the identity operator
of the B-model.  It has been shown in \cite{CV1}
that in the conformal case $\Pi_{i,1}$ is a holomorphic function
of the couplings, i.e. it is independent of the coefficients
of the superpotential ${\overline W}$ and depends only on
the holomorphic couplings appearing in $W$. In the non-conformal
case, this is no longer true.  However even in the massive
case one can obtain a holomorphic object by expanding near the conformal
limit.  This means formally considering ${\overline W} \rightarrow
{\overline \lambda} {\overline W}$ and taking the limit $\overline
\lambda
\rightarrow 0$.  From this point on, when we refer to $\Pi_{i,a}$
we have in mind this limit.
It has been shown in \cite{CV1} that in such a case
 $\Pi_{i,1}$ has a simple integral expression:
\beq
\Pi_{i,1}=\int_{\gamma_i}{\rm exp}(-W)dx_1...dx_n.
\eeq
This is the Landau-Ginzburg analog of the BPS mass
for the D-brane given by $\gamma_i$.
Moreover, with a good choice of basis for chiral fields of
the B-model (called the topological or flat coordinates) one has
\cite{CV1}
\beqa
&\partial_{t^a}\Pi_{i,1}=\Pi_{i,a}=\int_{\gamma_i}\phi_a\exp(-W)dx_1...dx_n,\\
&\partial_{t^a}\partial_{t^b}\Pi_{i,1}=\partial_{t^b}\Pi_{i,a}=
C_{ab}^c\Pi_{i,c},
\label{lgper}
\eeqa
where $\phi_a$ denotes the chiral field corresponding to the deformation
given
by $t^a$.  Note that this implies that the periods $\Pi_{i,a}$
have the full information about the chiral ring structure
constants.
 
Note that in the limit $W=0$ the above expression for the
period is the usual integral of the holomorphic $n$-form
 (which for ${\bf C}^n$ is $\Omega =dx_1...dx_n$)
on a cycle.  The appearance of $e^{-W}$ reflects the fact that
in the presence of superpotential the supercharges get modified
and field which corresponds to D-brane boundary conditions
in addition to the delta function forcing $x_r$'s to the subspace
$\gamma_i$ includes an additional factor $e^{-W}$, i.e. $e^{-W}\delta
(x-\gamma )$ is needed to be  $Q_{ac}=Q_- +{\overline
Q}_+$ invariant.

\subsection{Picard-Fuchs Equations for Periods}
Applying the previous discussion to the mirror of the non-compact
toric varieties we have discussed, we can compute
the periods of the branes in the mirror LG theory. 
We have the periods being given as
\beq
\Pi= \int \prod_{i,b} dY_{i} \prod_b
 \delta(\sum_i Q_{ib} Y_{i}-t_b){\rm exp}(-\sum_{i}  e^{-Y_{i}}),
\eeq
where we have suppressed the indices for cycles
$\gamma_i$ and the identity operator associated with $\Pi$.
Consider instead 
\beq
\Pi(\mu_{j},t_b)=
 \int \prod_{i,b} dY_{i} \prod_b
 \delta(\sum_i Q_{ib} Y_{i}-t_b){\rm exp}(-\sum_{i} \mu_{i} e^{-Y_{i}}),
\eeq
which satisfies for each $b$ the equations
\beq
[\prod_{Q_{ib}>0}({\partial \over \partial \mu_{i}})^{Q_{ib}}]
\Pi(\mu_{j},t_b)=
e^{-t_b}
[\prod_{Q_{ib}<0}({\partial \over \partial \mu_{i}})^{-Q_{ib}}]
\Pi(\mu_{j},t_b).
\label{picca}
\eeq

On the other hand by a shift in $Y_{i}$ by ${\rm log}
\mu_{i}$ we can get rid of the
$\mu_{i}$ dependence above, except for a shift in the delta function
constraint.  In other words we have
\beq
\Pi(\mu_{i},t_b)=\Pi_{i,1}(1,t_b-{\rm log}\prod_i \mu_{i}^{Q_{ib}}),
\eeq
which is the original periods we are interested in computing.
This in particular means that (\ref{picca}) can be rewritten as differential
operators involving the $t_b$ parameters. Together
with the boundary conditions for large $t_b$'s these give
an effective way of computing the periods.
\subsection*{\it ${\cal O}(-3)$ over $\CP^2$}

Just as an example of the previous equations, we can consider
the equations satisfied by the periods of the mirror of
sigma model on ${\cal O}(-3)$ over $\CP^2$.  This is given
by a $U(1)$ gauge theory with 4 matter fields with charges
$(-3,1,1,1)$. Using (\ref{picca}) we have
\beq
{\partial \over \partial \mu_2}{\partial \over \partial \mu_3}
{\partial \over
\partial \mu_4}\Pi=e^{-t}
{\partial^3 \over \partial \mu_1^3} \Pi.
\eeq

 Defining $\theta=-d/dt$ and noting that $\Pi$
 depends on $\mu_i$ in the combination $t-{\rm log}[\mu_2\mu_3\mu_4/\mu_1^3]$
 we obtain
\beq
\theta^3 \Pi=e^{-t}(3\theta +2)(3\theta +1)(3\theta)\Pi.
\eeq

\subsection{Prepotential for ${\cal O}(-1)+{\cal O}(-1)$
bundle over $\CP^1$}

We will now use the periods of the mirror to compute
the prepotential for the non-compact Calabi-Yau threefold
given by ${\cal O}(-1)+{\cal O}(-1)$
bundle over $\CP^1$.  The prepotential for this model is
known to be
given by trilogarithm function
\beq
F(t)=P_3(t)+\sum_{n>0}e^{-nt}/n^3
\label{prepno}
\eeq
where $P_3(t)$ is a polynomial of degree 3 in $t$
(some of whose coefficients are ambiguous).  The physical
meaning of $F(t)$ as far as D-branes are concerned
is as follows: $\Pi_2=t$ denotes the (complexified) volume of
D2-brane wrapping
$\CP^1$, whereas 
\beq
\Pi_4={\partial F\over \partial t} ,
\eeq
 which is the dual
period, corresponds to the complexified quantum volume of a non-compact
D4 brane intersecting the $\CP^1$ at one point.  The ambiguity
in the coefficients of $P_3(t)$ reflects the infinity of the
volume of this D4 brane.  

{}From the discussion of the BPS masses of D-branes, and
the mirror for this model, which we found in the form of
the LG model
\beq 
W=X_2+X_3+X_4+e^t X_3 X_4/X_2
\eeq
we are led to consider the appropriate periods given by
integrals
\beq
\Pi=\int {dX_2dX_3 dX_4\over X_2X_3X_4} e^{-W}
\eeq
where the measure is obtained by noting that the correct
variables are the $Y_i$'s.  It is more convenient to consider
$\partial \Pi /\partial t$ and integrating over $X_4$:
\beq
{ \partial \Pi \over \partial t}=e^t \int {dX_2 dX_3dX_4\over X_2^2}
e^{-(X_2+X_3+X_4+e^t X_3 X_4/X_2)}=e^t\int {dX_2 dX_3\over X_2^2}
\delta(1+e^t{X_3\over X_2})e^{(-X_2-X_3)}
\eeq
performing the integral over $X_3$ we have
\beq
{ \partial \Pi \over \partial t}=\int {dX_2\over X_2}e^{-X_2(1-e^{-t})}
\eeq
There are two cycle one can consider here: Integrating around the
origin of $X_2$ we obtain 1.  This corresponds to the D2 brane
BPS mass, namely $\partial \Pi_2/\partial t =1$. The other period
corresponds to integrating from $0$ to $\infty$ on the $X_2$ plane,
which is the dual cycle corresponding to D4 brane.  To do that consider
taking another derivative of $\Pi$:
\beq
{\partial ^2 \Pi\over \partial t^2}={\partial^3F \over
\partial t^3 }=-e^{-t} \int dX_2 e^{-X_2(1-e^{-t})}=-{e^{-t}\over (1-e^{-t})}
\eeq
In agreement with the known result for $F(t)$ (\ref{prepno}).

\subsection{Compact versus Non-Compact}

As we discussed in section 4 compact complete intersections
in toric varieties are closely related to the corresponding
non-compact ones.  Moreover all the quantities that are unaffected
by the $cc$ deformations in the compact theory, should be
computable in the non-compact theory.  This is true at the
level of embedding of twisted chiral operators of the compact
theory in the non-compact one and the example is
the relation (\ref{MPrelation}) by \cite{MP} presented in section 4.  
The same should be true about the overlap of states with
B-type boundary condition with the ground state vacua, which
according to \cite{OOY} is independent of $cc$ deformations.

{}From the above discussion of periods, one natural
state to consider is the one corresponding to the
identity operator in the context of the compact theory.  As discussed
in section 4 this corresponds to the state 
$|1\rangle_{Compact}\leftrightarrow|\delta \rangle_{Non-Compact}$
in the non-compact theory.
We now consider the pairing between B-type boundary
states and the ground states (represented using A-model)
in the sigma model.  The B-type boundary states
in the non-compact theory $\langle \widetilde{\gamma}_i|$
will flow to some boundary
states in the compact theory $\langle \gamma_i|$.  By the $\epsilon$
independence we thus have
\beq
\langle \gamma_i|...\rangle_{Compact}=\langle{\tilde \gamma_i}|...
|\delta_1...\delta_k
\rangle_{Non-Compact}
\eeq
In particular we learn that the BPS mass (which corresponds
to the overlap of the state associated to identity operator
with D-brane boundary state) for the compact theory is given by
\beq
\Pi_{i,1}=\langle \gamma_i|1\rangle_{Compact}=
\langle {\tilde \gamma_i}|\delta_1...
\delta_k \rangle_{Non-Compact}
\eeq
Since we have found the mirror of non-compact theory
in terms of an LG model, from our previous discussion
it follows that
\beq
\Pi_{i,1}=\int_{{\tilde \gamma_i}}\delta_1...\delta_k
{\rm exp}(-W)
\label{funda}
\eeq
Note that this result, for the case of linear sigma models 
corresponding to degree $d$ hypersurfaces in toric varieties
described
by
a single $U(1)$ gauge theory can also be written as
\beq
\Pi_{Compact}=d{\partial\over \partial t}\Pi_{Non-Compact}
\label{vfund}
\eeq
This result is well known in the context of local mirror
symmetry for Calabi-Yau manifolds \cite{kkv,kmv,klyz}
and should be viewed as a generalization of it.
The equations (\ref{funda}) and
its specialization (\ref{vfund}) will be the fundamental results
 we need in completing
our discussion for deriving the mirror of hypersurfaces
(and complete intersections) in toric varieties.

Note that using the fact that the periods of the non-compact
theory satisfy appropriate Picard-Fuchs equations, as derived
in (\ref{picca}) we can use the above result (\ref{vfund})
to write a Picard-Fuchs equation satisfied by the periods of the
mirror for the compact case.  
  
\section{Mirror Symmetry for Complete Intersections}

In this section we complete our derivation of the mirror
theory corresponding to complete intersections in
toric varieties.  First we discuss the simple case
of degree $d$ hypersurfaces in $\CP^{N-1}$ to illustrate
how the idea works.  Then we show
how a subset of complete intersection sigma
models have a similar mirror in the form
of a simple Landau-Ginzburg orbifolds in flat
space.  We then show that the most
general construction can also be described in terms
of LG theory on non-compact Calabi-Yau manifolds.

\subsection{Hypersurfaces in $\CP^{N-1}$}

Consider a degree $d$ hypersurface in $\CP^{N-1}$.
As discussed before, all the ring structure and BPS
masses can be computed in the associated non-compact
theory.  We have $N+1$ dual matter fields $Y_i$ with $i=1,..,N$
and $Y_P$ and one $\Sigma$ field. From  (\ref{vfund}) we 
can compute the BPS masses for the compact theory in the form
\beqa
\Pi&=&
d{\dd\over \dd t}\int\dd \Sigma\,\dd Y_P\,
\prod_{i=1}^N\dd Y_i\,
\exp\left(-\widetilde{W}\right)
\nonumber\\
&=&
d{\dd\over \dd t}\int\dd Y_P\,
\prod_{i=1}^N\dd Y_i\,
\delta\left(\sum_{i=1}^NY_i-dY_P-t\right)\,
\exp\left(-\sum_{i=1}^N\e^{-Y_i}-\e^{-Y_P}\right).
\eeqa
Constraint $\sum_iY_i-dY_P=t$ can be solved, as discussed earlier,
by
\beqa
&&\e^{-Y_i}\,=\,X_i^d,\\
&&\e^{-Y_P}\,=\,\e^{t/d}X_1\cdots X_N.
\eeqa
The map from $X_i$ to $\e^{-Y_i}$ and $\e^{-Y_P}$ is
one-to-one up to the action of
$(\Z_d)^{N-1}$ on $X_i$ defined by
\beq
X_i\mapsto \omega_iX_i,~~~\omega_i^d=1,\,~\omega_1\cdots\omega_N=1.
\label{Gamm}
\eeq
Then, we obtain
\beqa
\Pi&=&
d{\dd\over \dd t}\int
\prod_{i=1}^N{\dd X_i\over X_i}\,
\exp\left(-\sum_{i=1}^NX_i^d-\e^{t/d}\prod_{i=1}^NX_i\right)
\nonumber\\
&=&
\e^{t/d}\int
\prod_{i=1}^N\dd X_i\exp
\left(-\sum_{i=1}^NX_i^d-\e^{t/d}\prod_{i=1}^NX_i\right).
\eeqa
This is the period for Landau-Ginzburg model with superpotential
\beq
W_{LG}=X_1^d+\cdots+X_N^d+\e^{t/d}X_1\cdots X_N,
\label{WLG}
\eeq
or more precisely the LG orbifold by the $(\Z_d)^{N-1}$ action
(\ref{Gamm}).  What we see here is that the compact
theory and non-compact theory both have the same Landau-Ginzburg
potential, but the measure for the fundamental fields,
which determine the measure are different in the two cases.
In other words the LG orbifold with fields $X_i$ as the fundamental
fields (rather than $Y_i$ as in the
non-compact case) has the same ring structure and BPS masses as the
degree $d$  hypersurfaces in $\CP^{N-1}$.

Let us examine the vacua of this LG orbifold.
The equation $\dd W_{LG}=0$ has
$(N-d)$ solutions at
\beq
X_1^d=\cdots=X_N^d=-{\e^{t/d}\over d}X_1\cdots X_N=:S,~~~~
S^{N-d}=(-d)^d\e^{-t}
\label{massivac}
\eeq
and, for $d\geq 2$, one solution at
\beq
X_1=\cdots =X_N=0.
\eeq
The $(N-d)$ critical points (\ref{massivac}) are all non-degenerate
and correspond to massive vacua that break spontaneously
the $\Z_{2(N-d)}$ axial R-symmetry to $\Z_2$.
The critical point at $X_i=0$ is degenerate for $d> 2$
and corresponds to
a non-trivial fixed point.
The behaviour of the theory in the IR limit depends on
the relation of $d$ and $N$, as follows.

\noindent
$\bullet$
For $d=1$ where $M=\CP^{N-2}$, there are only $N-1$ massive
vacua.
If we integrate out $X_N$, we obtain the constraint
$X_{N-1}=-\e^{-t}/(X_1\cdots X_{N-2})$ and the effective
superpotential
for the remaining fields is
\beq
W=X_1+\cdots+X_{N-2}-\e^{-t}/(X_1\cdots X_{N-2}).
\eeq
This is nothing but the $A_{N-2}$ affine Toda superpotential.
Thus, we have reproduced the mirror symmetry of
$\CP^{N-2}$ model and affine Toda theory.

\noindent
$\bullet$
For $2\leq d<N$ where $M$ has positive first Chern class
and the sigma model is asymptotically free,
there are $(N-d)$ massive vacua and a vacuum at $X_i=0$.
At $X_i=0$,
the last term $\e^{t/d}X_1\cdots X_N$ of the potential (\ref{WLG})
is irrelevant compared to the first $N$-terms.
This can be seen by computing its dimension
in the superconformal field theory we will study below.
(More intuitively, this is because $t$
flows at low energies to large negative values.)
Thus, the theory at $X_i=0$ corresponds to the
$(\Z_d)^{N-1}$-orbifold of the LG model
with the quasi-homogeneous superpotential
\beq
W_{IR}=X_1^d+\cdots+X_N^d.
\label{WIR}
\eeq
In particular, we see the enhancement of the
axial $\Z_{2(N-d)}$ R-symmetry to
$U(1)_A$ symmetry.
For $d=2$ the critical point at $X_i=0$ is non-degenerate
and does not corresponds to a non-trivial fixed point.
However, because of the orbifolding, it could correspond to
multiple vacua.

\noindent
$\bullet$
For $d=N$ where $M$ is a Calabi-Yau manifold,
there are no massive vacua but one massless vacuum at $X_i=0$.
At $X_i=0$, the last term of (\ref{WLG})
is equally relevant compared to the first $N$ terms.
Thus $t$ remains as the parameter
describing a marginal deformation of the SCFT.

\noindent
$\bullet$
For $d>N$ where $M$ has negative first Chern class,
there are $(d-N)$ massive vacua and one massless vacuum at $X_i=0$.
At $X_i=0$,
the first $N$ terms of (\ref{WLG})
are irrelevant compared to the last term.
Thus, the IR fixed point is described by the
theory with superpotential
\beq
W_{IR}=\e^{t/d}X_1\cdots X_N.
\eeq
The vacuum equation $\dd W_{IR}=0$ is solved if two of $X_i$'s
vanish. Namely, the theory is a free SCFT on $\C^{N-2}$.
This is expected since if $c_1(M)<0$ the sigma model is
IR free.

Thus, we have seen that the LG orbifold
with the superpotential (\ref{WLG}) and group $(\Z_d)^{N-1}$
captures the physics
of the sigma model for all values of $d$.
Also, it
describes {\it both} the massive vacua {\it and} the massless vacuum
that flows to a non-trivial (or trivial) IR fixed point.
Below, we shall analyze the spectrum of the SCFT
with the superpotential (\ref{WIR})
that appears as the non-trivial
fixed point for hypersurfaces with $c_1>0$.

\newcommand{\bq}{\overline{q}}

\subsubsection{The Spectrum of Chiral Primary Fields for Hypersurfaces
in $\CP^{N-1}$}

We can use the method developed in \cite{LGO,IV} to analyze the
spectrum of the chiral primary fields.
We are considering the LG model with the superpotential
\beq
W=X_1^d+\cdots +X_N^d
\eeq
divided by the orbifold group $(\Z_d)^{N-1}$
acting on $X_i$'s as 
$X_i\mapsto \omega^{\alpha_i}X_i$ where
\beq
\omega=\e^{2\pi i/d},~~~\sum_{i=1}^N\alpha_i=0~~({\rm mod}~~d).
\eeq
Recall that we are in the non-standard convention of chirality
as the LG model;
the fields $X_i$
are {\it twisted chiral} superfields
of R-charge $-q_i=\bq_i=1/d$.\footnote{$q$ and $\bq$ 
are the left/right R-charges related to
the vector/axial R-charges by $q_V=\bq+q$ and $q_A=\bq-q$.}
The model before orbifolding has only
the $ac$ ring given by
\beq
{\cal R}=\C[X_1,\ldots,X_N]/\partial_iW
=\bigoplus_{0\leq p_i\leq d-2}\C~ X_1^{p_1}\cdots
X_N^{p_N},
\label{unorb}
\eeq
and the central charge is $c/3=\sum_i(1-2/d)=N(d-2)/d$.

\subsection*{\it Supersymmetric Ground States}

The supersymmetric (or Ramond)
ground states of the $(\Z_d)^{N-1}$ orbifold
come both from untwisted and twisted sectors.

The Ramond ground states in the untwisted sector are build from
the canonical ground state $|0\rangle^1_R$ of R-charge
$\bq=-q=-c/6=-N(d-2)/2d$ by multiplying the $(\Z_d)^{N-1}$ invariant
elements of (\ref{unorb}).
Obviously they are the powers of $X_1\cdots X_N$.
Thus, there are $d-1$ Ramond ground states in the untwisted sector;
\beq
(X_1\cdots X_N)^p|0\rangle^1_R,~~~p=0,1,\ldots,d-2.
\eeq
The R-charge is just the sum of $-c/6$ and $Np(1/d)$;
\beq
\bq_p=-q_p=
{N\over d}\left(p-{d-2\over 2}\right).
\eeq

Let us next consider the $h$-twisted sector for
$h=(\omega^{\alpha_1},\ldots,\omega^{\alpha_N})\in (\Z_d)^{N-1}$.
The $h$-invariant fields are $X_i$ with $\omega^{\alpha_i}=1$.
There is a (not necessarily physical) ground state $|0\rangle_R^h$
in the sector on which $g\in(\Z_d)^{N-1}$ acts
as the multiplication by $\det g|_h$ where $g|_h$ is the action of
$g$ restricted on $h$-invariant fields.
($\det g|_h=1$ if there is no $h$-invariant field).
The Ramond ground states in the $h$-twisted sector are
$(\Z_d)^{N-1}$ invariant states of the form
$X_{i_1}^{p_1}\cdots X_{i_s}^{p_s}|0\rangle_R^h$
where $X_{i_a}$'s are $h$-invariant fields and $0\leq p_a\leq d-2$.
If $\omega^{\alpha_i}=1$ for some $i$ (say $i=1$), then
$\omega^{\alpha_j}\ne 1$ for some other $j$ (say $j=N$) if $h\ne 1$.
Then $X_1$ is $h$-invariant but $X_N$ is not. For
$g=(\omega,1,\ldots,1,\omega^{-1})$, we have $\det g|_h=\omega$.
But then $g$ acts on the state $X_1^{p_1}\cdots |0\rangle^h_R$
by multiplication by $\omega^{p_1+1}\ne 1$.
So there is no $(\Z_d)^{N-1}$-invariant state.
Thus, Ramond ground states exist only for $h$ such that
$\omega^{\alpha_i}\ne 1$ for all $i$.
For such an $h$, there is no $h$-invariant field and therefore there is a
unique Ramond ground state
\beq
|0\rangle^h_R.
\eeq
Its R-charge can be obtained from the formula (3.2) in \cite{IV}
\beq
\bq_h=q_h=
-\left({\sum_i\alpha_i\over d}-{N\over 2}\right).
\eeq
Since $\sum_i\alpha_i=0$ mod $d$, we have $\sum_i\alpha_i=n_hd$
where $n_h$ is an integer in the range
$N/d\leq n_h\leq N(1-1/d)$.
Thus, the vector R-charge
$q^V=\bq_h+q_h$ is an integer $N-2n_h$ in the range
\beq
|q^V_h|\leq c/3< N-2.
\label{qrange}
\eeq
Let us count the number of such $h\in (\Z_d)^{N-1}$.
Let ${\cal C}_N$ be the set of such $h$'s.
Let us consider a sequence
$(\omega^{\alpha_1},\ldots,\omega^{\alpha_{N-1}})$ such that
any entry is not equal to $1$. If $\alpha_*:=\sum\alpha_i$ is
equal to $0$ mod $d$, this
determines an element of ${\cal C}_{N-1}$. Otherwise this
determines a unique element
$(\omega^{\alpha_1},\ldots,\omega^{\alpha_{N-1}},\omega^{-\alpha_*})$
of ${\cal C}_N$. Thus we obtain the recursion relation
$(d-1)^{N-1}=\sharp({\cal C}_{N-1})+\sharp({\cal C}_N)$
which is solved by
\beq
\sharp({\cal C}_N)=(-1)^N{(1-d)^N-(1-d)\over d}.
\eeq
This is the total number of Ramond ground states from twisted sectors.

Let us compute the Witten index ${\rm Tr}(-1)^F$.
We normalize $(-1)^F=1$ on $|0\rangle^1_R$.
Then the ground states from untwisted sector has $(-1)^F=(-1)^N$
\cite{IV}.
Thus, the index of the SCFT is
\beq
{\rm Tr}_{{}_{\rm SCFT}}(-1)^F=(d-1)+{(1-d)^N-(1-d)\over d}
={(1-d)^N+d^2-1\over d}.
\eeq
Together with the $(N-d)$ massive vacua having $(-1)^F=1$,
we obtain the total index
\beq
{\rm Tr}(-1)^F=(N-d)+{(1-d)^N+d^2-1\over d}={(1-d)^N+Nd-1\over d}.
\eeq
This indeed agrees with the Euler number of the hypersurface $M$.

Note that our original sigma model possesses the unbroken
$U(1)_V$ R-symmetry and it counts $-p+q$ of the Hodge number
$(p,q)$ of the harmonic form representing a vacuum.
For a hypersurface in $\CP^{N-1}$ the off diagonal Hodge number is
non-zero only for $p+q=N-2$.
Thus, the ground state $|0\rangle_R^h$
 with non-zero $q^V_h=N-2n_h$
corresponds to a harmonic $(p,q)$-form with
$q-p=q_h^V$ and $p+q=N-2$.
In other words, the number of such $h$'s
for a fixed $(p,q)$ must be equal to the Hodge number
$h^{p,q}$.
This is indeed easy to check
explicitly for small values of 
$N$ and $d$ (e.g. using a formula for the Hodge numbers in \cite{Hirz}).
For illustration, let us present the result for two cases
$N=4$, $d=3$ and $N=5$, $d=3$:

\noindent
\underline{\it N=4, d=3}

$M$ is the cubic surface in $\CP^3$ which is
known as $E_6$ del Pezzo surface.
It has $h^{0,0}=h^{2,2}=1$, $h^{1,1}=7$ and the off-diagonal Hodge numbers
are all zero.
Thus, classically there are one ground
state with $(-q,\bq)=(\pm 1,\pm 1)$ and
seven ground states with $(-q,\bq)=(0,0)$.
In the quantum theory 
there is a single massive vacuum and a massless vacuum.
The massless vacuum
corresponds to a SCFT of $c/3=4/3$.
There are one untwisted ground state
with $(-q,\bq)=(\pm 2/3,\pm 2/3)$
and six twisted ground states with $(-q,\bq)=(0,0)$.

\noindent
\underline{\it N=5, d=3}

$M$ is a cubic hupersurface in $\CP^4$ which
has $h^{p,p}=1$ for $p=0,1,2,3$ and $h^{2,1}=h^{1,2}=5$.
Thus, classically there are one ground state
with $(-q,\bq)=(p-3/2,p-3/2)$ for $p=0,1,2,3$,
five with $(-q,\bq)=(1/2,-1/2)$ and five with $(-q,\bq)=(-1/2,1/2)$.
In the quantum theory, there are two massive vacua and a massless vacuum.
The massless vacuum corresponds to a SCFT with $c/3=5/3$.
There are one untwisted ground state with
$(-q,\bq)=(\pm 5/6,\pm 5/6)$, five twisted states
with $(-q,\bq)=(1/2,-1/2)$ and five twisted states with
$(-q,\bq)=(-1/2,1/2)$.

\subsection*{\it $ac$ Primaries}

Since all the vector R-charges are integral,
the $ac$ primary states are in one to one correspondence with
the Ramond ground states by spectral flow.
The spectral flow simply changes the R-charges by
$\bq_{\it ac}=\bq_R+c/6$ and $q_{\it ac}=q_R-c/6$.
Thus, we have the following $ac$ primary states with
R-charges: From the untwisted sector
\beq
(X_1\cdots X_N)^p|0\rangle_{\it ac}^1;~~~
\bq_p=-q_p={Np\over d},~~~~~~~~~~
\eeq
($p=0,\ldots,d-2$), and from the twisted sectors
\beq
|0\rangle_{\it ac}^h;~~~~~
\bq_h=q_h+c/3=N\left(1-{1\over d}\right)-n_h.
\eeq
where $n_h$ is the integer in the range $N/d\leq n_h\leq N(1-1/d)$
defined above.

\subsection*{\it $cc$ Primaries}

Since the axial R-charges are not necessarily integers,
we must separately consider $cc$ primary states.
The $cc$ primaries in the $h$-twisted sector
for
$h=(\omega^{\alpha_1},\ldots,\omega^{\alpha_N})$ (including $h=1$)
can be found in the same way as
the search for Ramond ground states
in $hj^{-1}$-twisted sector where $j=(\omega,\ldots,\omega)$
\cite{IV}.
There is a unique $cc$ primary
\beq
|0\rangle_{\it cc}^h,
\label{ccpr}
\eeq
for each $h$ such that
$\beta_i:=\alpha_i-1$ obey
$\sum_{i=1}^N\beta_i=-N$ and
$\beta_i\ne 0$ (${\rm mod}~d$).
One can choose $\beta_i$ in the range $1\leq \beta_i\leq d-1$
and then $\sum_i\beta_i=-N+m_hd$ where $m_h$ is an integer
in the range $2N/d\leq m_h\leq N$.
The R-charge of the $cc$ primary state is then
\beq
\bq=q={c\over 6}-\sum_{i=1}^N\left({\beta_i\over d}-{1\over 2}\right)
=N-m_h.
\label{RR1}
\eeq
If $N$ is divisible by $d$, there are extra $cc$ primaries
from the $h=(\omega,\ldots,\omega)$-twisted sector;
\beq
(X_1\cdots X_N)^p|0\rangle_{\it cc}^h,
~~~p=0,\ldots,d-2.
\eeq
which have R-charges
\beq
{\bq\choose q}={c\over 12}\mp{c\over 12}\pm {Np\over d}.
\label{RR2}
\eeq
The R-charges (\ref{RR1}) and (\ref{RR2}) are integers
in the range $0\leq q,\bq\leq c/6$.

\subsection*{\it Marginal Deformations}

The marginal deformation of the theory preserving the
$(2,2)$ superconformal symmetry
is done by an $ac$ primary field of $\bq=-q=1$
or a $cc$ primary field of $\bq=q=1$.

{}From the above list, it is easy to see that
there are no such $ac$ primaries except the special case
$N=6,d=3$ where twenty $n_h=3$ twisted 
states does correspond to the marginal deformation.
This is the case which corresponds to the conformal
field theory of $K3$ \cite{vafamirrvol}.

On the other hand, there are always
$cc$ primaries with $\bq=q=1$; they are the states (\ref{ccpr})
with $m_h=N-1$. This corresponds to $\alpha_i$'s
such that $\sum_i(d-\alpha_i)=d$ for $0\leq d-\alpha_i\leq d-2$.
The number of such $\alpha_i$'s is the same as the number of
independent polynomial deformations of degree $d$
equation in $N$ variables. As we will see below,
they correspond to the complex structure
deformations of the hypersurface $M$.

Incidentally
all the cases were $d$ divides $N$ gives rise
to a conformal field theory in the IR which is mirror to a Calabi-Yau
manifold.  Let $N/d=k$ be an integer.  Then we obtain
a conformal field theory corresponding to a Calabi-Yau
with complex dimension $(d-2)k$.  In fact it is
mirror to a Calabi-Yau which is an orbifold
of $k$ copies of degree $d$ hypersurface
in $\CP^{d-1}$.  For example, as noted in \cite{vafamirrvol}
the case $N=9,d=3$ is the mirror to a rigid Calabi-Yau
threefold.  For such cases we can now find a geometric
interpretation of the mirror:  {\it The
mirror of these rigid Calabi-Yau manifolds
can be viewed as the infrared
limit of sigma model on the corresponding hypersurface
with positive } $c_1$.

\subsubsection{Mirror Symmetry of Orbifold Minimal Models as IR Duality}

An LG description of the IR fixed point in this model
is actually available
also in the original linear sigma model
\cite{QFTM} (see also \cite{Eguchiet}).
This is the extension of the basic argument
for CY/LG correspondence \cite{phases} to the sigma model of $c_1>0$
hypersurfaces in $\CP^{N-1}$.

For $d<N$, the Kahler parameter $r$ flows at low energies to
large negative values $r\ll 0$.
There we find $(N-d)$ massive vacua
at large values of $\sigma$; $\sigma^{N-d}=\e^{-t}(-d)^d$.
Actually, there is also a vacuum at $\sigma=0$
but $|p|^2=|r|/d$. (Presumably the equation for $\sigma$
should be replaced by $\sigma^{N-1}=\e^{-t}(-d)^d\sigma^{d-1}$,
which is the chiral ring relation
for the sigma model on the hypersurface $M$
\cite{jinzenji}. This indeed has a solution at $\sigma=0$ for $d>1$.)
Since $p$ has electric charge $-d$,
the $U(1)$ gauge symmetry is broken to $\Z_d$ by the Higgs mechanism.
The superfield $P$ is massive while
other fields of unit charge
$\Phi_i$ are massless and become the relevant
fields to describe the low energy theory.
Because of the expectation value of $P$, there is a non-trivial
superpotential for $\Phi_i$'s
\beq
W=\sqrt{|r|/d}\left(\Phi_1^d+\cdots+\Phi_N^d\right).
\label{WitLG}
\eeq
Namely, the theory flows in the IR limit to
$(N-d)$ empty theories and the $\Z_d$ orbifold
of the LG model with the superpotential (\ref{WitLG}).
This is very similar to the structure found in our effective theory
(\ref{WLG}) except that the group is now a single $\Z_d$.
Since the two LG orbifolds,
one by $\Z_d$ and the other by $(\Z_d)^{N-1}$,
arise as IR fixed points of
the same theory, they must agree with each other.
Namely, the $\Z_d$-orbifold and the
$(\Z_d)^{N-1}$-orbifold 
must be mirror to each
other.\footnote{$\Phi_i$ are {\it chiral} superfields and
the model (\ref{WitLG}) has the
standard convention of chirality.}
This mirror symmetry is actually the special case of
the mirror symmetry between orbifolds of minimal models
mentioned in section 2.

It is straightforward to compute the spectrum of chiral primary
fields in the $\Z_d$ orbifold model.
The result is of course in agreement with the one
for the $(\Z_d)^{N-1}$ orbifold.
Untwisted sector states in the $\Z_d$ orbifold corresponds to
twisted sector states in the $(\Z_d)^{N-1}$ orbifold, and vice versa.
For instance,
the untwisted
$cc$ primary field $\Phi_1^{p_1}\cdots\Phi_N^{p_N}$
($0\leq p_i\leq d-2$)
corresponds to
the $h=(\omega^{\alpha_1},\ldots,\omega^{\alpha_N})$-twisted
$cc$ primary state (\ref{ccpr})
in the $(\Z_d)^{N-1}$ orbifold where
$p_i=d-\alpha_i$.
This is actually what is expected since $\Phi_i$ and
$\e^{-Y_i}=X_i^d$ are dual to each other and a momentum (power)
of one corresponds to a winding (twist) of the other.

The interpretation of the marginal deformation by $cc$ primaries 
is clear in this picture.
The marginal $cc$ deformation
corresponds to deformation of the LG model (\ref{WitLG})
by $\Phi_1^{p_1}\cdots\Phi_N^{p_N}$
with $\sum_ip_i=d$ and $0\leq p_i\leq d-2$.
This corresponds to the deformation of the superpotential
by $P\Phi_1^{p_1}\cdots\Phi_N^{p_N}$ in the original
linear sigma model for $M$ which is
nothing but the polynomial deformation of the defining equation of $M$.
Actually, any deformation of the complex structure
is of this form.\footnote{This can be seen as follows (we thank
R. Pandharipande for explanation).
Using the long exact sequence for
$0\rightarrow T_M\rightarrow T_{\CP^{N-1}}|_M\rightarrow {\cal O}_M(d)
\rightarrow 0$,
it is enough to show $H^1(M,T_{\CP^{N-1}}|_M)=0$.
{}From $0\to {\cal O}\to {\cal O}(1)^{\oplus N}\to T_{\CP^{N-1}}\to0$,
this reduces to $H^1(M,{\cal O}_M(1))=0$ and
$H^2(M,{\cal O}_M)=0$. These further reduce via
$0\to {\cal O}(-d)\to{\cal O}\to{\cal O}_M\to 0$
to $H^{1+i}(\CP^{N-1},{\cal O}(1-i))=0$
and $H^{2+i}(\CP^{N-1},{\cal O}(1-d-i))=0$ for $i=0,1$.
That this holds for $d<N$ is a standard fact
(e.g. by Kodaira-Nakano vanishing theorem \cite{GrH}).
Within $d\leq N$, the only case this fails
is $N=d=4$, the famous example $M=K3$.}
Thus, the $cc$ deformation in the sigma model on a $c_1>0$ projective
hypersurface is in one to one correspondence with
the complex structure deformation, as in the case of Calabi-Yau sigma
models. It would be interesting to better understand why
this holds in this case and investigate whether this is a general fact.

\subsection{Complete Intersections}

LG orbifold description of the mirror is possible also for a class of
complete intersections in toric varieties.
Let $X$ be a toric variety defined by the charge matrix
$Q_{ia}$ ($i=1,\ldots,N$, $a=1,\ldots,k$).
We consider a complete intersection $M$ in $X$ defined by
the equations $G_b=0$ ($b=1,\ldots,k$)
as many as
the number of the $U(1)$ gauge groups to define $X$.
Here $G_b$ is a polynomial of
$\Phi_i$ of a certain ``degree'' $d_{ba}$
(that is, $G_b$ has charge
$d_{ba}$ for the $a$-th $U(1)$ gauge group)
where we assume $d_{ba}$ to be an invertible matrix.

The dual of the non-compact theory 
is described by $N+k$ twisted chiral fields
$Y_i$ and $Y_{P_b}$ dual to $\Phi_i$ and $P_b$ and $k$
field strengths $\Sigma_a$.
It has the twisted superpotential
\beq
\widetilde{W}
=\sum_{a=1}^k\Sigma_a\left(\sum_{i=1}^NQ_{ia}Y_i
-\sum_{b=1}^kd_{b a}Y_{P_{b}}-t_a\right)
+\sum_{i=1}^N\,\e^{-Y_i}+\sum_{b=1}^k\,\e^{-Y_{P_{b}}}.
\eeq
The period integral relevant for the compact theory is
given by 
\beq
\Pi=\int\prod_{a=1}^k\dd\Sigma_a\,\prod_{i=1}^N\dd Y_i\,
\prod_{b=1}^k\dd Y_{P_{b}}\,\delta_1\cdots\delta_k
\,\exp\left(-\widetilde{W}\right),
\eeq
where
$\delta_{b}=\sum_{a=1}^kd_{b a}\Sigma_a$.
Then, the period is expressed as follows (a quick derivation
of this is given in the next subsection in more general cases);
\beqa
\Pi&=&
\int\prod_{i=1}^N\dd Y_i\,
\prod_{b=1}^k\dd Y_{P_{b}}\,\,
\e^{-Y_{P_1}}\cdots\e^{-Y_{P_k}}
\nonumber\\
&&\times \prod_{a=1}^k\delta\left(\sum_{i=1}^NQ_{ia}Y_i
-\sum_{b=1}^kd_{b a}Y_{P_{b}}-t_a\right)
\times\exp\left(-\sum_{i=1}^N\,\e^{-Y_i}-\sum_{b=1}^k\,\e^{-Y_{P_{b}}}
\right).
\eeqa
Suppose 
there are matrices $m_{ji}$, $n_{jb}$ of non-negative integers
such that
\beq
\sum_{i=1}^Nm_{ji}Q_{ia}=\sum_{b=1}^kn_{jb}\,d_{ba}.
\label{condQd}
\eeq
Then, the constraints for $Y_i$ and $Y_{P_b}$
can be solved as
\beq
\e^{-Y_i}=\prod_{j=1}^NX_j^{m_{ji}},~~~
\e^{-Y_{P_b}}=\e^{d^{-1}_{ab}t_a}
\prod_{j=1}^NX_j^{n_{bj}}.
\eeq
Furthermore, if $m_{ji}$ is an invertible matrix
and $\sum_bn_{jb}=1$ for each $j$ (thus $n_{jb}$ is either $0$ or $1$),
the period integral can be expressed as
\beq
\Pi=\int\prod_{i=1}^N\dd X_i\,\exp\left(-\widetilde{W}\right),
\eeq
for
\beq
\widetilde{W}=\sum_{i=1}^N\prod_{j=1}^NX_j^{m_{ji}}
+\sum_{b=1}^k\e^{d^{-1}_{ab}t_a}\prod_{j=1}^NX_j^{n_{jb}}.
\label{wwww}
\eeq
Thus, under the condition that
the charges $Q_{ia}$ and the degrees $d_{ba}$
admit such matrices
$m_{ji}$ and $n_{bj}$ as the solution to (\ref{condQd}),
the period of
the dual theory is the same as that of the LG orbifold
with the twisted chiral superpotential (\ref{wwww}).
The orbifold group $\widetilde{\Gamma}$ is generated by
$X_j\to\omega_jX_j$
where $\omega_j$ are phases satisfying
$\prod_{j}\omega_j^{m_{ji}}=1$
and $\prod_j\omega_j^{n_{jb}}=1$.

As in the example of hypersurfaces in $\CP^{N-1}$,
one might expect that the last $k$ terms in (\ref{wwww})
are irrelevant at low energies and the IR fixed point is described by
\beq
\widetilde{W}_{IR}=\sum_{i=1}^N\prod_{j=1}^NX_j^{m_{ji}}.
\label{WWIR}
\eeq
However, as we will see, an extra condition is required for this.
Intuitively, the last $k$ terms are irrelevant when
$d^{-1}_{ab}t_a\to -\infty$ at low energies.
The flow of $t_a$ is determined by
$\beta_a=\sum_{i=1}^NQ_{ia}-\sum_{b=1}^kd_{ba}$ as
$t_a(\mu)=\beta_a\log(\mu/\Lambda)$.
Then, the condition of irrelevance is
$d^{-1}_{ab}\beta_a>0$ or equivalently
\beq
\sum_{i,a}d^{-1}_{ab}Q_{ia}> 1.
\label{cccond}
\eeq
More precisely, let $2q_i$ be the axial charge of
$X_i$ so that each of the first $N$ terms in (\ref{wwww}) has charge $2$,
namely $\sum_{j=1}^Nm_{ji}q_j=1$.
Then, from (\ref{condQd}) we find
$\sum_{i=1}^NQ_{ia}=\sum_{b,j}q_jn_{jb}d_{ba}$.
Thus the condition (\ref{cccond}) means
$\sum_{j=1}^Nq_jn_{jb}>1$, which is nothing but
the irrelevance of $\prod_{j=1}^NX_j^{n_{jb}}$.
Thus, under the condition (\ref{cccond}), the IR fixed point
is described by the LG orbifold with superpotential (\ref{WWIR}).
The central charge of this model is
$c/3=N-2\sum_{i,j}m^{-1}_{ij}$.
What we have said remains true when the inequalities are
relaxed to allow equalities.
If an equality holds, the corresponding term in
(\ref{wwww}) is a marginal operator and should be kept.

In the original linear sigma model, on the other hand,
under the same condition
we can take $G_b$ to be
\beq
G_b=\sum_{j=1}^N n_{jb}\prod_{i=1}^N\Phi_i^{m_{ji}},
\eeq
where the sum is over $j$ such that $n_{jb}=1$.
The D-term equations for the chiral fields
are expressed as
\beq
\sum_{i,a}d^{-1}_{ab}Q_{ia}|\phi_i|^2-|p_b|^2
=\sum_{a}d^{-1}_{ab}t_a.
\label{PhiPt}
\eeq
At low energies, under the condition (\ref{cccond}),
the right hand side flows to large negative
values for all $b$.
Under the same condition, the coefficients of $|\phi_i|^2$
are all positive.
Then, the equation (\ref{PhiPt}) implies all $p_b$ are non-vanishing.
The vacuum equation also requires $\sum_{b=1}^kp_b\partial_iG_b=0$.
It follows from this that $\prod_i\Phi_i^{m_{ji}}=0$
for all $j$ and that
$\sum^{\prime}_{j_1,b}n_{j_1b}
p_b\prod_{j_2\ne i}\Phi_{j_2}^{m_{j_1j_2}}=0$
for all $i$
where the sum $\sum^{\prime}$ is over such $(j_1,b)$ that $m_{j_1i}=1$. 
We assume that this implies $\Phi_i=0$ for all $i$.
(We do not attempt to prove it here.
It is possible that in general an extra condition
is required.)
Then
the gauge group $U(1)^k$ is broken to its subgroup
$\Gamma$ (generated by $g_a$ such that $\prod_ag_a^{d_{ba}}=1$,
$\forall b$)
which is a discrete subgroup since $d_{ba}$ is invertible,
and the massless degrees of freedom
are $\Phi_i$'s only.
Thus, we expect that
the theory flows in the IR limit
to the $\Gamma$-orbifold of the LG model with the superpotential
\beq
W=\sum_{j,b}n_{jb}\langle p_{b}\rangle
\prod_{i=1}^N\Phi_i^{m_{ji}}
\eeq
where $\langle p_b\rangle$ is the expectation value
of the massive field $p_b$.
One can relax the inequality in (\ref{cccond})
to admit equality. If an equality holds, we can {\it choose}
the value of $t_a$'s such that $d^{-1}_{ab}t_a$ are all negative,
and the same conclusion holds.
The central charge of the model is
$c/3=N-2\sum_{i,j}m^{-1}_{ij}$, the same as the one for
(\ref{WWIR}).

Since the two LG orbifold models appear as the IR limit
of the same theory,
they must be equivalent, or mirror to each other.  The
equivalences of this type between LG models have
been noted before \cite{Schimmrigk,grpch}.

For illustration, let us present an example.
We consider a complete intersection in
$\CP^{N-1}\times\CP^{M-1}$
($N\geq M$) defined by two equations of bi-degree
$(d_1,0)$ and $(1,d_2)$ respectively.
It has a non-negative first Chern class if
$N\geq d_1+1$ and $M\geq d_2$.
The equations for the homogeneous coordinates
$S_i$ ($i=1,\ldots,N$) and $T_j$ ($j=1,\ldots,M$) are
\beq
G_1=\sum_{i=1}^NS_i^{d_1},~~~
G_2=\sum_{j=1}^M S_jT_j^{d_2}.
\eeq
Under the condition\footnote{Note
that this condition is stronger than the non-negativity of
the first Chern class.}
\beq
N\geq d_1+M/d_2,~~~ M\geq d_2,
\label{condNM}
\eeq
$d^{-1}_{ab}t_a$ are (or can be chosen) large negative
at low energies,
and we find $p_1$ and $p_2$ are non-vanishing.
Then, we can show in this case that
$\sum_{b=1,2}p_b\partial G_b=0$ implies
that all $S_i$ and $T_j$ must be vanishing.
Thus, we find an LG orbifold description of the low energy theory
where the superpotential is
\beq
W=\sum_{j=1}^M(S_j^{d_1}+S_jT_j^{d_2})+\sum_{k=M+1}^NS_k^{d_1},
\eeq
and the orbifold group is the subgroup
of $U(1)_1\times U(1)_2$
defined by $g_1^{d_1}=1$ and $g_1g_2^{d_2}=1$.
On the other hand, the dual theory is described by the
LG orbifold
for $N+M$ twisted chiral superfields $U_i$ and
$V_j$
with the twisted superpotential
\beq
\widetilde{W}=
\sum_{j=1}^M(U_j^{d_1}V_j
+V_j^{d_2})+\sum_{k=M+1}^NU_k^{d_1}
+\e^{{t_1\over d_1}-{t_2\over d_1d_2}}\prod_{i=1}^NU_i
+\e^{{t_2\over d_2}}\prod_{j=1}^MV_j,
\label{LGLG}
\eeq
and the group acting on fields as
\beqa
&V_j\to\gamma_jV_j,~~~j=1,\ldots,M,\\
&U_j\to \eta_jU_j,~~~j=1,\ldots,M,\\
&U_k\to \omega_kU_k,~~~k=M+1,\ldots,N,
\eeqa
where $\omega_k^{d_1}=1$, $\gamma_j^{d_2}=1$,
$\gamma_j\eta_j^{d_1}=1$, and
$\prod_k\omega_k\prod_j\eta_j=\prod_j\gamma_j=1$.
Under the condition (\ref{condNM}), the last two terms of
(\ref{LGLG}) are
marginal or irrelevant depending on whether the equality
holds or not.
The two LG orbifolds must be mirror to each other.
Indeed both have a central charge
$c/3=(1-1/d_1)(N+M-2M/d_2)$.

\subsection{General Mirror Description}

As noted before, for the most general case, what
we will find is that the mirror theory can be expressed
as an LG theory on a non-compact Calabi-Yau manifold.  
To see how this works we begin with the case already
discussed, i.e. hypersurfaces
of degree $d$ in $\CP^{N-1}$ and provide an alternative description
of the mirror.
This reformulation sets the stage for the most general description
which follows it.

As noted before, the relevant periods (i.e.
D-brane `masses') are given in this case by
\beq
\Pi=d\int\dd\Sigma\,\dd Y_P\prod_{i=1}^N\dd Y_i\,
\Sigma \,\exp\left(-\widetilde{W}\right)
\eeq
Since $d\Sigma$ is given by $\partial /\partial Y_P$
of the linear terms in $\widetilde{W}$ we have
\beqa
\Pi&=&\int\dd\Sigma\,
\prod_{i=1}^N\dd Y_i\,\dd Y_P
{\partial\over\partial Y_P}\left[
\exp\left(-\Sigma\Bigl(\sum_{i=1}^NY_i-d Y_P-t\Bigr)\right)\right]
\exp\left(-\sum_{i=1}^N\e^{-Y_i}-\e^{-Y_P}\right)
\nonumber\\
&=&\int \dd \Sigma\,
\prod_{i=1}^N\dd Y_i\,\dd Y_P\,\e^{-Y_P}\,
\exp\left(-\Sigma\Bigl(\sum_{i=1}^NY_i-d Y_P-t\Bigr)\right)
\exp\left(-\sum_{i=1}^N\e^{-Y_i}-\e^{-Y_P}\right)
\nonumber\\
&=&\int \prod_{i=1}^N\dd Y_i
\,\dd Y_P\,\e^{-Y_P}\,
\delta\left(\sum_{i=1}^NY_i-d Y_P -t\right)
\exp\left(-\sum_{i=1}^N\e^{-Y_i}-\e^{-Y_P}\right)
\eeqa
We make the following change of variables
\beqa
&&\e^{-Y_P}=\widetilde{P},\\
&&\e^{-Y_i}= \widetilde{P}U_i,~~\mbox{for}~~i=1,\ldots,d,\\
&&\e^{-Y_j}=\,U_j,~~~~\mbox{for}~~j=d=1,\ldots,N.
\eeqa
Then,
\beqa
\Pi&=&\int \prod_{i=1}^N{\dd U_i\over U_i}\,
\dd \widetilde{P}\,
\delta\left(\log\Bigl(\prod_{i=1}^NU_i\Bigr)+t\right)
\exp\left(-\widetilde{P}
\Bigl(\sum_{i=1}^dU_i+1\Bigr)-\sum_{i=d+1}^NU_i\right)
\nonumber\\
&=&\int\prod_{i=1}^N{\dd U_i\over U_i}\,
\delta\left(\log\Bigl(\prod_{i=1}^NU_i\Bigr)+t\right)
\delta\left(\sum_{i=1}^dU_i+1\right)
\exp\left(-\sum_{i=d+1}^NU_i\right).
\label{per}
\eeqa
Thus we have obtained a submanifold $\widetilde{M}^{\circ}$
of $(\C^{\times})^N$ defined by
\beqa
&&\prod_{i=1}^NU_i=\e^{-t},
\label{prodU}\\
&&\sum_{i=1}^dU_i+1=0.
\label{Eqn}
\eeqa
This is a non-compact manifold of dimension $N-2$.
The expression (\ref{per}) is identical to the period of
an LG model on $\widetilde{M}^{\circ}$ with superpotential
\beq
W_{\widetilde{M}^{\circ}}=\sum_{i=d+1}^N U_i.
\label{WtilM}
\eeq
This model is the mirror of the sigma model on $M$,
at least when twisted to topological field theory.

The special case is the case $d=N$ in which $M$ is a
compact Calabi-Yau manifold. In this case, the superpotential (\ref{WtilM})
is trivial and the mirror is simply the non-linear sigma model on
$\widetilde{M}^{\circ}$.
The mirror manifold $\widetilde{M}^{\circ}$
is actually an open subset of a Calabi-Yau manifold
$\widetilde{M}$ which is familiar to us.
That is,
the orbifold of the hypersurface in $\CP^{N-1}$
\beq
G(Z_1,\ldots,Z_N)=Z_1^N+\cdots+Z_N^N+\e^{t/N}Z_1\cdots Z_N=0,
\eeq
by the $(\Z_N)^{N-2}$ action given by
\beq
Z_i\mapsto \gamma_iZ_i,~~~\gamma_i^N=1,~~\gamma_1\cdots\gamma_N=1.
\eeq
To see this, we note that
\beq
U_i=\e^{-t/N}{Z_i^N\over Z_1\cdots Z_N}
\eeq
is invariant under the $\C^{\times}\times (\Z_N)^{N-2}$ action
and
solves the first equation (\ref{prodU}). The second equation
(\ref{Eqn}) becomes $G(Z_i)=0$.
If $Z_i$ and $Z^{\prime}_i$ yields the same $U_i$,
it is easy to see that $Z_i^N=Z_i^{\prime N}$ and
$Z_1\cdots Z_N=Z_1^{\prime}\cdots Z_N^{\prime}$ modulo $\C^{\times}$
action. Then, this means $Z_i=Z_i^{\prime}$ modulo
the $\C^{\times}\times(\Z_N)^{N-2}$ action.
Thus, the map from $[Z_i]$
to $U_i$ is one to one.

Under this identification,
we have
\beqa
\Pi&=&\int {1\over {\rm vol}(\C^{\times})}
\prod_{i=1}^N{\dd Z_i\over Z_i}\,
\delta\left({G(Z_1,\ldots,Z_N)\over Z_1\cdots Z_N}\right)
\nonumber\\
&=&\int \prod_{i=1}^{N-1}\dd Z_i\,
\delta\left(G(Z_1,\ldots,Z_{N-1},1)\right)
\nonumber\\
&=&\int \left(\prod_{i=1}^{N-2}\dd Z_i\left/
\left.{\partial G|_{Z_N=1}\over \partial Z_{N-1}}
\right|_{G=0}\right.\right)
\,=\,\int\Omega,
\eeqa
which is the period of the holomorphic differential of
the Calabi-Yau manifold $\widetilde{M}$.
Thus, we have shown that the A-twisted sigma model on $M$
is equivalent to the B-twisted sigma model on $\widetilde{M}$.

This tempts us to propose that the sigma model on $M$
and the LG model on $\widetilde{M}^{\circ}$ with the superpotential
$W_{\widetilde{M}^{\circ}}$ are mirror to each other
as (2,2) quantum field theories, not just as topological theories.
This is certainly true for $d=1$ case with
$M=\CP^{N-2}$ where
$\widetilde{M}^{\circ}$ is the algebraic torus $(\C^{\times})^{N-2}$
and $W_{\widetilde{M}^{\circ}}$ is the affine Toda superpotential.
However, for $d\geq 2$ we must partially compactify
$\widetilde{M}^{\circ}$ for the model to
be the mirror of the sigma model on $M$.
The reason is that the superpotential
$W_{\widetilde{M}^{\circ}}$ for $d>1$ has a run-away direction
which yields a continuous spectrum,
a property that we do not expect for a sigma model on a compact
smooth manifold $M$.
The typical case is $d=N$; the sigma model on a non-compact manifold
has a continuous spectrum.
As we have seen above,
$\widetilde{M}^{\circ}$ can indeed be compactified to
a compact CY manifold $\widetilde{M}$
and we can claim that the sigma model on $M$ is mirror to
the sigma model on $\widetilde{M}$.
For $2\leq d<N$, we must compactify only partially since
a non-trivial superpotential is
not allowed on a compact complex manifold.
In fact, this partial compactification can be found
as a simple generalization of the $d=N$ case.
Let us solve the first equation (\ref{prodU}) for $U_i$'s as follows
\beqa
&&U_i=\e^{-t/d}{Z_i^d\over Z_1\cdots Z_N},~~~
i=1,\ldots, d,\\
&&U_j=Z_j^d,~~~~~~~~~~~~~
j=d+1,\ldots,N.
\eeqa
As in the $d=N$ case one can see that the map from $Z_i$ to
$U_i$ is one to one modulo the $\C^{\times}\times (\Z_d)^{N-2}$ action
given by
\beqa
&&Z_i\mapsto \lambda\gamma_iZ_i,~~~
i=1,\ldots, d,\\
&&Z_j\mapsto \gamma_jZ_j,~~~
j=d+1,\ldots,N,
\eeqa
where $\lambda\in \C^{\times}$
and $\gamma_i^d=\gamma_j^d=1$ and $\gamma_1\cdots\gamma_N=1$.
The second equation (\ref{Eqn}) is then expressed as
\beq
Z_1^d+\cdots+Z_d^d+\e^{t/d}
Z_1\cdots Z_d\cdot Z_{d+1}\cdots Z_N
=0.
\eeq
This is the equation for a Calabi-Yau hypersurface in
$\CP^{d-1}$ with the $\psi$ parameter $\psi=\e^{t/d}(Z_{d+1}\cdots Z_N)$.
Thus, the manifold $\widetilde{M}^{\circ}$
is partially compactified to a manifold $\widetilde{M}$
which is the $(\Z_d)^{N-2}$ quotient of
the total space of the family of CY hypersurface in $\CP^{d-1}$
parametrized by $\C^{N-d}$ via $\psi=\e^{t/d}(Z_{d+1}\cdots Z_N)$.
Now the superpotential $W_{\widetilde{M}^{\circ}}$ (\ref{WtilM})
on $\widetilde{M}^{\circ}$
extends to $\widetilde{M}$ as
\beq
W_{\widetilde{M}}=Z_{d+1}^d+\cdots +Z_N^d.
\label{WMtil}
\eeq
Repeating what we have done in the $d=N$ case,
we can see that the period is expressed as
\beq
\Pi=\int \Omega_{d-2}\wedge \dd Z_{d+1}\wedge\cdots\wedge\dd Z_N\,
\exp\left(-W_{\widetilde{M}}\right),
\eeq
where $\Omega_{d-2}$ is the holomorphic $(d-2)$-form of
the CY hypersurface in $\CP^{d-1}$.

Now, the superpotential $W_{\widetilde{M}}$ on $\widetilde{M}$
has no run-away direction;
$\widetilde{M}$ includes the limiting points of the
run-away direction in $\widetilde{M}^{\circ}$.
In particular, we expect that the theory has a discrete spectrum.
Thus, we claim that the sigma model on $M$ is mirror to
the LG model on $\widetilde{M}$ with the
superpotential (\ref{WMtil}).

The superpotential (\ref{WMtil}) has $(N-d)$ non-degenerate
critical points at
\beqa
&&{Z_i^d\over Z_1\cdots Z_N}=-{\e^{t/d}\over d},~~~i=1,\ldots,d,\\
&&Z_j^d=U,~~~~~j=d+1,\ldots,N;
~~~~U^{N-d}=(-d)^d\e^{-t},
\eeqa
and a critical manifold at
\beq
Z_{d+1}=\cdots =Z_N=0,
\label{critz}
\eeq
which is the CY hypersurface $\sum_{i=1}^dZ_i^d=0$
of $\CP^{d-1}$.
For $d>2$, this critical CY manifold has dimension $>0$
and also the superpotential (\ref{WMtil}) is degenerate there.
Thus, we expect that the theory for $d>2$ flows in the IR limit to
a non-trivial fixed point.
This must be equivalent to the non-trivial fixed point
studied in section 7.1.
For $d=2$, the critical CY manifold is actually a point and
the superpotential is non-degenerate. Thus, we expect that the theory
has a mass gap for $d=2$. However, because of the orbifolding,
the critical point (\ref{critz}) may correspond to
multiple vacua. From the result of section 7.1.1 the actual number
of vacua there is
$2$ for even $N$ and $1$ for odd $N$.

\subsection*{\it Complete Intersection in Toric Variety}

Let $X$ be the toric variety defined by the charge matrix $Q_{ia}$
($i=1,\ldots,N$, $a=1,\ldots,k$).
We consider the submanifold $M$ of $X$ defined by
the equations
\beq
G_{\beta}=0,~~~~\beta=1,\ldots,l,
\eeq
where $G_{\beta}$ are polynomials of $\Phi_i$ of charge
$d_{\beta a}$ for the $a$-th $U(1)$ gauge group.
The sigma model on $M$
can be realized as the linear sigma model of gauge group $U(1)^k$
with chiral superfields $\Phi_i$ of charge $Q_{ia}$ and 
$P_{\beta}$ of charge $-d_{\beta a}$ which has a superpotential
\beq
W=\sum_{\beta=1}^lP_{\beta}G_{\beta}(\Phi).
\eeq
The theory without this superpotential is the same as the
sigma model on a non-compact
toric variety $V$ (defined by charge matrix $(Q_{ia},-d_{\beta a})$)
and has the dual description in terms of
the twisted chiral superfield
$\Sigma_a$, $Y_i$ and $Y_{P_{\beta}}$ where
$\Sigma_a$ is the field strength of the $a$-th gauge group and
$Y_i$ and $Y_{P_{\beta}}$
are the dual variables of $\Phi_i$ and $P_{\beta}$.
The dual theory has the twisted superpotential
\beq
\widetilde{W}
=\sum_{a=1}^k\Sigma_a\left(\sum_{i=1}^NQ_{ia}Y_i
-\sum_{\beta=1}^ld_{\beta a}Y_{P_{\beta}}-t_a\right)
+\sum_{i=1}^N\,\e^{-Y_i}+\sum_{\beta=1}^l\,\e^{-Y_{P_{\beta}}}.
\eeq

As noted before, the period integral for the compact theory
is given by
\beq
\Pi=\int\prod_{a=1}^k\dd\Sigma_a\,\prod_{i=1}^N\dd Y_i\,
\prod_{\beta=1}^l\dd Y_{P_{\beta}}\,\delta_1\cdots\delta_l
\,\exp\left(-\widetilde{W}\right),
\eeq
where
\beq
\delta_{\beta}=\sum_{a=1}^kd_{\beta a}\Sigma_a.
\eeq
We note that this can be expressed as
\beq
\delta_{\beta}=-{\partial\over\partial Y_{P_{\beta}}}
\sum_{a=1}^k\Sigma_a\left(\sum_{i=1}^NQ_{ia}Y_i
-\sum_{\beta=1}^ld_{\beta a}Y_{P_{\beta}}-t_a\right).
\eeq
Then, via partial integration we obtain
\beqa
\Pi&=&\int\prod_{a=1}^k\dd\Sigma_a\,\prod_{i=1}^N\dd Y_i\,
\prod_{\beta=1}^l\dd Y_{P_{\beta}}\,\,
\e^{-Y_{P_1}}\cdots\e^{-Y_{P_l}}
\nonumber\\
&&\times\exp\left(-
\sum_{a=1}^k\Sigma_a\left(\sum_{i=1}^NQ_{ia}Y_i
-\sum_{\beta=1}^ld_{\beta a}Y_{P_{\beta}}-t_a\right)\right)
\exp\left(-\sum_{i=1}^N\,\e^{-Y_i}-\sum_{\beta=1}^l\,\e^{-Y_{P_{\beta}}}
\right)
\nonumber\\
&=&\int\prod_{i=1}^N\dd Y_i\,
\prod_{\beta=1}^l\e^{-Y_{P_{\beta}}}\dd Y_{P_{\beta}}\,
\prod_{a=1}^k\delta\left(\sum_{i=1}^NQ_{ia}Y_i
-\sum_{\beta=1}^ld_{\beta a}Y_{P_{\beta}}-t_a\right)
\nonumber\\
&&\times
\exp\left(-\sum_{i=1}^N\,\e^{-Y_i}-\sum_{\beta=1}^l\,\e^{-Y_{P_{\beta}}}
\right).
\eeqa
This is the expression for the BPS mass for the most general toric
complete intersection.
Now let us consider the case where one can find
$n_{\beta}^i=0$ or
$1$ such that
\beq
\sum_{i=1}^Nn_{\beta}^i\,Q_{ia}=d_{\beta a}.
\eeq
If we make the following change of variables
\beqa
&&\e^{-Y_{P_{\beta}}}=\widetilde{P}_{\beta},\\
&&\e^{-Y_i}=U_i\prod_{\beta=1}^l\widetilde{P}_{\beta}^{n_{\beta}^i},
\eeqa
the period integral is expressed as
\beqa
\Pi\!\!&=&\!\!
\int\prod_{i=1}^N{\dd U_i\over U_i}
\prod_{\beta=1}^l\dd\widetilde{P}_{\beta}\,
\prod_{a=1}^k\delta\left(\log
\Bigl(\prod_{i=1}^NU_i^{Q_{ia}}\Bigr)+t_a\right)
\exp\left(-
\sum_{\beta=1}^l\widetilde{P}_{\beta}
\Bigl(\sum_{n_{\beta}^i=1}U_i+1\Bigr)
-\sum_{n_{\beta}^j=0,\forall\beta}U_j\right)
\nonumber\\
&=&
\int\prod_{i=1}^N{\dd U_i\over U_i}\,
\prod_{a=1}^k\delta\left(\log
\Bigl(\prod_{i=1}^NU_i^{Q_{ia}}\Bigr)+t_a\right)
\prod_{\beta=1}^l\delta\left(\sum_{n_{\beta}^i=1}U_i+1\right)
\times\exp\left(-\sum_{n_{\beta}^j=0,\forall\beta}U_j\right)
\label{peri}
\eeqa
Thus, we have obtained a submanifold $\widetilde{M}^{\circ}$
of $(\C^{\times})^N$ defined by
\beqa
&&\prod_{i=1}^NU_i^{Q_{ia}}=\e^{-t_a},
\label{mirrrr}\\
&&\sum_{n_{\beta}^i=1}U_i+1=0,
\label{sumUi}
\eeqa
where the sum in (\ref{sumUi}) is over $i$ such that $n_{\beta}^i=1$.
This is a non-compact manifold of dimension $(N-k-l)$
which is the same as the dimension of $M$.
The period (\ref{peri}) is identical to the period of an LG model on
$\widetilde{M}^{\circ}$ with superpotential
\beq
W_{\widetilde{M}^{\circ}}=\sum_{n_{\beta}^j=0,\forall\beta}U_j,
\label{supmir}
\eeq
where the sum here is over $j$ such that $n_{\beta}^j=0$ for all
$\beta$.
Thus, we have shown that the mirror of the sigma model on $M$ is given by
this LG model on $\widetilde{M}^{\circ}$, at least when twisted
to topological field theory.
The expression (\ref{peri}) of the period
on the mirror manifold (\ref{mirrrr})-(\ref{sumUi})
with the superpotential (\ref{supmir})
is equivalent to the one that appears in \cite{G}.

\section{Directions for Future Work}

In this paper we have seen how mirror symmetry, formulated
as a duality between pairs of 2d QFT's can be proven using
rather simple physical ideas.  Not only have we
recovered the known formulation for mirror symmetry,
but we have also generalized it to classes which were
not known before.  In particular, we have shown
that a theory and the mirror will give rise to the
same ``BPS mass'' for the D-branes. 

It is natural to ask
to what extent one can derive dualities between QFT's
in higher dimensions.  One idea along this line
is their reduction to 2 dimensions.  For example,
it was shown in \cite{bjsv} how certain results about 4
dimensional $N=2$ Yang-Mills
theory compactified on a Riemann surface to 2 dimensions
can be related to sigma models, and computed via mirror
symmetry.  Given that we have now understood how mirror
symmetry can be derived, we can go back and ask to what extent
all the quantum corrections of $N=2$ theories can be understood
using results about 2d QFT's. It would not be surprising
if all the BPS data \cite{SW}, which is all we know at the present
about the $N=2$ theories in 4 dimensions, can be derived
in this way.  This in fact fits with the idea that in supersymmetric
theories certain exact
results can be computed rather easily.  For example 
computation of Witten's index
can be reduced to an ordinary integral.  It is thus
conceivable that the computations related to exact BPS
aspects for $N=2$ theories in 4 dimensions can be related
to derivable facts about 2 dimensional quantum field theories.
This program, we believe, is an interesting one to investigate.

At least in one case it already seems to work:  Consider pure
$N=1$ supersymmetric Yang-Mills theory in 4 dimensions.
It is known that upon compactification to 3 dimensions one
obtains an effective theory captured by a superpotential
of affine Toda type \cite{SW3d,kva,vatod,horet}.  The same superpotential
survives in the compactification down to 2 dimensions.  This
can now be derived in our setup as follows:  Consider compactification
of the four dimensional theory on $T^2$, down to two dimensions.
The effective theory in two dimensions will be a sigma model
whose target space is the moduli space of flat connection
of the corresponding group on $T^2$.  These spaces
have been studied \cite{Loujenga} and is found to be
the weighted projective space
$\C{\rm WP}^r_{(1,g_1,\ldots,g_r)}$
where $r$ is the rank of the gauge group and $g_1,\ldots,g_r$
are its coroot integers.
The $(2,2)$ sigma model on this space is realized as a $U(1)$
gauge theory with chiral matter fields of charge $(1,g_1,\ldots,g_r)$. 
Using the mirror symmetry results of this paper, we obtain
the mirror affine Toda potential
\beq
\widetilde{W}=\Lambda\left(\e^{\Theta_1}+\cdots +\e^{\Theta_r}
+\e^{-\sum_{i=1}^rg_i\Theta_i}\right),
\eeq
in agreement with
\cite{SW3d,kva,vatod,horet}.
It would be interesting to extend these results to other
$N=1$ theories in four dimensions\footnote{For another
relation between moduli of flat connections on $T^2$ and
mirror symmetry, see \cite{Mayrf}.}.

Another aspect of our work which may find further applications
is the issue of the meeting of Higgs and Coulomb branch
in 2 dimensional gauge theories.  This issue has
been studied in \cite{higgsc}.  The dual of gauge theory
coupled to matter that we have found in this paper
seems potentially useful for studying such questions.
In particular the dual of the matter fields (whose vevs
mark the Higgs branch) and the scalar in the vector
multiplet (whose vev marks the Coulomb branch) are
in the same type of multiplet and the superpotential
we have found captures certain aspects
of the gauge dynamics involving these fields.  Another
application of our work may be to the large $N$
behavior of Chern-Simons theory on $S^3$, which has
been conjectured to be dual to topological string on 
${\cal O}(-1)+{\cal O}(-1)$ over $\CP^1$ \cite{gopav}.

In appendix A we have presented a conjectured
generalization of the duality in (2,2) supersymmetric
gauge theories in 2 dimensions for non-abelian groups,
motivated by considering a generic point on the Coulomb
branch of these theories.  It would be interesting
to check the validity of this conjecture.  This conjecture
leads to the computation of BPS structure for
complete intersections defined on arbitrary flag varieties.

Certain aspects of our work may also be relevant for the
coupling of topological sigma model to topological gravity.
In fact, it is tempting to conjecture that some matrix version
of the toda theories we have found would be relevant for that
reformulation.  This would be interesting to study further.

\section*{Acknowledgement}

We would like to thank K. Intriligator, A. Iqbal, S. Katz,
A. Klemm, B. Lian, D. Morrison,
H. Ooguri, R. Pandharipande, R. Plesser, N. Seiberg, R. Thomas
and S.T. Yau
for valuable discussions. 
K.H. would like to thank the Institute for Advanced Study, Princeton and
Duke University where a part of this work was carried out.
The research of K.H. is supported in
part by NSF-DMS 9709694.  The research of C.V.
is supported in part by NSF grant PHY-98-02709.

\appendix{ Non-Abelian Gauge Theories and
Complete Intersections in Grassmannians--- A Conjecture}

In this paper we have been mainly dealing with linear
sigma models realized as complete
intersection in toric varieties.  We have used abelian
 gauge theories in analyzing them.
It is natural to ask if we can compute analogous
quantities for the case of complete intersections in
Grassmannians.  In fact this class can also be realized
as gauge theories as well \cite{phases}, but in this
case it will involve a $U(N)$ gauge theory coupled
to matter.  So we would need do know the analog
of the dual formulation for this gauge system.

Our argument for abelian gauge theory is not applicable
to non-abelain gauge theories. 
Here we attempt to make a conjecture for what the analog dual is,
which we motivate using the results already
obtained for the product of $U(1)$'s.  We also make some
checks for the validity of the conjecture.

We consider the $U(N)$ gauge theory
coupled to  matter in a (not necessarily
irreducible) representation ${\cal R}$.\footnote{
One can generalize this conjecture to arbitrary groups}.
The $U(N)$ gauge supermultiplet contains a complex scalar
in the adjoint representation of $U(N)$. The ``generic''\footnote{
This is why we cannot prove our conjecture:  The places
where some $\Sigma_i $ and $\Sigma_j$ are equal
yields a non-abelian unbroken group and we are assuming
that this does not cause any problems for our dualization.} 
configuration which survives in the infrared is
the one corresponding to the complex scalar
given by a diagonal matrix. This is the generic
point on the Coulomb branch of the theory.  
Let $\Sigma_i$ denote the fields corresponding
to these diagonal elements.  These are well defined
up to permutation.  In other words the invariant
fields involve symmetric polynomials in $\Sigma_i$.  
In the generic configuration for $\Sigma_i$ the theory becomes a
$U(1)^N/S_N$ gauge theory.  So we could then apply our
results for the product of $U(1)$ gauge theories to obtain the
dual, modulo taking into account the permutation action
on the groups.  There will be $dim{\cal R}$ fields $Y^{\alpha}$
obtained by dualizing the matter field in representation $R$.  Each
$Y_\alpha$ corresponds to a weight of $U(N)$ lattice, and we can
associate the corresponding $U(1)^N$ charges $Q_i^{\alpha}$
to them.  Consider the superpotential
\beq
W= \sum_i \Sigma_i (Q_i^{\alpha}Y^{\alpha}-t)+\sum_{\alpha}
e^{-Y_{\alpha}}
\eeq
where $t$ denotes the FI parameter for the $U(N)$ gauge theory.
This is almost what we propose as the dual, except
that we now have to consider Weyl invariant (i.e. $S_N$ invariant) combination
of fields.  The $S_N$ acts on $\Sigma_i$ by permutation,
as already noted.  It acts on $Y^{\alpha}$ by the permutation
induced on the weights of the representation ${\cal R}$
by the action of the Weyl group.  Clearly the above
action is invariant under the Weyl group action.  We then consider
the theory given by
\beq
W_{invariant}=W//S_N
\label{conjun}
\eeq
where by this we mean that the fundamental fields
of the theory are to be written in terms
of the $S_N$ invariant combinations (note that this
is not the same as orbifolding the theory).  

Concretely what this would mean in computing the periods
(i.e. D-brane masses) is as follows:  We will consider integrals
of the form
\beq
\int \prod_i d\Sigma_i  \prod_{\alpha}dY^{\alpha}
\prod_{i<j}(\Sigma_i-\Sigma_j)\ e^{-W}
\eeq
The insertion of $\Delta=\prod_{i<j}(\Sigma_i-\Sigma_j)$ is to
make the measure correspond to the symmetric measure, and
is reminiscent of the $\delta$ insertion discussed in the
context of getting hypersurfaces from non-compact
toric varieties.  Note that this also agrees with the
dimension count for the space.  Our proposed
dual has infrared degrees of freedom which is too big:
$dim{\cal R}-N$ whereas it should have had $dim{\cal R}-N^2$.
Insertion of $\Delta^2$
in the correlation functions (which
is equivalent to the insertion of $\Delta$ in the periods)
 changes the dimension count by $N(N-1)$, and makes up for the
 discrepancy.  This is very similar to how
the insertion of $\delta$'s in the non-compact toric
case was used to embed the compact cohomology ring
in the non-compact one.

The above periods can be easily computed
from the corresponding non-compact toric varieties:
The insertion of $\Delta$ is the same as the action
of $\prod_{i<j}(\partial/\partial t_i-\partial /\partial t_j)$
acting on the periods of the corresponding non-compact
toric case (and substituting $t_i=t_j=t$ at the end).  Similarly
the case of hypersurfaces in Grassmannians would be obtained
from the non-compact version of bundles over Grassmannians by
inclusions of extra insertions similar to $\delta$.  Also this
conjecture naturally extends to the case of flag manifolds
and complete intersections in them.

The above conjecture would be interesting to verify.
For the case of Grassmannian itself, this
conjecture gives the
correct ring structure \cite{CV}, which was
one of the motivations for the above conjecture.  It would
also be interesting to connect the above formulation
with the corresponding LG model proposed for the
Grassmannian \cite{EHY}
which was further generalized in
\cite{giventalflag,kimetal} for other homogeneous spaces
and complete intersections therein.

\appendix{Supersymmetry transformation}

\newcommand{\bepsilon}{\overline{\epsilon}}

We record here the supersymmetry transformation
of the vector and chiral multiplet fields
(in the Wess-Zumino gauge) \cite{phases,grassm}.
This is obtained by dimensional reduction of the formulae in
\cite{WB} for $N=1$ supersymmetry transformation in $3+1$
dimensions. The reduction is in $x^1,x^2$ directions
and the scalars in the
vector multiplet is defined as
$\sigma=(v_1-iv_2)$ and $\bsigma=(v_1+iv_2)$.
The time coordinate is still $x^0$ but we
rename the spacial coordinate $x^3$
as $x^1$.
(The normalization of vector multiplet fields
used in this paper differs from
the one in \cite{phases,grassm} by factors of $\sqrt{2}$:
if we denote the latter as $\Sigma^{\prime}$, $\sigma^{\prime}$, etc,
the relations are $\Sigma=\sqrt{2}\Sigma^{\prime}$,
$\sigma=\sqrt{2}\sigma^{\prime}$,
$\lambda_{\pm}=\sqrt{2}\lambda_{\pm}^{\prime}$,
$D=D^{\prime}$ and $F_{01}=F_{01}^{\prime}$.)

The four supersymmetry generators are combined as
\beq
\delta=\epsilon^{\alpha}Q_{\alpha}+\bepsilon_{\alpha}\bQ^{\alpha}=
\epsilon_+Q_--\epsilon_-Q_+-\bepsilon_+\bQ_-+\bepsilon_-\bQ_+,
\eeq
where $\epsilon^{\pm}$ and $\bepsilon^{\pm}$ are anti-commuting
spinorial parameters
($\epsilon^{\mp}=\pm\epsilon_{\pm}$ etc).
The transformation of the vector multiplet fields is
\beqa
&&\delta v_{\pm}=\sqrt{2}i\bepsilon_{\pm}\lambda_{\pm}
+\sqrt{2}i\epsilon_{\pm}\blambda_{\pm},
\nonumber\\
&&\delta\sigma=-\sqrt{2}i\bepsilon_+\lambda_-
-\sqrt{2}i\epsilon_-\blambda_+,
\nonumber\\
&&\delta\bsigma=-\sqrt{2}i\epsilon_+\blambda_-
-\sqrt{2}i\bepsilon_-\lambda_+,
\nonumber\\
&&\delta D={1\over \sqrt{2}}\Bigl(
-\bepsilon_+D_-\lambda_+-\bepsilon_-D_+\lambda_-
+\epsilon_+D_-\blambda_++\epsilon_-D_+\blambda_-\Bigr.
\nonumber\\
&&~~~~~~~~~~~~~\Bigl.+\epsilon_+[\sigma,\blambda_-]
+\epsilon_-[\bsigma,\blambda_+]
-\bepsilon_-[\sigma,\lambda_+]
-\bepsilon_+[\bsigma,\lambda_-]\Bigr),
\label{vectSUSY}\\
&&\delta\lambda_+=
\sqrt{2}i\epsilon_+(D+iF_{01}+{i\over 2}[\sigma,\bsigma])
+\sqrt{2}\epsilon_-D_+\bsigma,
\nonumber\\
&&\delta\lambda_-=
\sqrt{2}i\epsilon_-(D-iF_{01}-{i\over 2}[\sigma,\bsigma])
+\sqrt{2}\epsilon_+D_-\sigma,
\nonumber\\
&&\delta\blambda_+=
-\sqrt{2}i\bepsilon_+(D-iF_{01}+{i\over 2}[\sigma,\bsigma])
+\sqrt{2}\bepsilon_-D_+\sigma,
\nonumber\\
&&\delta\blambda_-=
-\sqrt{2}i\bepsilon_-(D+iF_{01}-{i\over 2}[\sigma,\bsigma])
+\sqrt{2}\bepsilon_+D_-\bsigma,
\nonumber
\eeqa
where $v_{\pm}=v_0\pm v_1$, $D_{\pm}=D_0\pm D_1$.
The transformation of the charged chiral multiplet fields
is given by
\beqa
&&\delta\phi=\sqrt{2}\epsilon_+\psi_--\sqrt{2}\epsilon_-\psi_+,
\nonumber\\
&&\delta\psi_+=\sqrt{2}i\bepsilon_-D_+\phi+\sqrt{2}\epsilon_+F
-\sqrt{2}\bepsilon_+\bsigma\phi,
\nonumber\\
&&\delta\psi_-=-\sqrt{2}i\bepsilon_+D_-\phi+\sqrt{2}\epsilon_-F
+\sqrt{2}\bepsilon_-\sigma\phi,
\label{chiralSUSY}\\
&&\delta F=
-\sqrt{2}i\bepsilon_+D_-\psi_+-\sqrt{2}i\bepsilon_-D_+\psi_-
\nonumber\\
&&~~~~~~~+\sqrt{2}(\bepsilon_+\bsigma\psi_-+\bepsilon_-\sigma\psi_+)
+\sqrt{2}i(\bepsilon_-\blambda_+-\bepsilon_+\blambda_-)\phi.
\nonumber
\eeqa

{}From (\ref{vectSUSY}) it is clear that turning on an expectation value of
the scalar component of the vector multiplet
$\sigma=\widetilde{m}$ and freezing the fluctuation
of the entire vector multiplet (by setting the coupling zero)
does not break any supersymmetry
provided $\widetilde{m}$ and its hermitian conjugate
$\overline{\widetilde{m}}$ commute with each other
\beq
[\widetilde{m},\overline{\widetilde{m}}]=0.
\eeq
Thus, twisted mass for an abelian subgroup of the flavor symmetry
preserves $(2,2)$ supersymmetry.
The same thing can be said also for holomorphic isometry of
a Kahler manifold (see the formulae (28)-(30) in \cite{BagW}).

\end{document}